\documentclass[fontsize=11pt, paper=a4, parskip=never]{scrartcl}

\usepackage{etoolbox}

\providetoggle{blackandwhite}
\togglefalse{blackandwhite}

\providetoggle{review}
\togglefalse{review}

\providetoggle{biblatex}
\toggletrue{biblatex}

\providebool{jinststyle} 
\ifbool{jinststyle}{
  \togglefalse{biblatex}
  \KOMAoptions{parskip=full}
  \usepackage{amsthm}
}{
}

\usepackage{tgheros} 
\usepackage{courier}
\usepackage{newtxtext,newtxmath} 

\usepackage[utf8]{inputenx}
\usepackage[T1]{fontenc}
\usepackage{textcomp}

\KOMAoptions{DIV=last}

\usepackage[top=2.5cm, bottom=3.5cm]{geometry}

\usepackage{xspace}

\usepackage[english]{babel}
\usepackage{csquotes}

\usepackage{amsmath}
\usepackage{amsfonts}
\usepackage[euler]{textgreek}

\usepackage{rotating}
\usepackage{float}
\floatstyle{plaintop}
\restylefloat{table}
\usepackage{multirow}
\usepackage{booktabs}

\usepackage{caption}
\usepackage{subcaption}
\usepackage{graphicx}
\usepackage{xcolor}

\usepackage{siunitx}
\DeclareSIUnit{\%}{\percent}
\DeclareSIUnit\ppm{\ensuremath{\mathrm{ppm}}}
\DeclareSIUnit\lightspeed{\ensuremath{c}}
\DeclareSIUnit\clight{\lightspeed}
\DeclareSIUnit\eVperc{\eV\per\clight}
\DeclareSIUnit\keVperc{\kilo\eVperc}
\DeclareSIUnit\MeVperc{\mega\eVperc}
\DeclareSIUnit\GeVperc{\giga\eVperc}
\DeclareSIUnit\TeVperc{\tera\eVperc}
\DeclareSIUnit\eVpercsq{\eV\per\clight\squared}
\DeclareSIUnit\keVpercsq{\kilo\eVpercsq}
\DeclareSIUnit\MeVpercsq{\mega\eVpercsq}
\DeclareSIUnit\GeVpercsq{\giga\eVpercsq}
\DeclareSIUnit\TeVpercsq{\tera\eVpercsq}

\definecolor{linkblue}{HTML}{0060BB}
\usepackage[
  colorlinks=true,
  urlcolor=blue,
  anchorcolor=blue,
  citecolor=blue,
  filecolor=blue,
  linkcolor=blue,
  menucolor=blue,
  pagecolor=blue,
  linktocpage=true,
  pdfa=true
]{hyperref}
\ifbool{jinststyle}{
}{
  \hypersetup{%
    breaklinks=true,
    unicode=true,
    naturalnames=true,
    colorlinks=false,
    linktocpage=true,
    linkcolor=linkblue,
    citecolor=linkblue,
    urlcolor=linkblue
  }
}
\usepackage[noabbrev]{cleveref}

\usepackage[shortcuts]{extdash}

\iftoggle{review}{
  \usepackage{lineno}
  \modulolinenumbers[5]
  \linenumbers
}{}

\ifbool{jinststyle}{
  \usepackage{jinstpub}
  \newcommand*{\myauth}[2][]{\author[#1]{#2,}}
  \newcommand*{\mylastauth}[2][]{\author[#1]{#2}}
  \newcommand*{%
    \myaffil}[3][]{\affiliation[#1]{#2,\\
                                    #3}
  }
}{
  \usepackage{authblk}

  \newcommand*{\myauth}[2][]{\author[#1]{#2}}
  \newcommand*{\mylastauth}[2][]{\author[#1]{#2}}
  \newcommand*{\myaffil}[3][]{\affil[#1]{#2, #3}}
  \providecommand{\affiliation}[2][]{\affil[#1]{#2}}
  \newcommand\note[2][]{%
  \if!#1!%
    \stepcounter{footnote}\footnotetext{#2}%
  \else%
    {\renewcommand\thefootnote{#1}%
    \footnotetext{#2}}%
  \fi%
  }
}

\iftoggle{biblatex}{
  \usepackage[backend=biber, style=numeric-comp, sorting=none]{biblatex}
  \addbibresource{main.bib}
  \ExecuteBibliographyOptions{%
    sorting=none,
    block=space,
    autolang=other,
    related=true,
    sortcites=false,
    abbreviate=true,
    giveninits=true,
    maxnames=3,
    minnames=2,
    alldates=year,
    eventdate=comp,
    urldate=comp,
    url=true,
    doi=true,
    eprint=true,
    isbn=false,
  }
  \NewBibliographyString{advisor,advisors,byadvisor}
  \DefineBibliographyStrings{english}{%
    page = \!,
    pages= \!,
    phdthesis = {doctoral thesis},
    advisor   = {superv.},
    advisors  = {superv.},
    byadvisor = {superv.\ by},
  }
  \DeclareDataInheritance{mvbook}{book}{\noinherit{volumes}\noinherit{url}}
  \AtEveryBibitem{%
    \clearfield{pagetotal}
    \ifentrytype{article}{\clearfield{number}}{}
    \iffieldundef{doi}{}{
      \ifentrytype{article}{\clearfield{url}}{}
      \ifentrytype{book}{\clearfield{url}}{}
      \ifentrytype{collection}{\clearfield{url}}{}
      \ifentrytype{incollection}{\clearfield{url}}{}
      \ifentrytype{inproceedings}{\clearfield{url}}{}
      \ifentrytype{report}{\clearfield{url}}{}
      \ifentrytype{thesis}{\clearfield{url}}{}
    }
    \ifentrytype{article}{%
      \iffieldundef{shortjournal}{}{
        \clearfield{journalsubtitle}%
        \savefield*{shortjournal}{\short}%
        \restorefield{journaltitle}{\short}%
      }
    }{}
  }
}{
}

\DeclareMathOperator\erf{erf}

\newcommand*{\dEdx}{\ensuremath{\mathrm{d}E/\mathrm{d}x}\xspace}
\newcommand*{\Xzero}{\ensuremath{X_0}\xspace}
\newcommand*{\chisquare}{\ensuremath{\chi^2}\xspace}

\setkomafont{disposition}{\normalfont\boldmath\bfseries}
\addtokomafont{title}{\large}
\setkomafont{subtitle}{\normalfont}

\title{Double\-/hit separation and \boldmath \dEdx resolution of a time projection chamber with GEM readout}

\newcommand{\myabstract}{%
    A time projection chamber (TPC) with micropattern gaseous detector (MPGD) readout is investigated as main tracking device of the International Large Detector (ILD) concept at the planned International Linear Collider (ILC).
    A prototype TPC equipped with a triple gas electron multiplier (GEM) readout has been built and operated in an electron test beam.
    The TPC was placed in a \SI{1}{\tesla} solenoidal field at the DESY II Test Beam Facility, which provides an electron beam up to \SI{6}{\GeVperc}.
    The performance of the readout modules, in particular the spatial point resolution, is determined and compared to earlier tests.
    New studies are presented with first results on the separation of close-by tracks and the capability of the system to measure the specific energy loss \dEdx.
    This is complemented by a simulation study on the optimization of the readout granularity to improve particle identification by \dEdx.
}
\newcommand{\mykeywords}{Time Projection Chambers~(TPC), Micropattern Gaseous Detectors~(MPGD), Gas Electron Multipliers~(GEM), International Linear Collider~(ILC), International Large Detector~(ILD)}

\ifbool{jinststyle}{
  \collaboration{%
    \includegraphics[height=17mm]{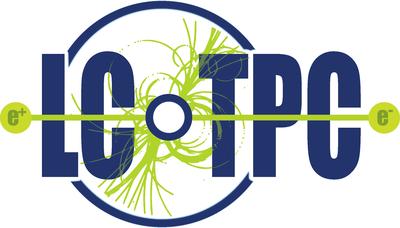}\\[6pt]
    The LCTPC Collaboration}
  \abstract{\myabstract}
  \toccontinuoustrue
}{
  \subtitle{\vspace{0.5\baselineskip}This document is a collaborative effort by the LCTPC Collaboration.}
  \date{}
}

\myauth[9]    {Y.~Aoki}
\myauth[2]    {D.~Atti\'e}
\myauth[4]    {T.~Behnke}
\myauth[1]    {A.~Bellerive}
\myauth[19]   {O.~Bezshyyko}
\myauth[2,32] {D.S.~Bhattacharya}
\myauth[18,24]{P.~Bhattacharya}
\myauth[18,*] {S.~Bhattacharya\note[*]{retired}}
\myauth[7]    {Y.~Chang}
\myauth[2]    {P.~Colas}
\myauth[6]    {G.~De~Lentdecker}
\myauth[4,30] {K.~Dehmelt}
\myauth[21]   {K.~Desch}
\myauth[4]    {R.~Diener} \ifbool{jinststyle}{\emailAdd{ralf.diener@desy.de}}{}
\myauth[1,20] {M.~Dixit}
\myauth[4,22]    {U.~Einhaus}
\myauth[4,22]    {O.~Fedorchuk}
\myauth[23]   {I.~Fleck}
\myauth[9]    {K.~Fujii}
\myauth[17]   {T.~Fusayasu}
\myauth[2]    {S.~Ganjour}
\myauth[17,28]{P.~Gros}
\myauth[1]    {P.~Hayman}
\myauth[17,26]   {K.~Ikematsu}
\myauth[13]   {L.~J\"onsson}
\myauth[21]   {J.~Kaminski}
\myauth[11]   {Y.~Kato}
\myauth[4]    {C.~Kleinwort}
\myauth[15]   {P.M.~Kluit}
\myauth[9]    {M.~Kobayashi}
\myauth[4,22,27]    {U.~Kr\"amer}
\myauth[13]   {B.~Lundberg}
\myauth[21]   {M.~Lupberger}
\myauth[18]   {N.~Majumdar}
\myauth[4,22]    {P.~Malek}
\myauth[9]    {T.~Matsuda}
\myauth[13]   {U.~Mj\"ornmark}
\myauth[18]   {S.~Mukhopadhyay}
\myauth[4]    {F.~M\"uller}
\myauth[4,25] {A.~M\"unnich}
\myauth[9]    {J.~Nakajima}
\myauth[8]    {S.~Narita}
\myauth[8]    {K.~Negishi}
\myauth[9]    {T.~Ogawa}
\myauth[13]   {A.~Oskarsson}
\myauth[13]   {L.~\"Osterman}
\myauth[3]    {D.~Peterson}
\myauth[7]    {H.~Qi}
\myauth[2]    {M.~Riallot}
\myauth[4]    {C.~Rosemann}
\myauth[16]   {S.~Roth}
\myauth[4]    {P.~Schade}
\myauth[4]    {O.~Sch\"afer}
\myauth[14]   {R.D.~Settles}
\myauth[23]   {A.N.~Shirazi}
\myauth[8]    {A.~Shoji}
\myauth[13]   {O.~Smirnova}
\myauth[17]   {A.~Sugiyama}
\myauth[5]    {T.~Takahashi}
\myauth[15]   {J.~Timmermans}
\myauth[2]    {M.~Titov}
\myauth[4]    {D.~Tsionou}
\myauth[4,31] {A.~Vauth}
\myauth[12]   {T.~Watanabe}
\myauth[23]   {U.~Werthenbach}
\myauth[4,29] {M.~Wu}
\myauth[6]    {Y.~Yang}
\myauth[6,26] {R.~Yonamine}
\myauth[7]    {Z.~Yuan}
\myauth[9]    {K.~Yumino}
\mylastauth[10]{F.~Zhang}

\myaffil[1] {Carleton~University, Department~of~Physics}{1125~Colonel~By~Drive, Ottawa, ON~K1S~5B6, Canada}
\myaffil[2] {CEA~Saclay, Institute~of~research~into~the~fundamental~laws~of~the~Universe~(IRFU)}{91191~Gif-sur-Yvette, France}
\myaffil[3] {Cornell~University, Laboratory~for~Elementary-Particle~Physics~(LEPP)}{Ithaca, NY~14853, USA}
\myaffil[4] {Deutsches~Elektronen-Synchrotron~DESY}{Notkestr.~85, 22607~Hamburg, Germany}
\myaffil[5] {Hiroshima~University, Graduate~School~of~Advanced~Sciences~of~Matter}{1-3-1~Kagamiyama, Higashi-Hiroshima, Hiroshima~739-8530, Japan}
\myaffil[6] {Inter~University~Institute~for~High~Energies~(ULB-VUB)}{Pleinlaan~2, 1050~Brussels, Belgium}
\myaffil[7] {Institute~of~High~Energy~Physics, Chinese~Academy~of~Sciences}{19B~Yuquan~Road, Shijingshan~District, Beijing~100049, P.R.~China}
\myaffil[8] {Iwate~University}{3-18-8~Ueda, Morioka, Iwate~020-8550, Japan}
\myaffil[9] {High~Energy~Accelerator~Research~Organization~(KEK)}{1-1~Oho, Tsukuba, Ibaraki~305-0801, Japan}
\myaffil[10]{Hubei~University~of~Technology}{28~Nanli~Road, Hong-shan~District, Wuchang, Wuhan, Hubei~Province~430068, P.R.~China}
\myaffil[11]{Kindai~University, Department~of~Physics}{3-4-1~Kowakae, Higashi-Osaka, Osaka~577-8502, Japan}
\myaffil[12]{Kogakuin~University, Laboratory~for~High~Energy~Physics}{2665-1~Nakano-machi, Hachioji, Tokyo~192-0015, Japan}
\myaffil[13]{Lund~University, Department~of~Physics, Div.~of~Particle~Physics}{Box~117, 221~00~Lund, Sweden}
\myaffil[14]{Max~Planck~Institute~for~Physics (Werner~Heisenberg~Institute)}{F\"ohringer~Ring~6, 80805~Munich, Germany}
\myaffil[15]{National~Institute~for~Subatomic~Physics~(Nikhef)}{P.O.~Box~41882, 1009~DB~Amsterdam, The~Netherlands}
\myaffil[16]{RWTH~Aachen~University, Department~of~Physics, Institute~III~B}{Otto-Blumenthal-Str., 52074~Aachen, Germany}
\myaffil[17]{Saga~University, Department~of~Physics}{1~Honjo-machi, Saga~840-8502, Japan}
\myaffil[18]{Saha~Institute~of~Nuclear~Physics, Applied~Nuclear~Physics~Division}{1/AF, Sector~1, Bidhan~Nagar, Kolkata~700064, India}
\myaffil[19]{Taras~Shevchenko~National~University~of~Kyiv}{64/13, Volodymyrska~Street, City~of~Kyiv~01601, Ukraine}
\myaffil[20]{TRIUMF}{4004~Wesbrook~Mall, Vancouver, BC~V6T~2A3, Canada}
\myaffil[21]{University~of~Bonn, Physikalisches~Institut}{Nu{\ss}allee~12, 53115~Bonn, Germany}
\myaffil[22]{University~of~Hamburg, Department~of~Physics, Institute~of~Experimental~Physics}{Luruper~Chaussee~149, 22761~Hamburg, Germany}
\myaffil[23]{University~of~Siegen, Department~of~Physics}{Emmy~Noether~Campus, Walter-Flex Str.~3, 57068~Siegen, Germany}
\myaffil[24]{\emph{now~at} Adamas~University, Department~of~Physics}{Adamas~Knowledge~City, Barasat-Barrackpore~Road, 24~Parganas North, Kolkata~700126, India}
\myaffil[25]{\emph{now~at} European~XFEL~GmbH}{Holzkoppel~4, 22869~Schenefeld, Germany}
\myaffil[26]{\emph{now~at} High~Energy~Accelerator~Research~Organization~(KEK)}{1-1~Oho, Tsukuba, Ibaraki~305-0801, Japan}
\myaffil[27]{\emph{now~at} National~Institute~for~Subatomic~Physics~(Nikhef)}{P.O.~Box~41882, 1009~DB~Amsterdam, The~Netherlands}
\myaffil[28]{\emph{now~at} Queen’s~University, Department~of~Physics,~Engineering~Physics~\&~Astronomy}{Stirling~Hall, 64~Bader~Lane, Kingston, Ontario~K7L~3N6, Canada}
\myaffil[29]{\emph{now~at} Radboud~University, Institute~for~Mathematics,~Astrophysics~and~Particle~Physics}{P.O.~Box~9010, 6500~GL~Nijmegen, The~Netherlands}                                                                                            
\myaffil[30]{\emph{now~at} State~University~of~New~York~at~Stony~Brook, Department~of~Physics~and~Astronomy}{100~Nicolls~Road, Stony~Brook, NY~11794-3800, USA}
\myaffil[31]{\emph{now~at} University~of~Hamburg, Department~of~Physics, Institute~of~Experimental~Physics}{Luruper~Chaussee~149, 22761~Hamburg, Germany}
\myaffil[32]{\emph{now~at} University~of~W\"urzburg, Institute~for~Theoretical~Physics~and~Astrophysics}{Am~Hubland, 97074~W\"urzburg, Germany}

\begin{document}

\maketitle

\ifbool{jinststyle}{}{
  \begin{abstract}
    \noindent \myabstract
  \end{abstract}

}

\flushbottom

\section{Introduction}
\label{sec:intro}

The LCTPC collaboration has built and operated a versatile prototype for the ILD TPC~\cite{ilc_tdr_detectors}.\footnote{See \url{https://www.lctpc.org} for more information about LCTPC.}
This prototype TPC, including the principles of its construction, its operation at a test beam and its performance, has been described in detail in~\cite{FMueller2017}.
In the present publication, results from a second test\-/beam campaign with an upgraded set of readout modules are presented.
The findings in the previous paper have been validated.
New studies are shown on measurements of the double\-/hit separation power and on the capabilities to do particle identification via a measurement of the specific energy loss \dEdx.

In \cref{sec:expsetup}, after a brief overview of the experimental setup, the updates in the experiment compared to the previous measurement campaign are discussed.
\Cref{sec:reco} gives an overview of the methods used in the reconstruction of the test\-/beam data.
Results from an analysis to re\-/establish the basic performance are shown in \cref{sec:tpcperformance}.
In \cref{sec:dhr} two new algorithms to optimize the double\-/hit separation are introduced.
In \cref{sec:dEdx_Res} a first determination of the resolution of the specific energy loss measurement is presented. 
These results are extrapolated to the proposed large TPC at the ILD and compared to theoretical expectations.
Furthermore, in a simulation study two approaches to measure the specific energy loss are compared, with a focus on their performance in dependence of the readout granularity.

\section{Experimental setup}
\label{sec:expsetup}

The TPC prototype consists of a light\-/weight field cage, providing an active gaseous volume of \SI{57}{\cm} length and \SI{72}{\cm} diameter, and an end plate which can house up to \num{7} identical readout modules, arranged in three staggered rows.
Each module covers the sector of and annulus with inner and outer radii of \SIlist{14.3;16}{\cm}, respectively, and an included angle of \ang{\sim 8.4}.
The prototype has been constructed such that it can be used to test the different readout technologies that are under discussion for use at the ILC\@.
Details about the construction of the prototype can be found in~\cite{Schade}.
The construction has been supported by the EUDET and AIDA projects.\footnote{For more information on EUDET and AIDA see \url{https://www.eudet.org/} and \url{https://aida2020.web.cern.ch/aida2020/node/283.html}, respectively.}

The TPC prototype is operated at the DESY~II Test Beam Facility~\cite{DIENER2019265}, which provides an electron beam with an adjustable momentum between \SI{1}{\GeVperc} and \SI{6}{\GeVperc}.
The facility hosts a superconducting solenoid magnet (PCMAG)~\cite{pcmag:magnet} that can provide a magnetic field of up to \SI{1}{\tesla}.
The dimensions of the prototype TPC have been chosen to exactly fit into the solenoid bore.
Before traversing the TPC, the electron beam has to pass through the magnet, which presents around \SI{20}{\percent} of a radiation length \Xzero.
A movable stage, specifically constructed for the demands of this type of research, can move the solenoid including the TPC horizontally and vertically perpendicular to the electron beam, and rotate the system relative to the beam around the vertical axis.

The measurements presented here were taken with the TPC prototype equipped with a triple GEM readout system.
The gas used was the so\-/called T2K gas mixture of \SI{95}{\percent} argon, \SI{3}{\percent} tetrafluoromethane ($\mathrm{CF}_4$), and \SI{2}{\percent} isobutane ($i\mathrm{C}_4\mathrm{H}_{10}$)~\cite{Abgrall201125}.
The chamber was operated at atmospheric pressure.
Unless stated otherwise, the presented measurements were taken with \SI{5}{\GeVperc} electron momentum, a magnetic field of \SI{1}{\tesla} and a drift field of \SI{240}{\V\per\cm}, which is close to the maximum of the electron drift velocity for this gas mixture.

\subsection{The GridGEM module}
\label{sec:module}

\begin{figure}
  \centering
  \includegraphics[
    width=\textwidth,
    height=0.25\textheight,
    keepaspectratio,
  ]{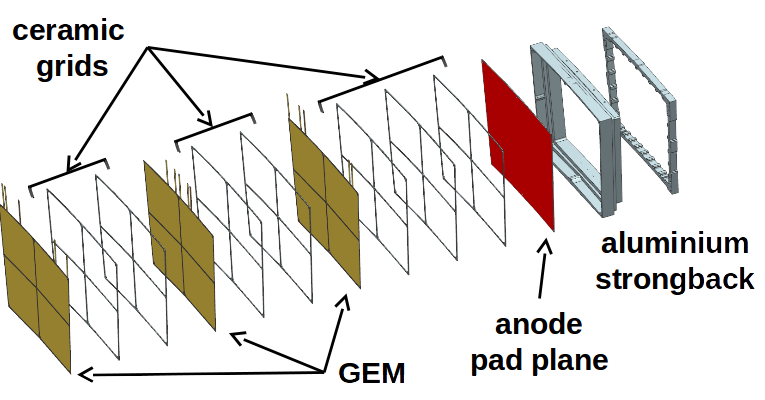}
  \caption{Explosion view of the module, showing the GEM foils, the ceramic mounting grids and the readout and back plane.}
  \label{fig:moduleexplosion}
\end{figure}

The defining characteristic of the system is the amplification by a triple GEM stack positioned on top of a finely segmented readout plane, as shown in \cref{fig:moduleexplosion}.
The GEM foils~\cite{GEM} are spaced and kept under tension by a system of thin ceramic frames.
The readout plane is segmented into pads with a pitch of \SI{1.26 x 5.85}{\mm}, arranged in \num{28} rows.
This results in a total of \num{4828} pads per module.
The readout electronics~\cite{LC-DET-2012-080} is based on the ALTRO ASIC~\cite{ALTROchip}, originally developed for the ALICE experiment at CERN and modified by the Lund University to be used for GEM readout.
The results presented in~\cite{FMueller2017} were based on a previous generation of GEM based readout modules.
Details on that module generation can be found in~\cite{Mueller:301339}.

When operating the first generation system, the following issues were identified and are subsequently addressed in the second generation system, presented here.
First, although the earlier system was operated stably in a test beam over several weeks, there was concern about potential weaknesses in its high voltage stability.
These were addressed by a number of measures.
The procedure to glue the GEMs on the frames was improved and the quality of the glue joint was closely monitored.
This ensures a continuous glue line with an average width of \SI{\sim0.5}{\mm} without interruptions.
The design of the interface between GEM foil and ceramic frame was changed such that there is always a minimum distance of \SI{0.15}{\mm} between a GEM hole and the ceramic frame.
This avoids glue spillage into GEM holes, which is suspected to lead to instabilities.
The electrode layout of the GEM foils and their connection scheme to the external high voltage supply were revised to minimize high field areas at the edges of the GEMs.


\begin{figure}
  \centering
  \includegraphics[
    width=\textwidth,
    height=0.25\textheight,
    keepaspectratio,
  ]{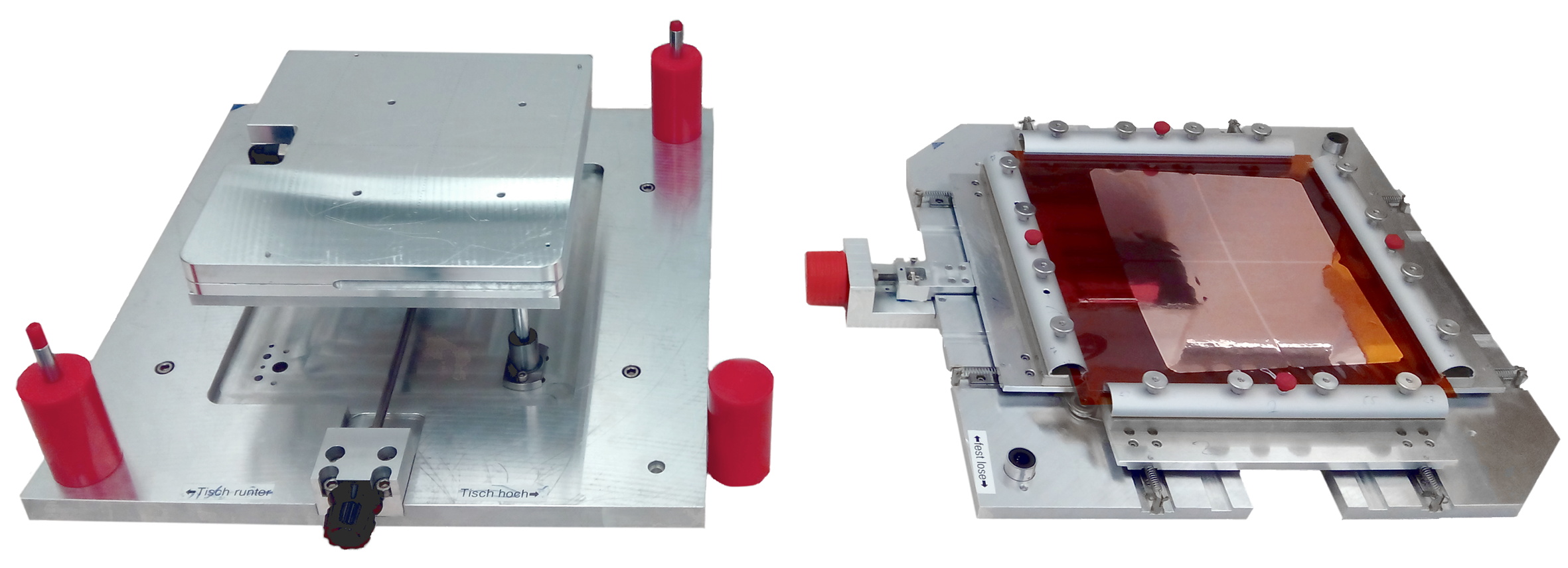}
  \caption{Picture of the stretching tool.
    The left part positions the ceramic frame, holding it with a vacuum system.
    The right part tensions the GEM foil and is positioned on top of the left part after the glue has been deposited on the ceramic frame by a robotized dispenser.
  }
  \label{fig:stretch}
\end{figure}

A second issue observed in the previous modules was related to the flatness of the GEM foils.
Measurements showed deflections of the GEM foils of up to \SI{300}{\um} peak\-/to\-/peak, which translate into significant gain inhomogeneities across a GEM module.
To provide a more controllable and reproducible procedure to tension the GEM foils while gluing, a new stretching tool, shown in \cref{fig:stretch}, was designed and built.
The tension is controlled by a system of springs that ensures a consistent and uniform force.
In addition, the tool allows for a precise and reproducible alignment between the GEM foil and the ceramic frame.
The tension is held during the gluing and curing process and is only released at the very end.
Using this approach, the GEM deflections could be reduced significantly, which directly improves the gain homogeneity.

\begin{figure}
  \begin{subfigure}[b]{0.5\textwidth}
    \includegraphics[width=\textwidth]{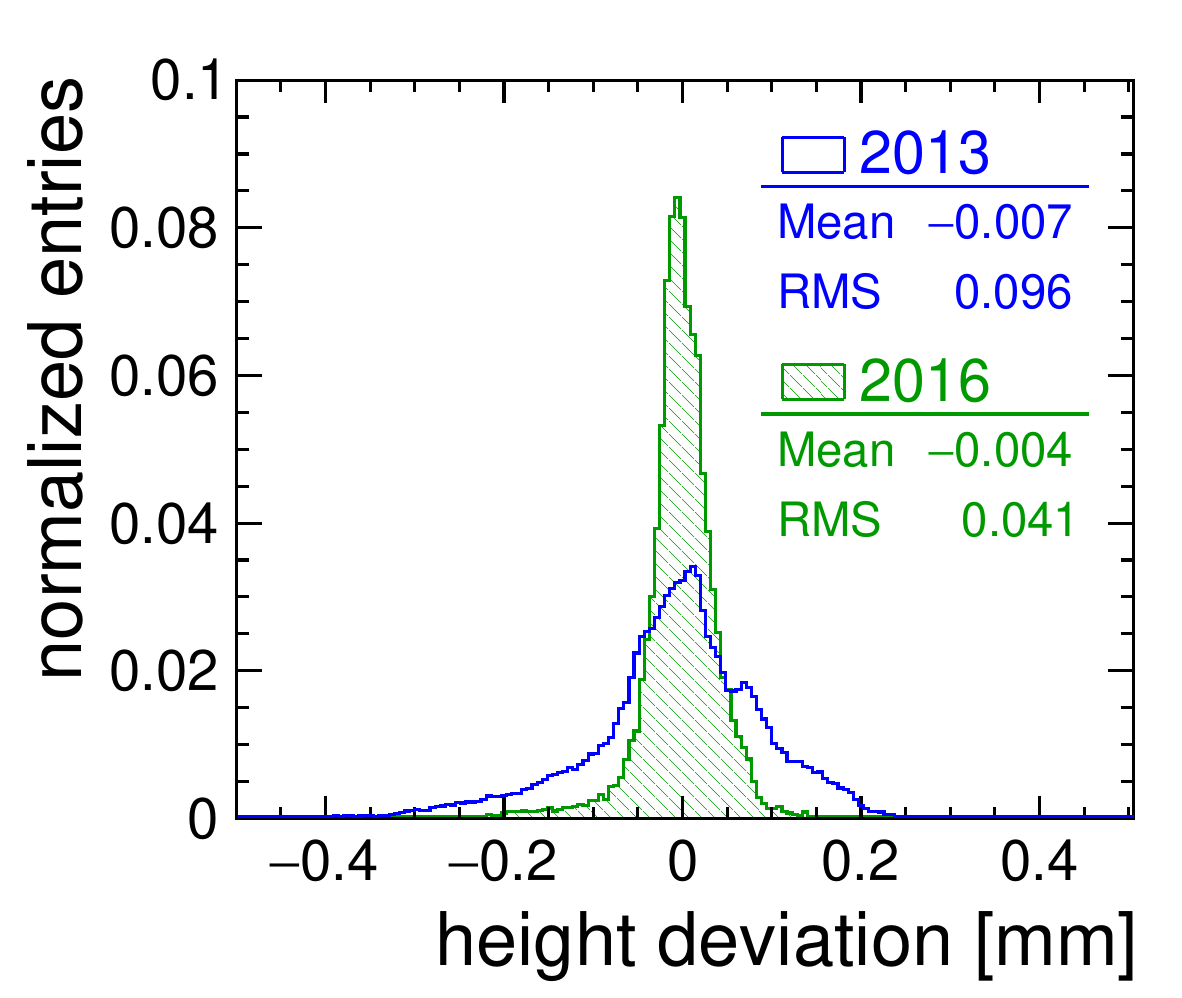}
    \caption{Measured GEM height deviations.}
    \label{fig:GEMheight}
  \end{subfigure}%
  \begin{subfigure}[b]{0.5\textwidth}
    \includegraphics[width=\textwidth]{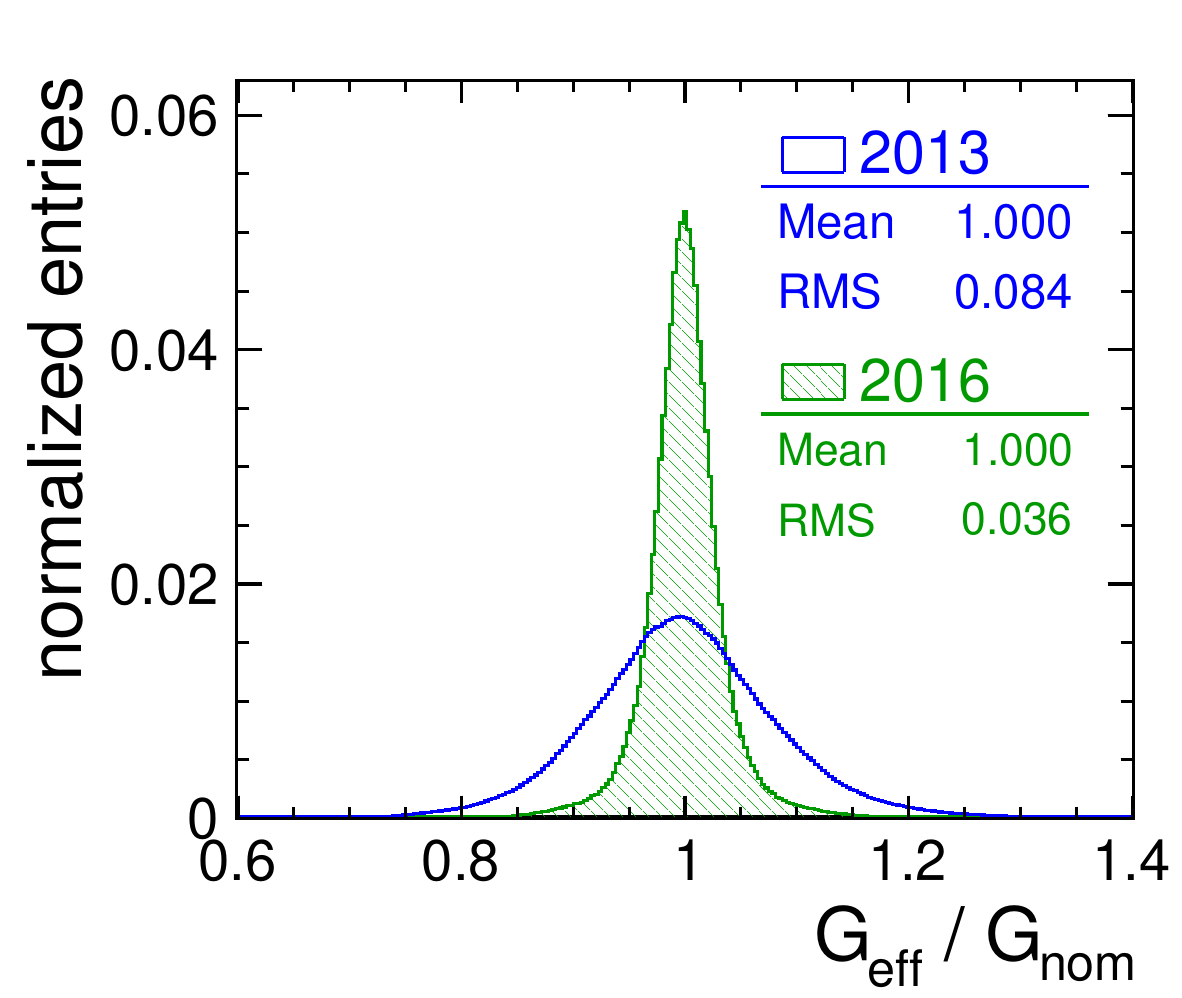}
    \caption{Calculated module gain deviations.}
    \label{fig:GEMeffGain}
  \end{subfigure}
  \caption{Effects of the new GEM mounting method.
    \protect\subref*{fig:GEMheight}~Distribution of the measured height deviation for the old (2013) and new (2016) GEM modules.
    \protect\subref*{fig:GEMeffGain}~The corresponding calculated effective gain of the system, normalized to the nominal gain for perfectly flat GEMs.
  }
  \label{fig:GEMresults}
\end{figure}

In \cref{fig:GEMheight} the distribution of the measured GEM height deviations is shown for the old and the new modules.
The scale of the deflections, given by the RMS of the distribution, has been reduced by about a factor of \num{2} from \SI{96}{\um} to \SI{41}{\um}.
A simulation, following the arguments and using the methods described in~\cite{HallermannPhD}, was used to translate the measured GEM deflections into an impact
on the effective gain of the system.
The resulting distribution of the normalized gain is shown in \cref{fig:GEMeffGain}.
The gain deviations, again given by the RMS, could also be reduced by more than a factor of \num{2} from \SI{8.4}{\%} to \SI{3.6}{\%}.

\subsection{Test-beam configuration}
\label{sec:tb_config}

Three new modules were built and installed into the end plate of the TPC prototype in a way that allows to measure tracks over a length of about \SI{50}{\cm} on a total of \num{84} pad rows.
To allow for a fair comparison of the results, the same operational parameters as in the previous test beam, described in detail in~\cite{FMueller2017}, were used.
The TPC with all three modules installed showed stable performance for the full three week test\-/beam period.
In total, \num{9.5e6} events were recorded, mostly at an electron momentum of \SI{5}{\GeVperc}.
The electron beam consisted of mostly single electrons at a rate in the order of \SI{100}{\hertz} and the data was recorded at an event rate around \SI{30}{\hertz}.
The spatial distribution of the beam electrons perpendicular to the beam direction follows roughly a two\-/dimensional Gaussian and has an RMS close to \SI{3}{\mm}.
The momentum spread of the beam is about \SI{150}{\MeVperc} resulting at \SI{5}{\GeVperc} in a relative uncertainty of \SI{\sim1.5}{\percent}.
The distance between the beam and the readout plane was varied between \SI{0}{\cm} and \SI{57}{\cm}.
For the data presented in this paper the beam was kept parallel to the readout plane, i.e.\ $\theta\approx\ang{90}$, and perpendicular to the pad rows of the readout.
The angular spread of the beam after passing through the magnet wall was about \ang{2}.

The objectives of this test beam were to evaluate the stability and reproducibility of the performance of the detector, measure the \dEdx resolution and to study the double\-/hit separation of the TPC prototype.
Therefore, two different data sets were taken:
One to study the performance of the detector under nominal conditions and one including a thin target in front of the TPC, to produce multiple particles out of a single beam electron to study the double\-/hit separation in multi\-/track events.
In addition to measurements using the default GEM voltage settings intended to provide high gain and stable operation, the former data set includes a subset using voltage settings intended to minimize ion\-/backflow.
According to~\cite{ZenkerPhD}, the gain of the default setting is estimated to be \num{\sim2000}.
The charge measurements with both GEM settings at the test beam show that the gain of the minimum\-/ion\-/backflow setting is about \SI{20}{\percent} lower.
As in previous measurements, the readout electronics was run at a sampling rate of \SI{20}{\mega\hertz} with a gain of \SI{12}{\milli\volt\per\femto\coulomb} and a peaking time of \SI{120}{\ns}.

\begin{figure}
  \begin{subfigure}[b]{0.47\textwidth}
    \includegraphics[width=\textwidth]{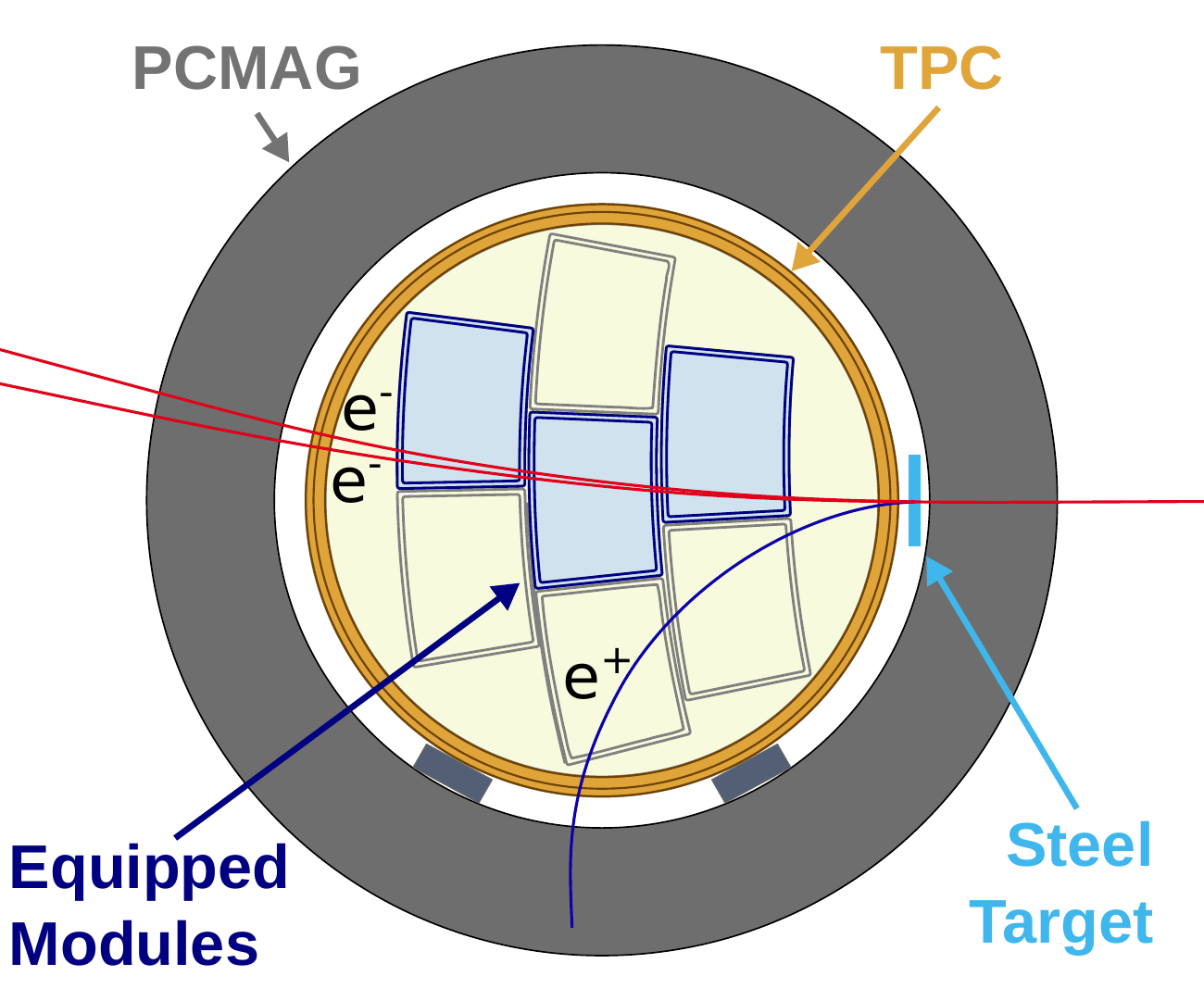}
    \caption{Sketch of the target location.}
    \label{fig:target_principle_simu}
  \end{subfigure}%
  \begin{subfigure}[b]{0.53\textwidth}
    \includegraphics[width=\textwidth]{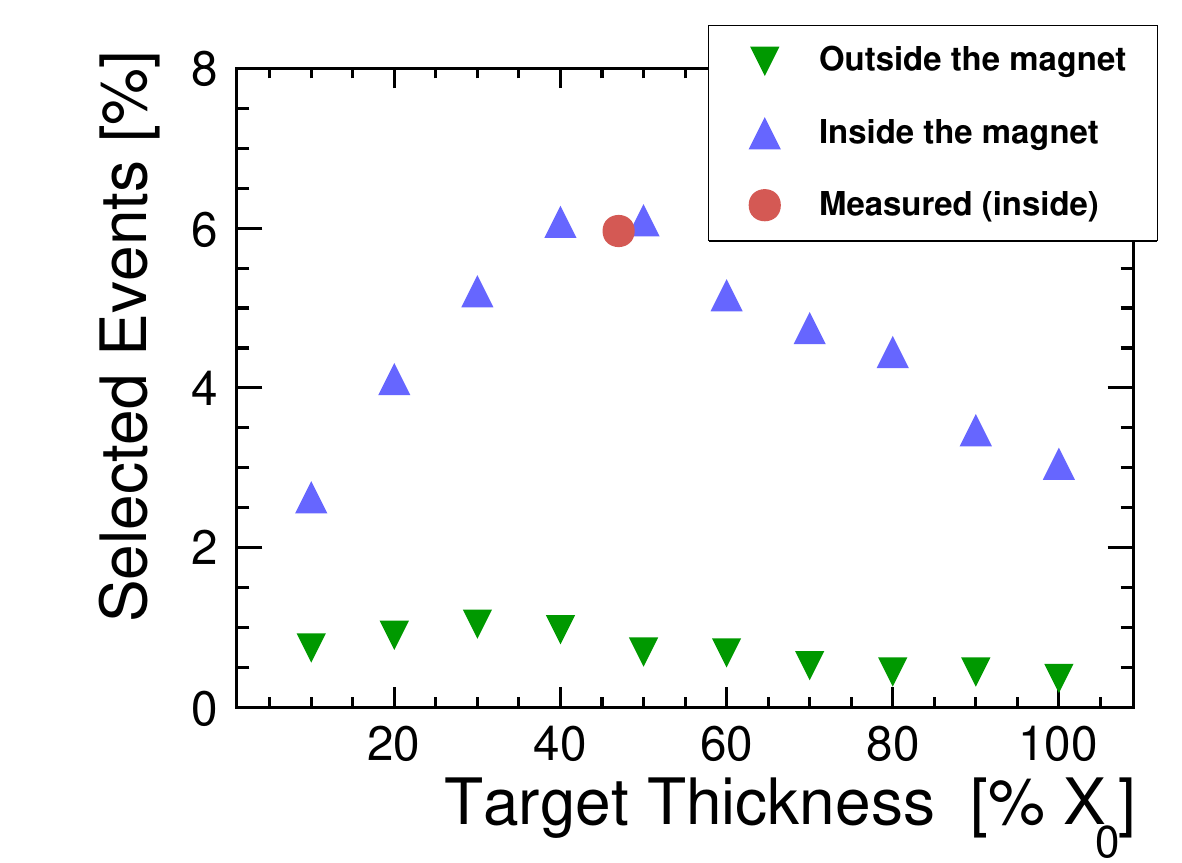}
    \caption{Yield of suitable track pairs.}
    \label{fig:targetx0study}
  \end{subfigure}
  \caption{Setup for the measurement of multi\-/track events.
    \protect\subref*{fig:target_principle_simu}~Sketch of the setup overlaid with an event from a \textsc{Geant4} simulation.
    \protect\subref*{fig:targetx0study}~Simulated yield of track pairs suitable for double\-/hit studies for the options to install the target outside the magnet or inside directly in front of the TPC prototype, in dependence on the target thickness as a fraction of a radiation length \Xzero.
    The fraction of suitable events determined from measured data is shown in comparison.
  }
  \label{fig:targetprinciple}
\end{figure}

For the second data set, a stainless steel target was installed inside the magnet in front of the TPC as shown in the sketch in \cref{fig:target_principle_simu}.
\Cref{fig:targetx0study} shows the fraction of suitable events as a function of the thickness of the target, for \SI{5}{\GeVperc} beam electrons.
For events to be included they need to contain at least two tracks reconstructed in the TPC\@.
The distance between two of the tracks has to be less than \SI{3}{\mm} along the first two or more pad rows at the entry point of the TPC, which is usually fulfilled for tracks originating from an interaction of a beam electron with the steel target.
This ensures at least two merged hits from two different tracks.
In addition, at least \SI{40}{\percent} of the remaining hits of these tracks have to be clearly separated.
This corresponds to a distance between the tracks of \SI{7}{\mm} or more.
The \lcnamecref{fig:targetx0study} shows results from \textsc{Geant4}~\cite{Agostinelli:2002hh,Allison:2006ve,Allison:2016lfl} simulations for two different locations of the target: one inside the solenoid magnet right in front of the TPC prototype and one outside of the magnet directly in front of it.


According to the simulation, the highest yield of events suitable for the double\-/hit separation studies is expected for a target with a thickness of about \SI{50}{\percent} \Xzero, placed inside the magnet right in front of the TPC prototype.
Therefore, a target made of \SI{8}{\mm} thick stainless steel, which corresponds to \SI{47}{\percent} \Xzero, was installed in the setup.
\Cref{fig:targetx0study} also shows the fraction of events fulfilling the selection criteria determined from the data measured with this setup.
Very good agreement between measurement and simulation is found.

\section{Reconstruction methods}
\label{sec:reco}

The MarlinTPC software package~\cite{MarlinTPC} was used to reconstruct and analyze the data.
This package is based on the linear collider software suite~\cite{lcsoft,Marlin,lcio}.
The individual steps of the reconstruction are introduced briefly in the following.
For a detailed description see~\cite{FMueller2017}.

\begin{figure}[htb]
  \centering
  \begin{subfigure}{0.6\linewidth}
    \centering
    \includegraphics[keepaspectratio, width=\linewidth]{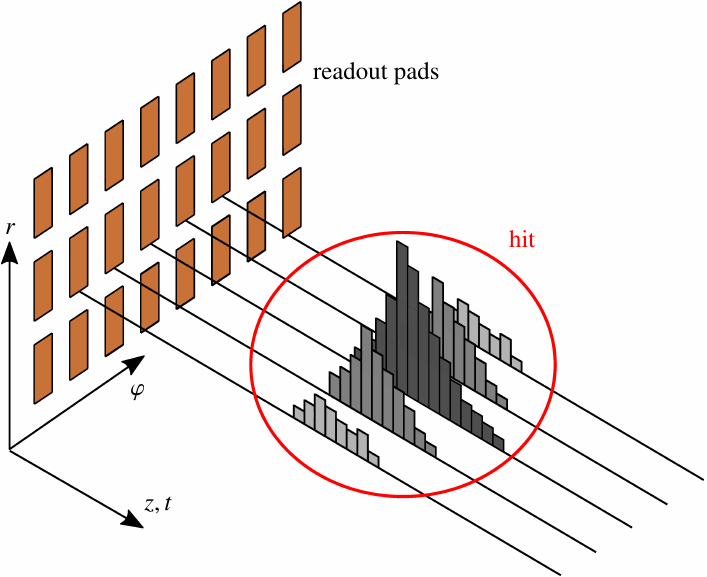}
    \caption{Hit identification.}
    \label{fig:pad_pulses}
  \end{subfigure}\\
  \begin{subfigure}[b]{0.5\linewidth}
    \centering
    \includegraphics[keepaspectratio, width=0.95\linewidth]{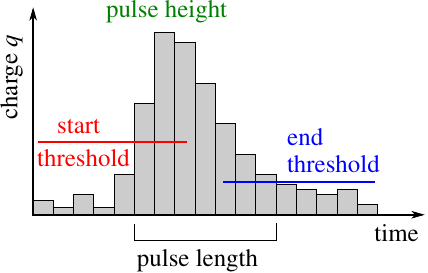}
    \caption{Pulse identification.}
    \label{fig:pulseshape}
  \end{subfigure}%
  \begin{subfigure}[b]{0.5\linewidth}
    \centering
    \includegraphics[keepaspectratio, width=0.95\linewidth]{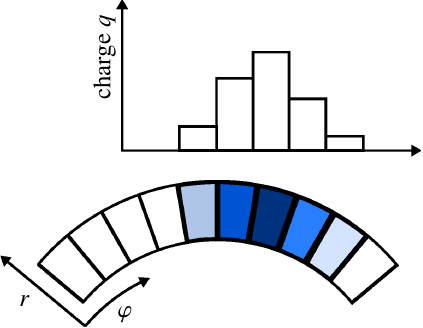}
    \caption{Hit charge distribution.}
    \label{fig:hitreco}
  \end{subfigure}%
  \caption{Hit reconstruction principle in the TPC prototype~\cite{Mueller:301339}.
    \subref*{fig:pad_pulses}~The measured charge is detected on the individual pads in time and peaks in the charge distribution are identified.
    Charge pulses from adjacent pads in one readout row are combined to hits.
    \subref*{fig:pulseshape}~The steering parameter for the pulse reconstruction.
    \subref*{fig:hitreco}~Charge distribution in a typical hit. Darker colors represents higher charge.
  }
  \label{fig:hitfinding}
\end{figure}

The amplified electron avalanche after the GEM stack is collected on the pads of the segmented readout anode, as sketched in \cref{fig:pad_pulses}.
The signal on a single pad is called a pulse.
The readout electronics measures the collected charge in time bins.
The charge of a pulse is the sum of the ADC counts in these bins from the rise of the signal over a start threshold to its fall under a stop threshold.
In addition to the thresholds, a minimum number of samples over threshold is required.
These parameters are indicated in \cref{fig:pulseshape}.
The arrival time of each pulse is derived from the inflection point of its rising edge.

A row\-/based clustering algorithm is used to combine several pulses on adjacent pads in one pad row, as indicated in \cref{fig:pad_pulses}.
Each group of pulses is called a hit.
The coordinate along the pad rows is determined using a charge\-/weighted center\-/of\-/gravity method.
The typical charge distribution in a hit is shown in \cref{fig:hitreco}.
The charge of the hit is the sum of the charge of the contributing pulses.
For the time of the hit, only the arrival time of the largest contributing pulse is used, since the time information of the adjacent pulses is affected by a number of systematic effects, which have been explored in~\cite{Mueller:301339}.
First, there is a charge dependence of the reconstructed arrival time, which causes smaller pulses to be systematically earlier than the leading pulse.
Secondly, there is an additional dependence on the distance of the corresponding pad to the pad of the leading pulse.

To combine the reconstructed hits to tracks, an iterative Hough transformation~\cite{KleinwortHough} is used in the following, unless stated otherwise.
The track parameters are determined by a fit using the General Broken Lines method~\cite{KleinwortGBL}.
As track model either a helix, for data taken with magnetic field, or a straight line, for data taken without magnetic field, is used.

Two coordinate systems are used in the reconstruction and analysis, one global for the whole prototype and one local to each readout module.
The global coordinate system is Cartesian.
The $z$\=/axis is perpendicular to the readout plane pointing from the
anode towards the cathode.
The $xy$\=/plane is parallel to the readout plane with the $y$\=/axis pointing upwards and the $x$\=/axis in the direction of the electron beam.
Since the shape of each module is an annular segment, the local coordinate system is cylindrical with its origin at the center of the annulus.
The $z$\=/direction is identical to that of the global coordinate system.
The position parallel to the readout plane is defined by the radial coordinate $r$
and the azimuth angle $\varphi$.
The coordinate $r\varphi$ describes the corresponding distance along the circumference of a circle with radius $r$.

\section{Overall performance of the TPC}
\label{sec:tpcperformance}

The performance of the new readout module generation has been studied and compared to the previous generation, based on the data taken in the new test\-/beam campaign and the data presented in~\cite{FMueller2017}, respectively.
For both data sets, the same settings were used in reconstruction and analysis, with the exception of an adjustment of the minimum number of pulse samples over threshold to account for different operating conditions in the new measurement.
As in the previous analysis, the used setup does not have the possibilities to perform a charge calibration into initial number of electrons and therefore the  reconstructed charge is given in units of ADC counts.
The data taken with the previous module generation have been re-analyzed with the current software version and the alignment and distortion corrections were performed as in the previous study.
The results differ slightly from the original analysis, since in the re-analysis the lower boundary of the fit range of the function describing the spatial resolution was changed from \SI{70}{\mm} to \SI{90}{\mm}.

\subsection{Drift velocity}
\label{sec:v_drift}

A key parameter for a TPC is the electron drift velocity in the drift volume.
It has been determined from the reconstructed mean arrival time for several measurement runs at various positions of the beam along the drift direction.
The position of the beam along the z\=/axis is determined from the position of the movable stage, since no absolute reference is available.

\begin{figure}
  \centering
  \includegraphics[
    width=\textwidth,
    height=0.33\textheight,
    keepaspectratio,
  ]{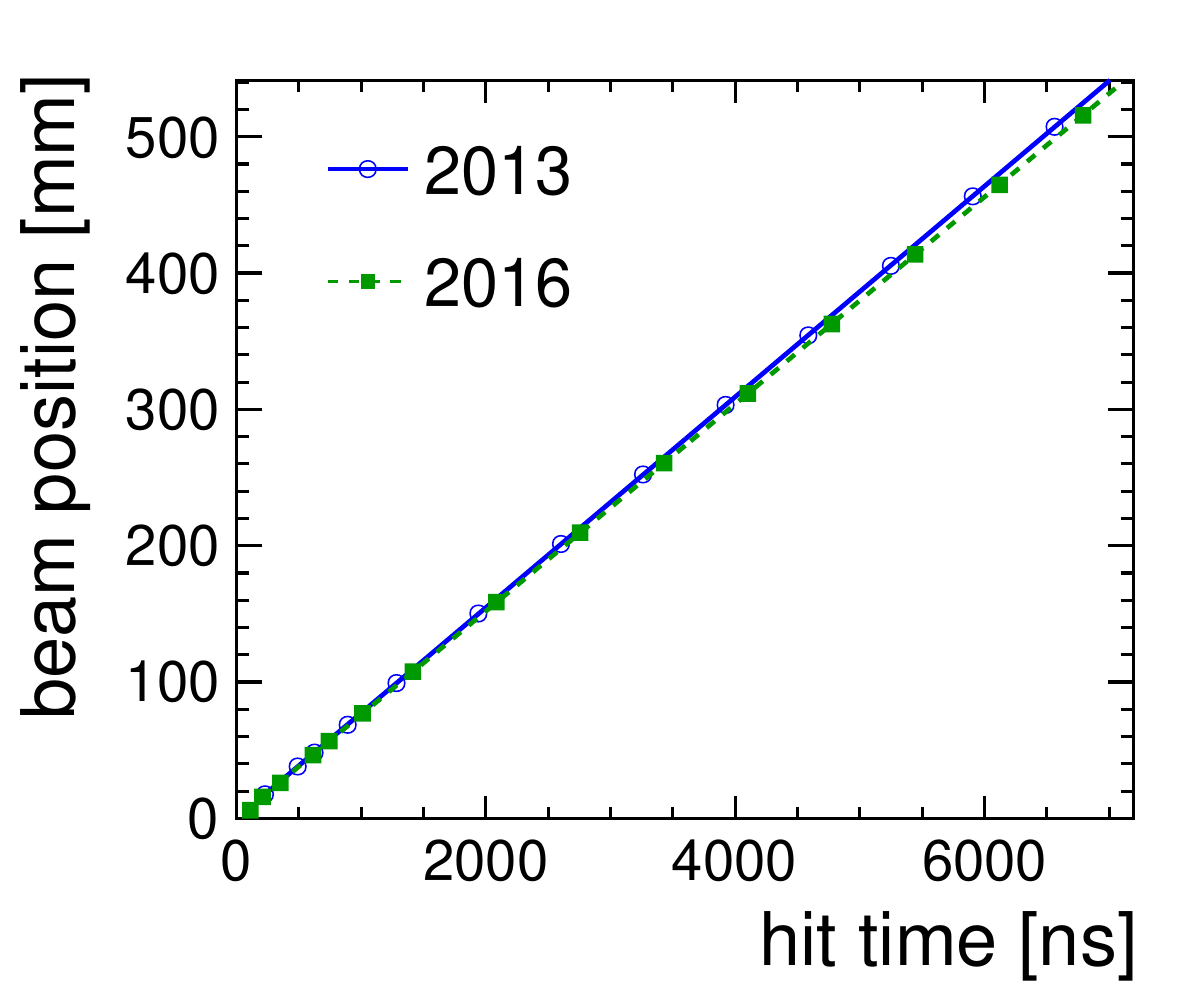}
  \caption{Drift velocity measurement for the new module generation (2016) and the previous one (2013).
    Shown is the beam position versus reconstructed arrival time of the hit signal for $B=\SI{1}{\tesla}$ at the default drift field of \SI{240}{\V\per\cm} including linear fits to determine the drift velocity.
  }
  \label{fig:driftvel}
\end{figure}

\Cref{fig:driftvel} shows the beam position inside the drift volume plotted against the signal arrival time.
The slope of a linear fit to the data points corresponds to the drift velocity.
The fit results are given in \cref{tab:vdrift}, including a comparison with values from a Magboltz~\cite{Magboltz} simulation.
The errors on the measured values originate mainly from the uncertainties of the stage position measurement and less from the uncertainty of the time measurement.
Other potential uncertainties are not considered in the fit.
The simulation errors are negligible in comparison to the errors of the measured values.
The measured values agree reasonably well with the corresponding results of the simulation and are similarly consistent between the two data sets.
The remaining discrepancy can be explained by uncertainties in the gas conditions, i.e.\ pressure, temperature and contaminants.
Using the resulting drift\-/velocity values, the time measurement of the reconstructed hits is converted to the spatial coordinate in the drift direction.

\begin{table}
  \centering
  \begin{tabular}{
      c
      S[table-format=2.2(1)]
      S[table-format=2.2]
  }
    \toprule
    data set & \multicolumn{2}{c}{$v_\mathrm{d}$ [$\si{\mm\per\us}$]} \\
               \cmidrule(l){2-3}
             & {measurement} & {simulation} \\
    \midrule
    2013     &     77.28(04) & 76.87 \\
    2016     &     75.97(03) & 76.45 \\
    \bottomrule
  \end{tabular}
  \caption{The resulting slope of the linear fits in \protect\cref{fig:driftvel} and the corresponding simulated values of the drift velocity.
    The statistical uncertainties of the simulation are negligible compared to the measurement errors and have therefore been omitted.
  }
  \label{tab:vdrift}
\end{table}

\subsection{Pad response function}
\label{sec:prf}

Another key parameter for the performance is the pad response.
The pad response function (PRF) describes the average shape of the charge distribution of a hit on the pad plane.
Its width is defined primarily by the diffusion of the primary electrons in the drift volume and the defocusing of the charge avalanche in the GEM stack.
Therefore, the shape of the distribution should follow a convolution of a Gaussian and a box function with the width of the pad pitch.
The latter accounts for the uniform probability distribution of the initial position of the electron relative to the pad.

\begin{figure}[htb]
  \begin{subfigure}[b]{0.5\textwidth}
    \includegraphics[width=\textwidth]{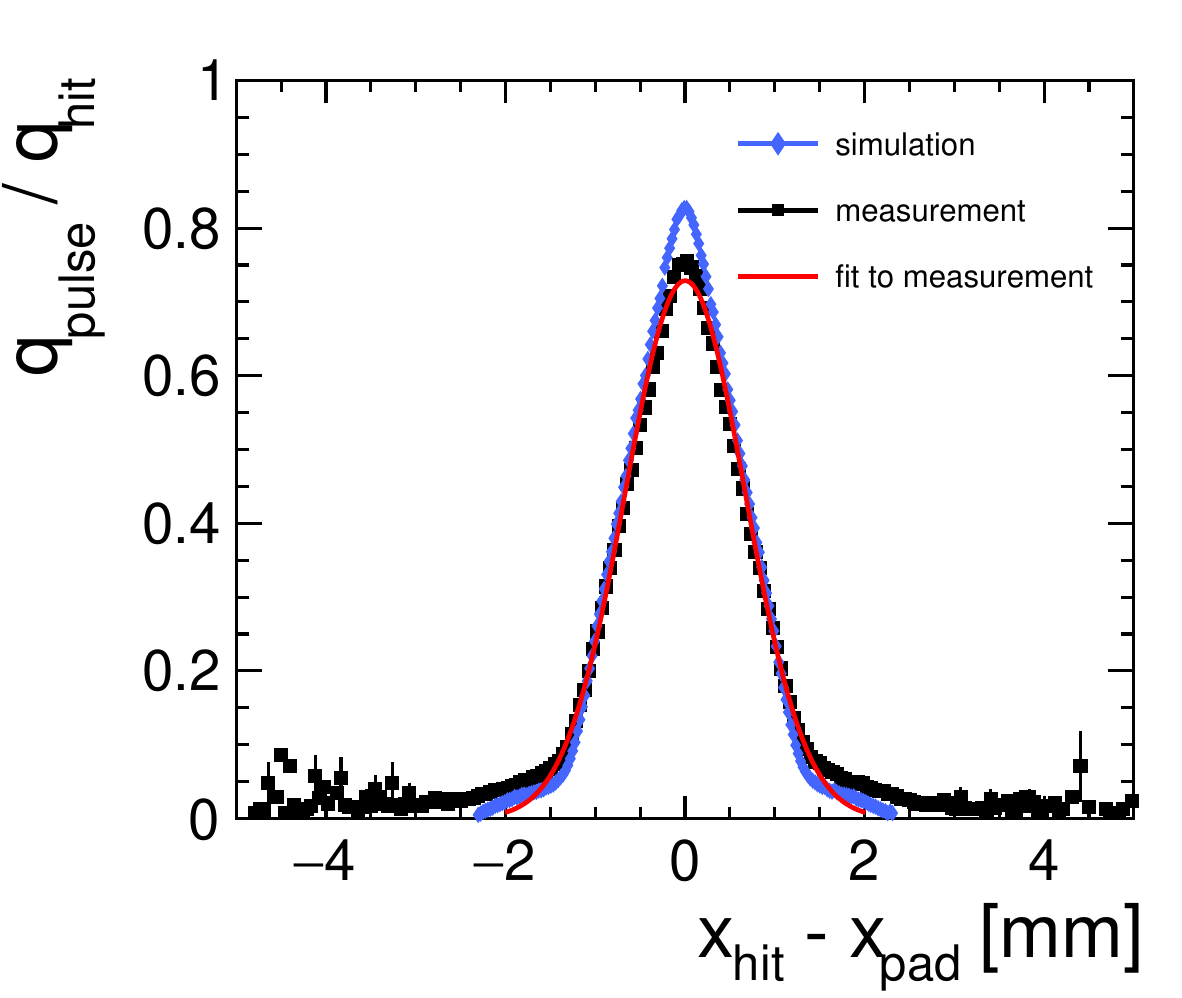}
    \caption{Shape of pad response function.}
    \label{fig:PRF_shape}
  \end{subfigure}%
  \begin{subfigure}[b]{0.5\textwidth}
    \includegraphics[width=\textwidth]{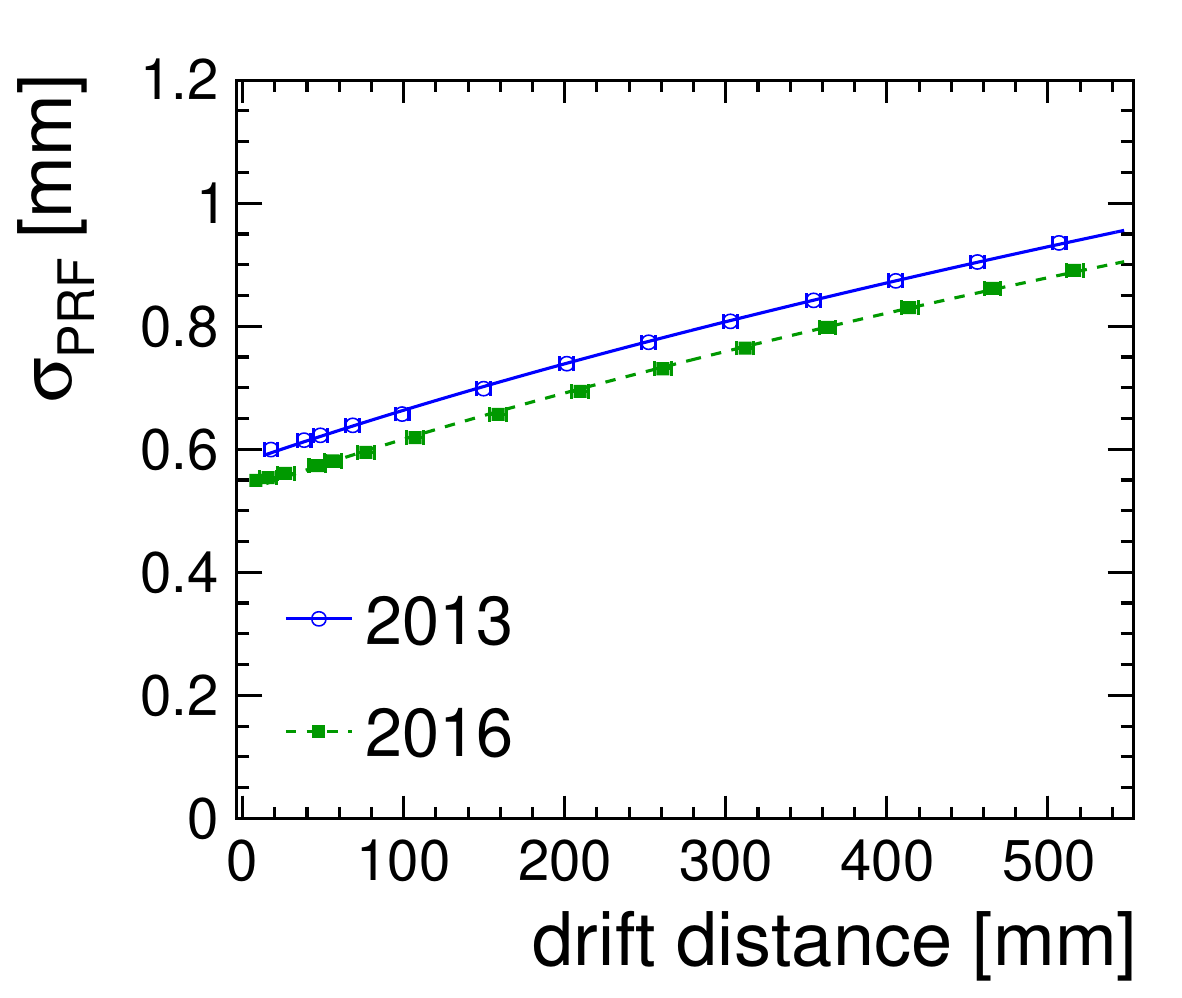}
    \caption{Transverse diffusion measurement.}
    \label{fig:sigma_PRF}
  \end{subfigure}
  \caption{Pad response function and diffusion measurement.
    \protect\subref*{fig:PRF_shape}~The PRF in simulation and measurement at \SI{\sim20}{\mm} drift distance in a \SI{240}{\volt\per\cm} drift field.
    Also shown is a fit to the measured data.
    \protect\subref*{fig:sigma_PRF}~The width of the PRF versus drift distance including a fit of \protect\cref{eq:prffit} to extract the transverse diffusion constant.
    Results are shown for the 2016 test beam using the new module generation and the 2013 test beam using the previous one.
  }
  \label{fig:prf_comp}
\end{figure}

In reality, the tails of this distribution are wider than expected from the above model, as can be seen in the measured PRF shown in \cref{fig:PRF_shape} for the example of a drift distance of \SI{\sim20}{\mm}.
This can be explained by induction from drifting electrons ending up on adjacent pads.
This induced signal has an equal positive and negative part but the ALTRO readout only records negative charge signals and cuts off the positive part.
Thus, a non-zero signal is recorded.
The size of this induction signal on the adjacent pads has been estimated to be around \SI{10}{\percent}.
As can be seen in the \lcnamecref{fig:PRF_shape}, a simulation including the induction effect describes the tails in the measured values only partially.
Therefore, to extract the width of the PRF, the fit of the simple model described above is performed only within a range of \num{+-3} standard deviations around the maximum, as shown in the \lcnamecref{fig:PRF_shape}.
As visible, this results in a realistic value for the width of the distribution.

\begin{table}
  \centering
  \begin{tabular}{
      c
      S[table-format=3.1(1)]
      S[table-format=3.2(1)]
      S[table-format=2.2(1)]
    }
    \toprule
    data set & {$\sigma_\mathrm{PRF,0}$ [\si{\um}]} & \multicolumn{2}{c}{$D_\mathrm{t}$ [$\si{\um\per\sqrt{\cm}}$]}   \\
                                                      \cmidrule(l){3-4}
             &                               & {measurement}         & {simulation}            \\
    \midrule
    2013 & 578.2(19)  & 102.87(38)  & 97.49(70)\\
    2016 & 530.8(20)  &  99.06(40)  & 95.48(63)\\
    \bottomrule
  \end{tabular}
  \caption{The resulting parameters of \protect\cref{eq:prffit} in the combined fits to the PRF width and the $r\varphi$\=/resolution, as described in the text.
    Also shown are the corresponding simulated values of $D_\mathrm{t}$.
  }
  \label{tab:prf}
\end{table}

In \cref{fig:sigma_PRF} the resulting PRF width $\sigma_\mathrm{PRF}$ is plotted against the drift distance.
The behavior can be described by the function~\cite{Blum:2008}
\begin{equation}
  \sigma_\mathrm{PRF}(z) = \sqrt{\bigl(\sigma_\mathrm{PRF,0}\bigr)^2 + \bigl(D_\mathrm{t}\bigr)^2 \cdot z} \enskip,
  \label{eq:prffit}
\end{equation}
which allows to determine the transverse diffusion constant $D_\mathrm{t}$.
Since this is also required as an input to the fit of the transverse point resolution $\sigma_{r\varphi}$, as is explained in \cref{sec:pointres}, a combined \chisquare{}\=/fit of $\sigma_\mathrm{PRF}$ with \cref{eq:prffit} and $\sigma_{r\varphi}$ with \cref{eq:zResolutionElectronLoss} was performed.
The fit results regarding the PRF width are summarized in \cref{tab:prf}, including simulated values of the corresponding diffusion constants, taking into account differences in measured gas pressure, temperature and contamination with air and water vapor.
As expected from the Magboltz simulation, the diffusion measured in the 2016 data set is smaller than in the 2013 data set, mostly due to a difference in the gas pressure.
However, the measured diffusion constants from both data sets are significantly larger than the corresponding simulated results.
As discussed in~\cite{Mueller:301339}, the simulation only describes the ideal case of perfectly homogeneous and parallel electric and magnetic fields.
The inhomogeneity of the magnetic field in the measurement setup introduces field transverse components, which increase the spread of the drifting electrons and thus can explain the systematic differences between measured and simulated results.

The intrinsic width $\sigma_\mathrm{PRF,0}$ at zero drift distance, which is caused by the defocusing of the electron avalanche in the GEM stack and signal induction on the pad plane, shows a significant difference between the two measurements, with a smaller width in the 2016 data.
Qualitatively this result is consistent with the reduced deflections of the GEMs used in the new modules.
Deflections of the GEMs lead to inhomogeneous electric fields in the transfer and induction regions between the GEMs and the pad plane, creating transverse field components.
Equivalent to the magnetic field inhomogeneity in the drift region, these field distortions increase the charge spread in the amplification region.
This results in an increased width of the PRF independent of the drift distance.


\subsection{Point resolution}
\label{sec:pointres}

The point resolution for the single\-/hit position is determined from the width of the residual distribution as explained in detail in~\cite{FMueller2017}.
The residual is the distance of a hit to the corresponding track.
Here, the resolution is evaluated in the directions $r\varphi$ and $z$, which correspond to the distance of a hit to its track along a pad row and along the drift direction, respectively.

The dependence of the point resolution $\sigma_{r\varphi/z}(z)$ on the drift distance $z$ is described by the equation~\cite{Yonamine_2014}
\begin{equation}
  \sigma_{r\varphi/z} (z) =\sqrt{\bigl(\sigma_{r\varphi/z,0}\bigr)^2+\frac{\bigl(D_\mathrm{t/l}\bigr)^2}{N_{\mathrm{eff}}\cdot e^{-Az}} \cdot z} \enskip,
  \label{eq:zResolutionElectronLoss}
\end{equation}
where $\sigma_{r\varphi/z,0}$ denotes the intrinsic resolution of the setup, due to diffusion in the GEM stack and the finite pad width, and $N_\mathrm{eff}$ the effective number of primary electrons.
The term $e^{-Az}$, with the electron attachment coefficient $A$, describes the loss of signal electrons during drift by attachment to gas molecules, primarily due to oxygen contamination of the gas.
This equation only holds for tracks parallel to the readout plane and perpendicular to the pad rows.

Since in the \lcnamecref{eq:zResolutionElectronLoss} the diffusion constants $D_\mathrm{t/l}$ and $N_\mathrm{eff}$ appear as the quotient
$\bigl(D_\mathrm{t/l}\bigr)^2/N_\mathrm{eff}$,
they are fully correlated in a fit and only the quotient can be determined.
Thus, the longitudinal diffusion constant $D_\mathrm{l}$ was provided by a Magboltz simulation and used to calculate $N_\mathrm{eff}$ from the fit result.
As described in the previous \lcnamecref{sec:prf}, the transverse diffusion constant $D_\mathrm{t}$ can also be determined by a fit of \cref{eq:prffit} to the PRF width.
Due to the observed systematic difference of this measurement to the simulated result, the measured value is assumed to be correct.
Therefore, to resolve the correlation of $D_\mathrm{t}$ and $N_\mathrm{eff}$, a combined fit of \cref{eq:prffit,eq:zResolutionElectronLoss} to the PRF width and the $r\varphi$\=/resolution, respectively, is performed.
Since \cref{eq:zResolutionElectronLoss} is not a good description of the $r\varphi$\=/resolution at short drift distances, it is fitted only to the data points with a drift distance larger than \SI{90}{\mm}.

The parameters $N_\mathrm{eff}$ and $A$ are also strongly correlated.
Therefore, it would be preferable to determine the attachment coefficient directly from the signal charge measurement.
However, in the 2013 data the charge measurement is affected by environmental effects and no charge calibration is available for this data.
Thus, this approach is not reliable here and both parameters are left free in the fits, to ensure equal treatment of the two data sets.

\begin{figure}[htb]
  \begin{subfigure}[b]{0.5\textwidth}
    \includegraphics[width=\textwidth]{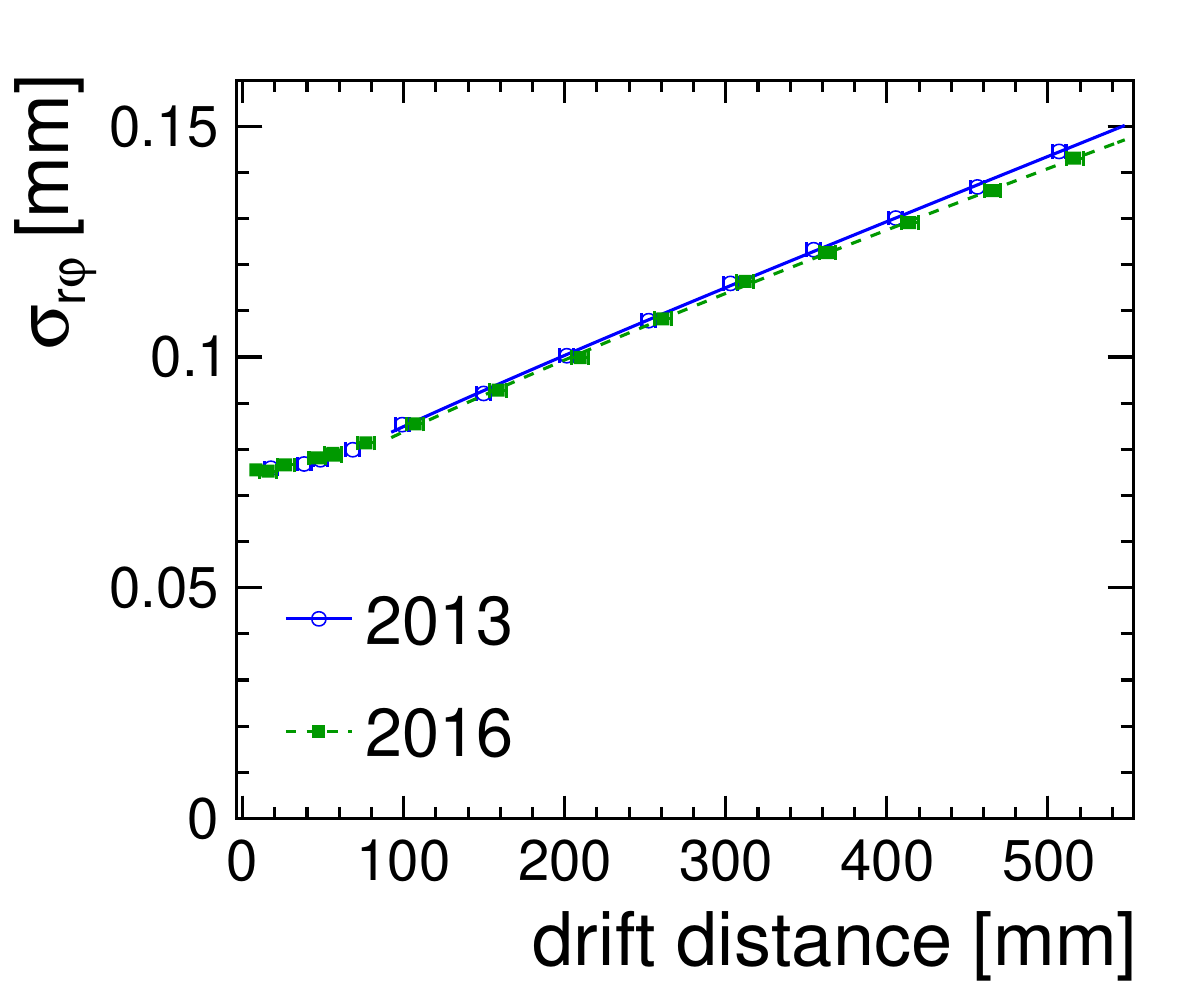}
    \caption{$r\varphi$-resolution.}
    \label{fig:pres_13-16_rphi}
  \end{subfigure}%
  \begin{subfigure}[b]{0.5\textwidth}
    \includegraphics[width=\textwidth]{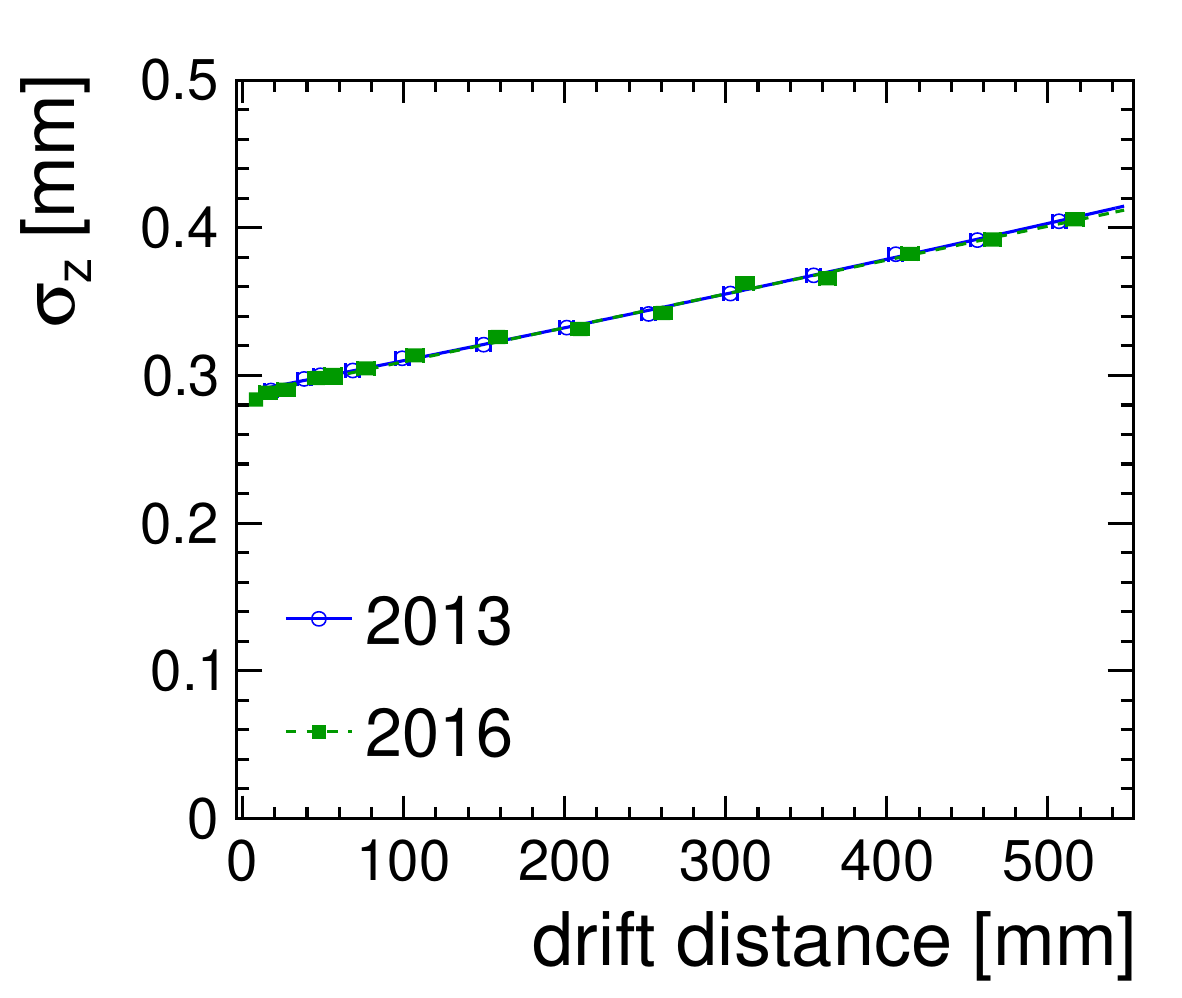}
    \caption{$z$-resolution.}
    \label{fig:pres_13-16_z}
  \end{subfigure}
  \caption{Comparison of the measured point resolution for current (2016) and previous (2013) module generation in \protect\subref*{fig:pres_13-16_rphi}~$r\varphi$- and \protect\subref*{fig:pres_13-16_z}~$z$\=/direction, both as a function of the drift distance.
    The lines represent fits of \protect\cref{eq:zResolutionElectronLoss}.
  }
  \label{fig:pointresolution}
\end{figure}

In \cref{fig:pointresolution}, comparisons of the point resolution in the $r\varphi$ and $z$ directions are shown between the 2013 test\-/beam campaign with the first\-/generation modules and the 2016 test\-/beam campaign with the new modules.
The plots include the fits of \cref{eq:zResolutionElectronLoss} to the measured resolution, as described above.
The fit results are summarized in \cref{tab:rphiresolution,tab:zresolution}.
A very good agreement between the two campaigns is found in the measurements and the fitted results, indicating a good stability and reproducibility of the performance of the system.
%

\begin{table}
  \centering
  \begin{tabular}{
      c
      S[table-format=2.1(1)]
      S[table-format=2.1(1)]
      S[table-format=1.4(1)]
      S[table-format=3.2(1)]
    }
    \toprule
    data set & {$\sigma_{r\varphi,0}$ [\si{\um}]} & {$N_\mathrm{eff}$} & {$A$ [\si{\per\cm}]} & {$D_\mathrm{t}$ [$\si{\um\per\sqrt{\cm}}$]}\\
    \midrule
    2013 & 68.1(16)  & 43.4(28)  & 0.0054(11) & 102.87(38)\\
    2016 & 66.2(21)  & 38.7(30)  & 0.0040(13) &  99.06(40)\\
    \bottomrule
  \end{tabular}
  \caption{The resulting parameters of \protect\cref{eq:zResolutionElectronLoss} in the combined fits to the PRF width and the $r\varphi$\=/resolution, as described in the text.
    The results for $D_\mathrm{t}$ are the same as already shown in \protect\cref{tab:prf}.
  }
  \label{tab:rphiresolution}
\end{table}

\begin{table}
  \centering
  \begin{tabular}{
      c
      S[table-format=3.1(1)]
      S[table-format=2.1(1)]
      S[table-format=1.4(1)]
      S[table-format=3.1(1)]
    }
    \toprule
    data set & {$\sigma_{z,0}$ [\si{\um}]} & {{$N_\mathrm{eff}$}} & {$A$ [\si{\per\cm}]} & {$D_\mathrm{l}^\mathrm{sim}$ [$\si{\um\per\sqrt{\cm}}$]}\\
    \midrule
    2013 & 288.4(06) & 39.9(14)  &  0.0050(06) & 221.6(16)\\
    2016 & 286.0(07) & 38.6(13)  &  0.0033(06) & 227.1(14)\\
    \bottomrule
  \end{tabular}
  \caption{The results of fits of \protect\cref{eq:zResolutionElectronLoss} to the $z$\=/resolution.
    $N_\mathrm{eff}$ is calculated from the fit parameters using the simulated values of $D_\mathrm{l}$ given in the last column.
  }
  \label{tab:zresolution}
\end{table}

\paragraph{Minimal ion backflow}

Ions from the GEM amplification flowing back into the sensitive TPC volume can cause field distortions and electron attachment.
A gating GEM system~\cite{KOBAYASHI201941} is foreseen for the ILD TPC that will capture the ions produced in the amplification while having a very high transparency to drifting electrons.
In addition to using the gate, the amount of ions flowing from the amplification into the drift volume will be minimized by the GEM\-/voltage settings described in the following.

\begin{figure}[htb]
  \begin{subfigure}[b]{0.52\textwidth}
    \raggedright
    \includegraphics[width=0.98\textwidth]{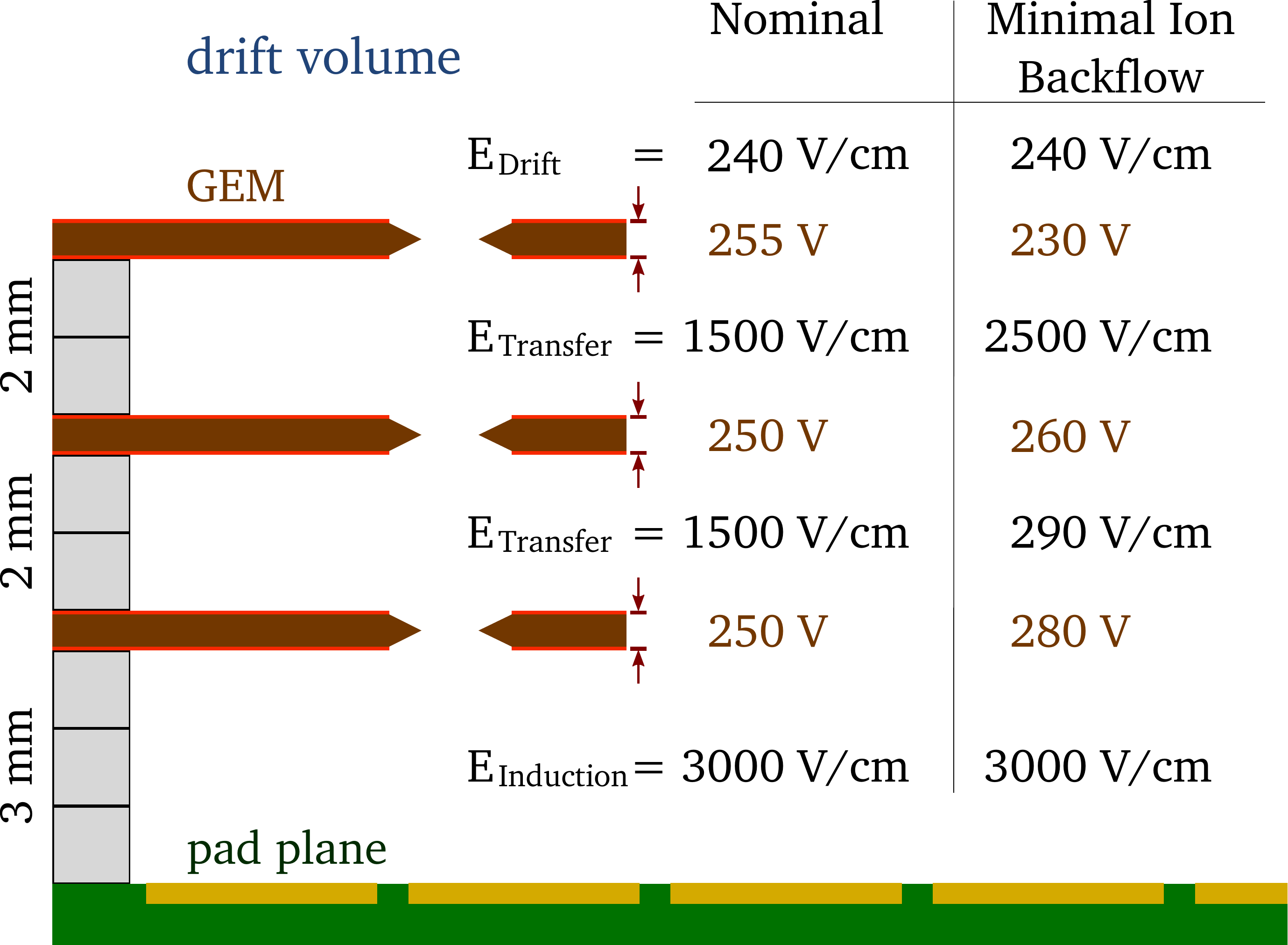}
    \caption{Comparison of module voltages.}
    \label{fig:voltages_nom-mib}
  \end{subfigure}%
  \begin{subfigure}[b]{0.48\textwidth}
    \raggedleft
    \includegraphics[width=0.98\textwidth]{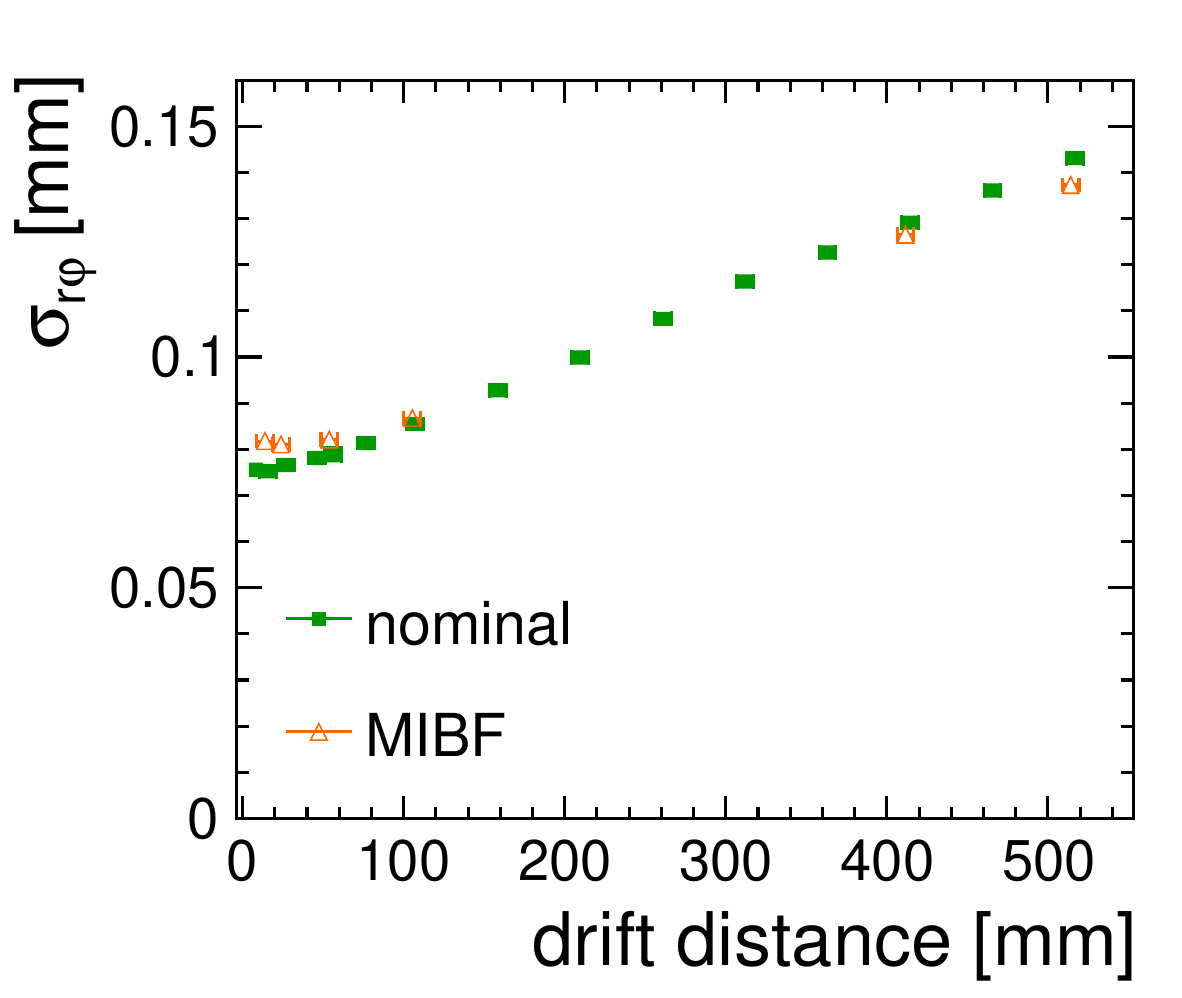}
    \caption{Comparison of $r\varphi$\=/resolution.}
    \label{fig:rphires_mibf}
  \end{subfigure}
  \caption{Minimum ion backflow studies:
    \protect\subref*{fig:voltages_nom-mib}~Comparison of the nominal GEM voltage settings used to collect most of the test\-/beam data and the settings corresponding to the MIBF conditions.
    \protect\subref*{fig:rphires_mibf}~Comparison of the point resolution in $r\varphi$ for nominal and MIBF GEM\-/voltage settings.
  }
  \label{fig:mibf}
\end{figure}

During the 2016 test\-/beam campaign a data set was taken with altered GEM voltage settings based on the studies in~\cite{ZenkerPhD}, as shown in \cref{fig:voltages_nom-mib}, which should minimize the ion backflow (MIBF).
The main idea is that the GEM facing the sensitive volume is operated at a lower voltage to minimize the ion production at this stage.
To compensate and ensure a sufficient total gain, the remaining two GEMs are operated at higher voltages.
The ions produced here will be mostly captured by the GEM facing the sensitive volume.
The average hit charge measured with the MIBF settings was about \SI{20}{\percent} lower than observed with the nominal voltages.

In \cref{fig:rphires_mibf} the point resolution in the $r\varphi$\=/direction is plotted against the drift distance.
Good agreement is observed between the measurements under MIBF conditions and the default voltage settings, except for a slightly worse resolution at small drift distances with the MIBF settings.
This is interpreted as a more pronounced hodoscope effect~\cite{YonamineJinst2014}, likely caused by a combination of the lower amplification and the slightly lower diffusion of the electron avalanche within the GEM stack due to the changed fields.
Still, this shows that the MIBF settings only have a small influence on the achievable point resolution and therefore represent a viable parameter set for the operation of the ILD TPC\@.

\paragraph{Extrapolation to the ILD TPC}

To judge the performance of a large TPC in the final detector, \cref{eq:zResolutionElectronLoss} is used to extrapolate the measured point resolution to the characteristics of the ILD TPC,
i.e.\ to a magnetic field of \SI{3.5}{\tesla} and a maximum drift length of \SI{2.35}{\m}.
As input to the extrapolation, $\sigma_{r\varphi,0}$ and $N_{\mathrm{eff}}$ are taken from the fit of \cref{eq:zResolutionElectronLoss} to the measured resolution of the 2016 test\-/beam data given in \cref{tab:rphiresolution}.
A Magboltz simulation has been performed to determine the transverse diffusion constant at the ILD magnetic field, giving $D_\mathrm{t}=\SI{31.8(4)}{\um\per\sqrt{\cm}}$.

\begin{figure}
  \centering
  \includegraphics[
    width=\textwidth,
    height=0.33\textheight,
    keepaspectratio,
  ]{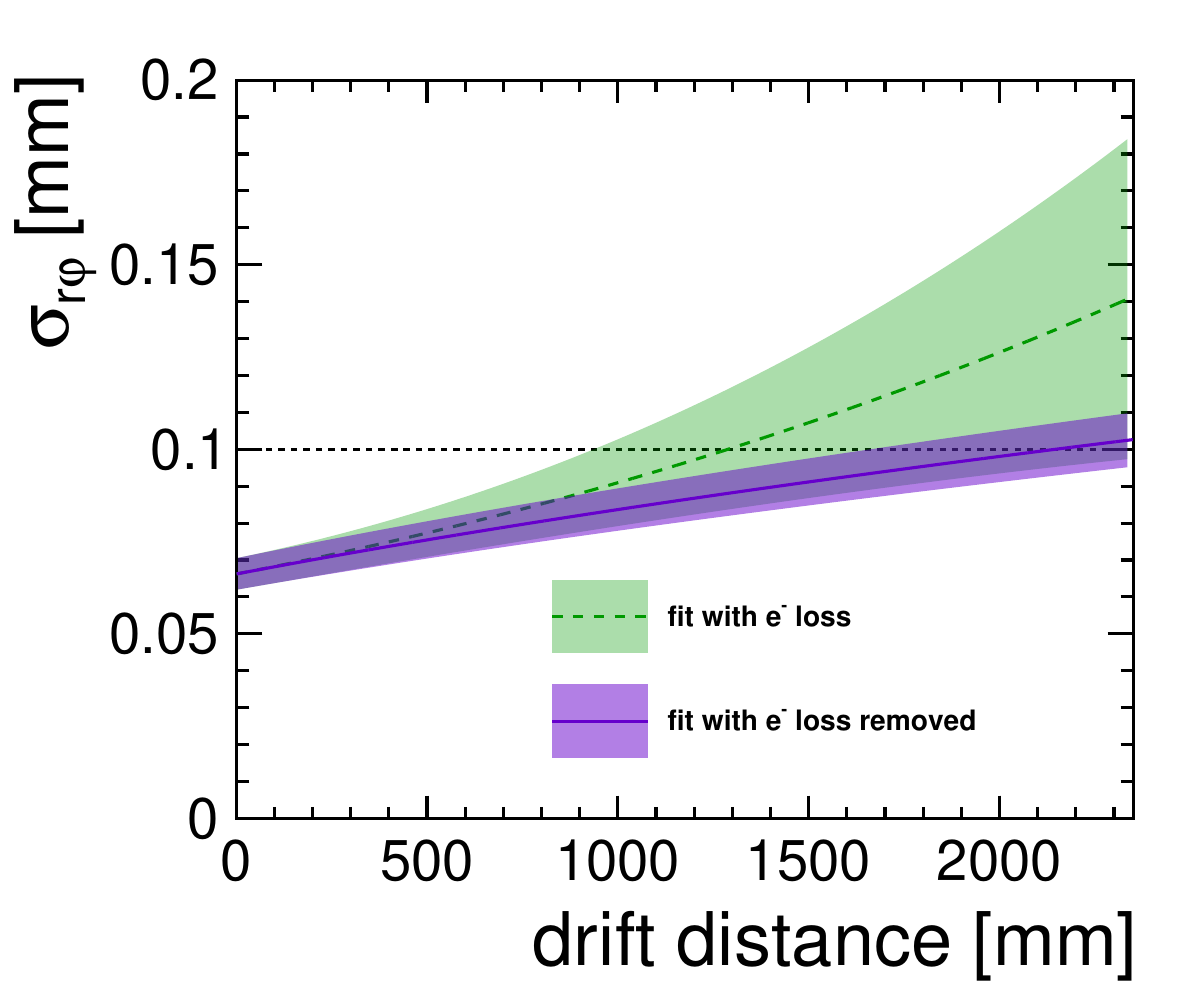}
  \caption{Extrapolation of the point resolution to the ILD TPC dimensions with a magnetic field of \SI{3.5}{\tesla} and a maximum drift length of \SI{2.35}{\m}.
    Each curve includes a $\SI{+-2}{\sigma}$ confidence band.
    The extrapolation based on the measured electron loss due to attachment is denoted \emph{fit with $e^-$ loss}.
    For the other graph (\emph{fit with $e^-$ loss removed}) the attachment coefficient was set to zero while keeping the other parameters unchanged.
  }
  \label{fig:ILD_rphi}
\end{figure}

The results of the extrapolation are shown in \cref{fig:ILD_rphi} for two cases.
The first case assumes the electron attachment rate
as determined from the measurement, where a nearly constant oxygen contamination of \SI{\sim65}{\ppm} was observed.
For the other case, the electron attachment rate is set to zero to show the ultimate resolution possible under perfect operating conditions.
For both cases $\SI{+-2}{\sigma}$ confidence bands are shown, which are derived from the errors of the fitted parameters and their correlations.
For the case in which the attachment is set to zero, its uncertainty is ---~in contrast to the previous analysis in~\cite{FMueller2017}~--- also set to zero, resulting in a lower uncertainty overall.


The extrapolation shows that ---~if contamination of the gas is minimized~--- a point resolution of \SI{100}{\um} can be reached over almost the full drift length in the ILD TPC, fulfilling the requirement set for the ILD TPC\@.
Experience of running experiments using a TPC, e.g.\ T2K~\cite{Abgrall201125} and ALICE~\cite{AliceUpgradeTDR}, indicates that an oxygen contamination of less than \SI{1}{\ppm} is a realistic assumption.

\section{Double-hit separation in \texorpdfstring{\boldmath$r\varphi$}{r\textphi}}
\label{sec:dhr}

In this \lcnamecref{sec:dhr}, the performance of the double\-/hit separation between two tracks is studied using different reconstruction techniques on the multi\-/track data set described in \cref{sec:tb_config}.

The \emph{default hit reconstruction} algorithm uses the charge deposited on each pad and computes its position in $r\varphi$ using a charge weighted center of gravity method.
It includes a simple local minimum algorithm to identify double\-/hit candidates before the track finding, but no hit separation is performed.
A hit is tagged as a double\-/hit candidate, if the charge distribution on the pads includes a dip in charge amplitude of at least one ADC count in comparison to the adjacent pads.
In this double\-/hit study, the \emph{triplet track finder}~\cite{kleinwort_triplet} ---~a local track finding algorithm using triplet chains~--- is used as the default algorithm on these hits.

In order to improve the double\-/hit and track separation in the pattern recognition, two new methods have been developed and implemented in MarlinTPC\@.
The first method, called \emph{pad pulse road search}, is a combined track and hit finding algorithm based on local road search with pulses.
The second method, called \emph{PRF hit separation}, is a hit splitting algorithm, where the pad response function is fitted to the charge distribution of a (double-) hit candidate.

\paragraph{Pad pulse road search}
\label{sec:ppr}

The pad pulse road search is a combined track and hit finding algorithm taking the global event topology into account.
It was adapted specifically for the data taken with the prototype TPC in order to improve the double\-/hit separation and works directly on the individual pulse signals on the pads of a module.

In a first step, seed pulses are identified that have the maximum absolute charge in their neighborhood, which is a region in a pad row of a size reflecting the average transverse diffusion.
Starting from these seed pulses, linear roads with a width of a few times the pad pitch are defined between pairs of yet unused seed pulses in different rows, provided they exceed a minimum distance.
Pulses inside a road are added and the road is iteratively refined using a linear or a helical fit, resulting in a so\-/called pad pulse segment.
Ambiguous pulses, which belong to more than one pad pulse segment, are removed and the segments refitted.
Then the ambiguous pulses are reassigned to the segment with the closest distance in the $r\varphi$ plane.

The hit finding is performed locally per row.
Pulses are added to the maximum pulse not yet assigned to a hit until another pad pulse segment starts or there is a gap.
The hit position is determined as in the default hit finding procedure described in \cref{sec:reco}.
The \emph{integrated hit separation} of the pad pulse finder assigns ambiguous pulses to hits based on proximity.
In the case of close-by seed pulses, the corresponding hits can only grow outwards, leading to biases in the position determination.

Track segments are defined and fitted based on all hits of a pad pulse segment.
In a last step, matching track segments on different readout modules are combined to track candidates, which are passed to the track fitting.
A detailed description of this combined track and hit finding method is given in~\cite{Kleinwort:395416}.

\paragraph{PRF hit separation}
\label{sec:prf_hit_sep}

\begin{figure}[htb]
  \centering
  \includegraphics[
    width=\textwidth,
    height=0.33\textheight,
    keepaspectratio,
  ]{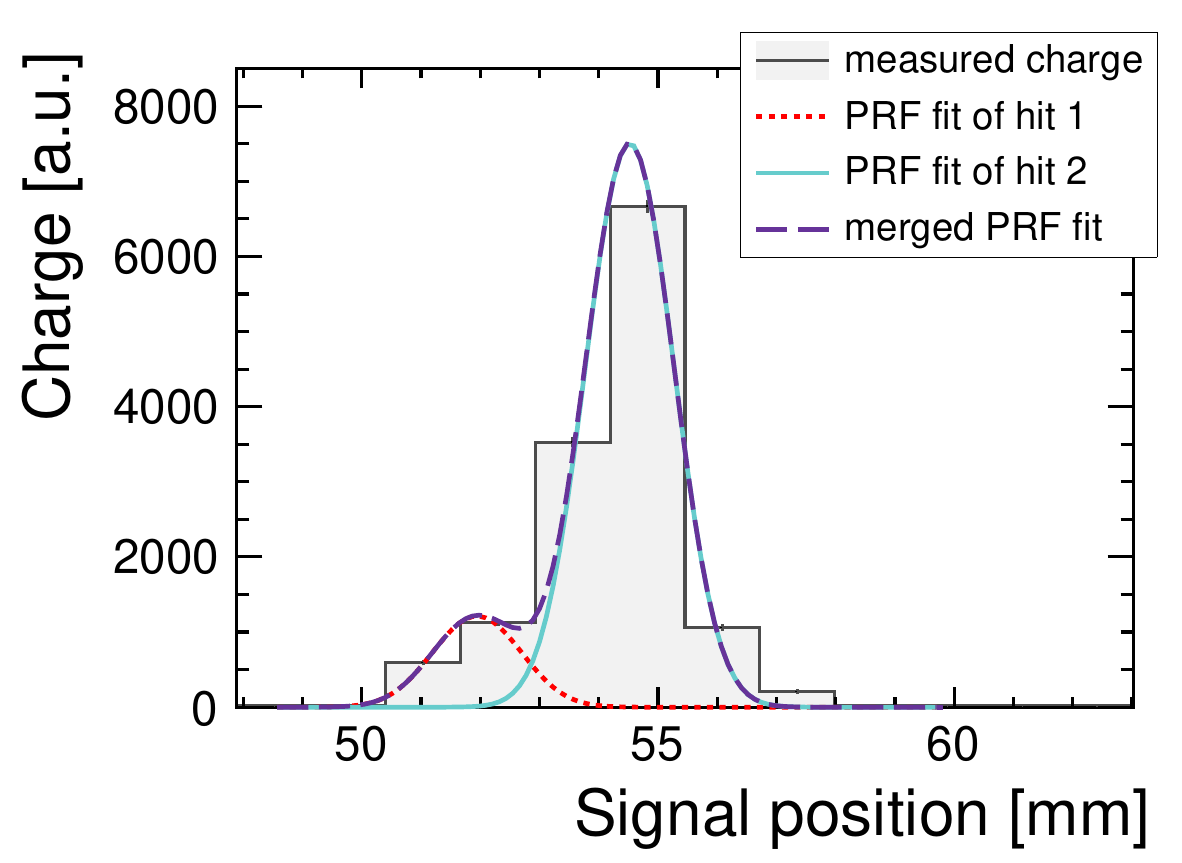}
  \caption{Schematic illustration of a fit of the sum of two pad response functions to the signal of a merged hit.}
  \label{fig:mergedhit}
\end{figure}

The PRF hit separation method is shown schematically in the example in \cref{fig:mergedhit}.
After a merged hit candidate has been identified, a fit is performed using the sum of two single\-/hit PRFs, which are convolutions of a Gaussian and a box function.
In this fit, the sigma of the two Gaussian functions is fixed according to the specific drift distance.
It is determined from a fit of \cref{eq:prffit} to the measured PRF width, as shown in \cref{fig:sigma_PRF}.
The width of the box functions is equal to the pad pitch.
The mean positions are free parameters, as are the amplitudes to adjust for different hit charges.


A series of requirements, listed in \cref{tab:cutsprfseparation}, are used in order to identify double\-/hit versus single\-/hit candidates and ensure a proper reconstruction.
To exclude signals from $\delta$\=/electrons, the fit is not performed if the hit charge is significantly larger than average.
If the charge of a hit after the separation is too small, it will be rejected.
If the number of pulses contributing is too small, the fit becomes unstable.
If it is too large, the signal most probably does not come from a single or double hit but from other effects like  $\delta$\=/electrons or more than two overlapping tracks.
Finally, the $\chi^2$ of the fit of the single- and the double\-/hit PRF are compared and there are limits on their accepted values.

\begin{table}
  \centering
  \begin{tabular}{ l c l }
    \toprule
    parameter & accepted range & comment \\
    \midrule
    charge of single hit $Q$ [a.u.]           & $\num{600} < Q < \num{20000}$       & average $Q \approx \num{2000}$ \\
    number of pulses $N$                      & $\num{4} < N < \num{15}$            & for single hit: $N \leq \num{7}$ \\
    $\chi^2$ of a fit of the single\-/hit PRF & $\chi^2_\mathrm{single} < \num{10}$ &  \\
    $\chi^2$ of a fit of the double\-/hit PRF & $\chi^2_\mathrm{double} < \num{52}$ &  \\
    \bottomrule
  \end{tabular}
  \caption{The cut values used in the PRF hit separation.}
  \label{tab:cutsprfseparation}
\end{table}

The PRF hit separation can be used either on the tagged double\-/hit candidates of the local hit search of the standard reconstruction or to improve the handling of ambiguous pulses in the hit finding of the pad pulse road search.

\paragraph{Performance comparison}
\label{sec:dhr_results}

The default reconstruction is compared to the results from the pad pulse road search.
In an additional step, the PRF hit separation is run on the double\-/hit candidates identified by the pad pulse road search to further improve the double\-/hit separation.

\begin{figure}
  \begin{subfigure}[b]{0.5\textwidth}
    \includegraphics[width=\textwidth]{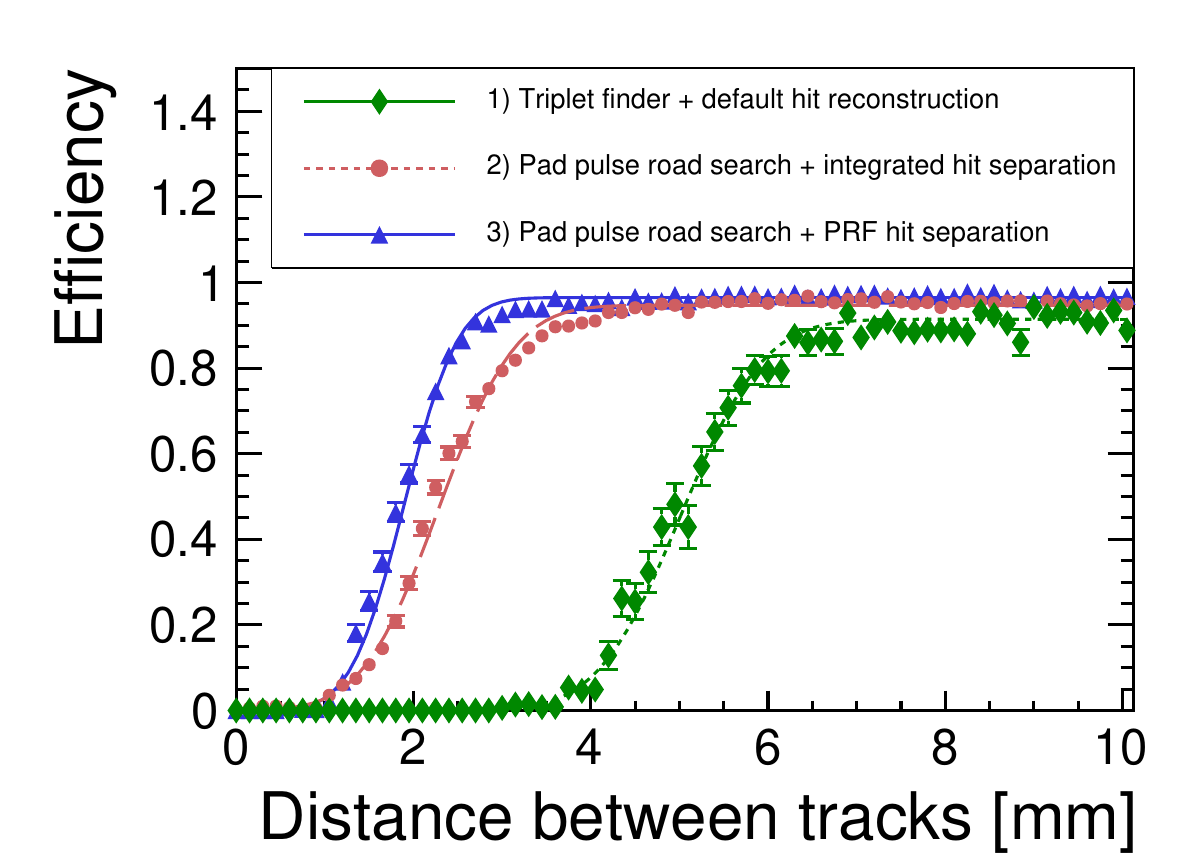}
    \caption{Measurement.}
    \label{fig:doublehitreso_data}
  \end{subfigure}%
  \begin{subfigure}[b]{0.5\textwidth}
    \includegraphics[width=\textwidth]{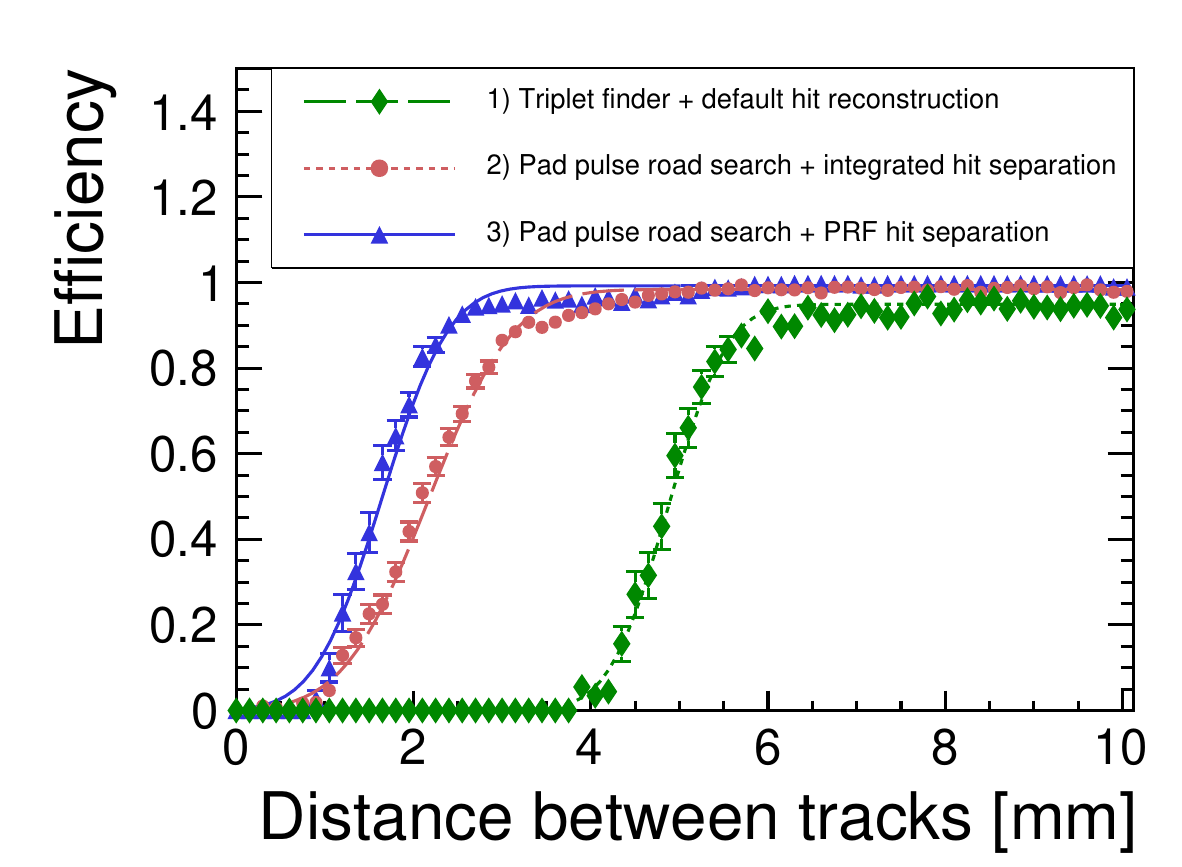}
    \caption{Simulation.}
    \label{fig:doublehitreso_MC}
  \end{subfigure}
  \caption{Comparison between different hit and track reconstruction algorithms and their effect on the double\-/hit separation \protect\subref*{fig:doublehitreso_data}~for measured test\-/beam data and \protect\subref*{fig:doublehitreso_MC}~for simulated tracks,
    both times for a drift distance of \SI{200}{\mm}.
    Shown are the triplet track finder with default hit reconstruction and the pad pulse road search with its integrated and with the PRF hit separation.
    The results of the fitted curves are listed in \protect\cref{tab:doublehitfitvalues}.
    The curves show the fits of \protect\cref{eq:errorfunc} to the data points.
  }
  \label{fig:doublehitreso}
\end{figure}

\Cref{fig:doublehitreso} shows the efficiency of the double\-/hit separation, using the methods described above, as a function of the distance between the two tracks at the location of the hit, for \subref{fig:doublehitreso_data}~test\-/beam data and \subref{fig:doublehitreso_MC}~tracks from a Monte Carlo simulation, both at a drift distance of \SI{\sim200}{\mm}.
The efficiency is defined as the ratio of the number of separated hits to that of all double\-/hit candidates considered at the respective distance between tracks.
The distance between two tracks corresponding to an efficiency of \SI{50}{\percent} defines the double\-/hit separation capability of the respective method.
These values were determined from a $\chi^2$ fit of the following function to the measured separation efficiency:
\begin{equation}
    f\left(\Delta_{r\varphi}\right) = \frac{A}{2} \left( 1 + \erf \left(\frac{\Delta_{r\varphi} - M}{\sqrt{2} \cdot \sigma} \right )\right) \enskip.
  \label{eq:errorfunc}
\end{equation}
Here, $\Delta_{r\varphi}$ is the distance between tracks in the $r\varphi$ direction, $\erf(x)$ is the error function and $A$, $M$ and $\sigma$ are fit parameters denoting the maximum efficiency at large track distance, the inflection point and the standard deviation, respectively.

The fitted values of the parameters for the examples in \cref{fig:doublehitreso} are given in \cref{tab:doublehitfitvalues}.
The results from the pad pulse road search algorithm show a significant improvement in comparison to the results from the default reconstruction.
The combination of the pad pulse road search with the PRF hit separation further improves the double\-/hit separation.
Overall, an improvement of the double\-/hit separation from around \SI{5}{\mm} to less than \SI{2}{\mm} is observed in data and simulation.
This is consistent with the studies using parallel laser tracks in~\cite{Karlen2005}, where efficient two\-/track separation was achieved for distances down to \num{\sim1.5} times the pad pitch.

\begin{table}
  \centering
  \sisetup{table-format=1.2}
  \begin{tabular}{ l S S S S S S }
    \toprule
              & \multicolumn{3}{c}{measured data} & \multicolumn{3}{c}{simulation} \\
                \cmidrule(lr){2-4}                  \cmidrule(l){5-7}
    algorithm & {$A$} & {$M$ [\si{\mm}]} & {$\sigma$ [\si{\mm}]}        & {$A$} & {$M$ [\si{\mm}]} & {$\sigma$ [\si{\mm}]} \\
    \midrule
    triplet track finder\\
      + default hit reconstruction & 0.91 & 5.02 & 0.74 & 0.95 & 4.87 & 0.52\\
    pad pulse road search\\
      + integrated hit separation & 0.95 & 2.29 & 0.68 & 0.98 & 2.17 & 0.75\\
    pad pulse road search\\
      + PRF hit separation        & 0.96 & 1.90 & 0.48 & 0.99 & 1.65 & 0.60\\

    \bottomrule
  \end{tabular}
  \caption{Parameter values of the fits of \protect\cref{eq:errorfunc} in \protect\cref{fig:doublehitreso} for the different hit and track reconstruction algorithms.}
  \label{tab:doublehitfitvalues}
\end{table}

\begin{figure}
  \begin{subfigure}[b]{0.5\textwidth}
    \includegraphics[width=\textwidth]{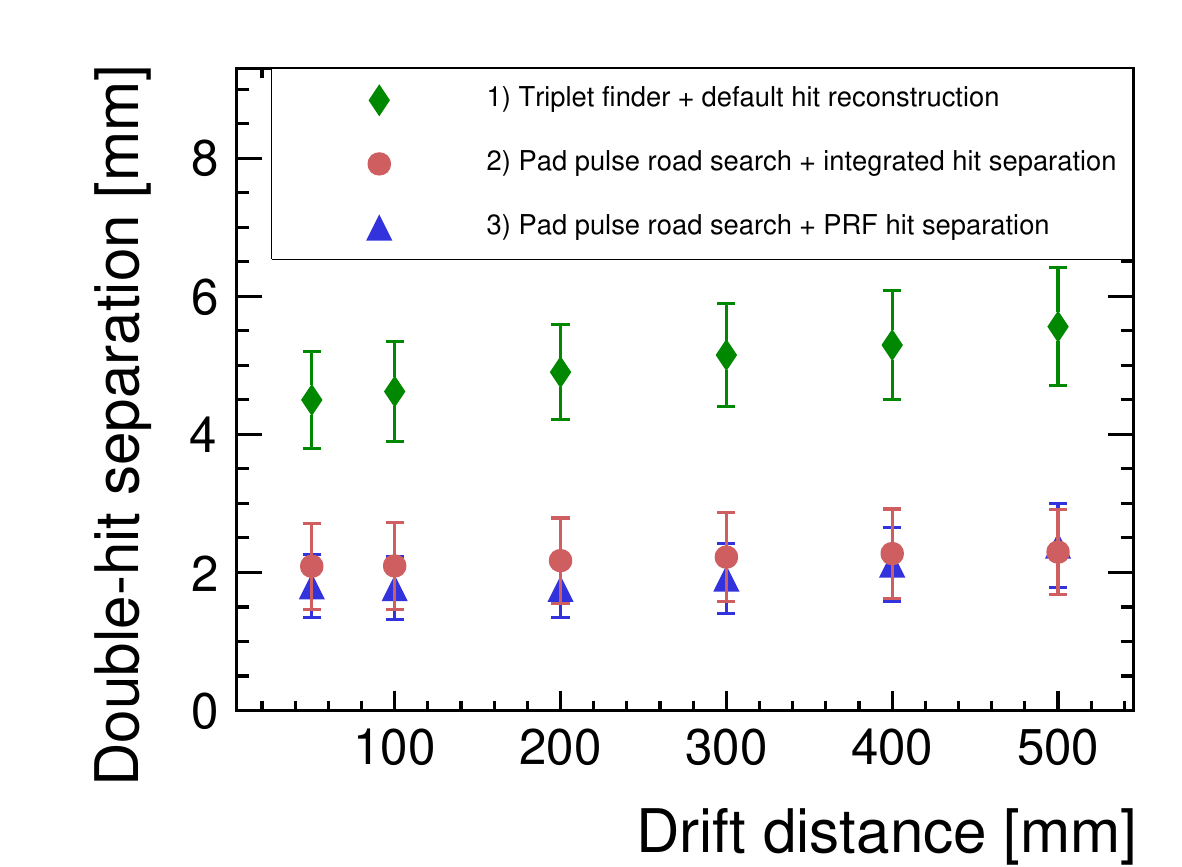}
    \caption{Measurement.}
    \label{fig:DHR_vs_drift_data}
  \end{subfigure}%
  \begin{subfigure}[b]{0.5\textwidth}
    \includegraphics[width=\textwidth]{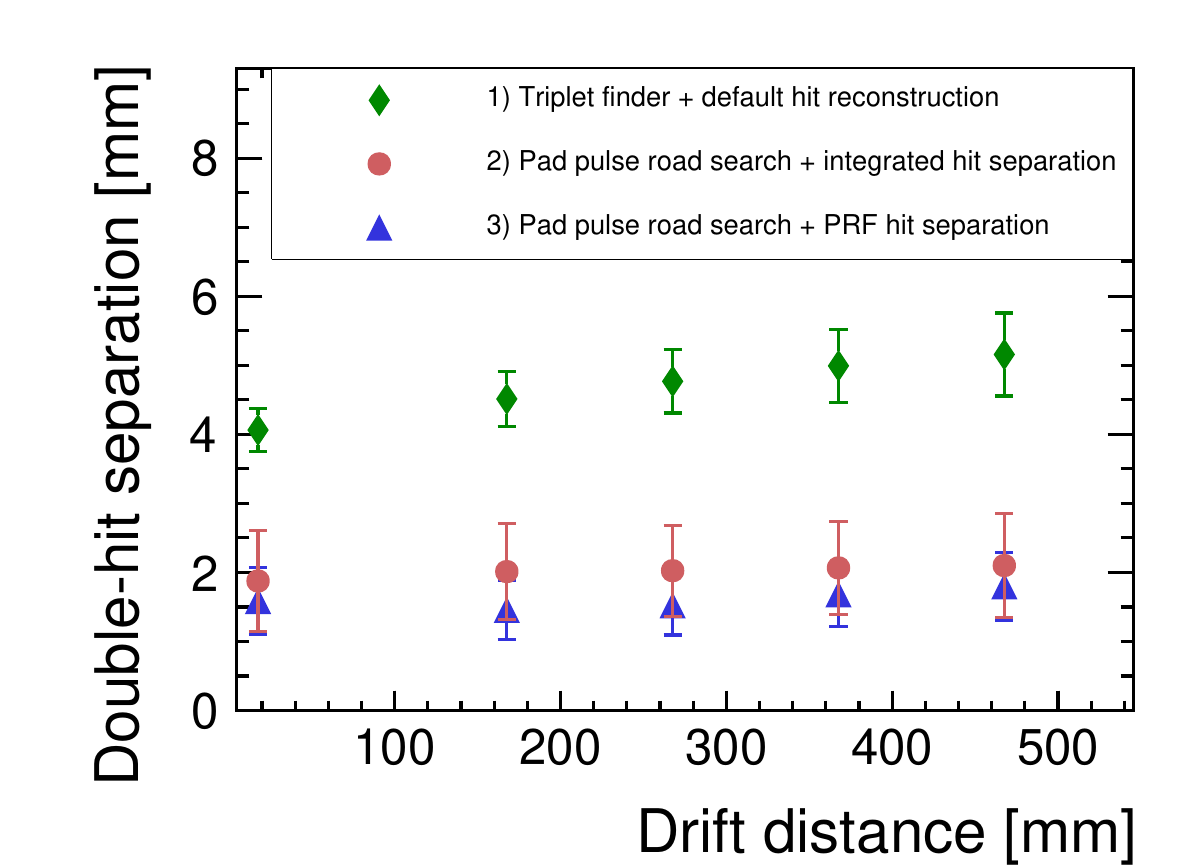}
    \caption{Simulation.}
    \label{fig:DHR_vs_drift_MC}
  \end{subfigure}
  \caption{Comparison of the double\-/hit separation for three different combinations of hit and track reconstruction algorithms \protect\subref*{fig:DHR_vs_drift_data}~for measured test\-/beam data and \protect\subref*{fig:DHR_vs_drift_MC}~for simulated tracks.
    Shown are the triplet track finder with default hit reconstruction and the pad pulse road search with its integrated as well as with the PRF hit separation.
    The bars depict the $\SI{+-1}{\sigma}$ width of \protect\cref{eq:errorfunc} describing the efficiency in dependence of the track distance from zero to full separation.
  }
  \label{fig:DHR_vs_drift}
\end{figure}

To study the impact of the charge diffusion in the TPC volume on the hit separation, the performance has been determined for different drift distances.
A comparison of the different methods is shown in \cref{fig:DHR_vs_drift} for \subref{fig:DHR_vs_drift_data}~data and \subref{fig:DHR_vs_drift_MC}~the simulation.
At short drift distances, the pad pulse road search performs better in combination with the PRF hit separation than with its integrated hit separation.
At larger distances, where the diffusion effect dominates, the performance of both methods is similar.
Overall, an improvement by more than a factor of two relative to the default hit reconstruction can be seen.
A more detailed description of this study is given in~\cite{FedorchPhD}.

\section{Resolution of the specific energy loss measurement}
\label{sec:dEdx_Res}

The specific energy loss per unit length (\dEdx) of a charged particle at a certain momentum depends on its rest mass.
Thus, by measuring the \dEdx and the momentum, the Bethe\-/Bloch curve can be used to infer the mass and identify the particle species~\cite{PhysRevD.98.030001}.
The specific energy loss is determined by measuring the charge deposited on each readout row, which is proportional to the energy loss.
The mean energy loss, as given by the Bethe\-/Bloch formula, is determined from the resulting Laudau\-/like distribution of all charge measurements along the trajectory.

In the analysis presented here, certain quality criteria are applied to ensure that only hits with an unbiased charge measurement contribute to the \dEdx estimation.
Before any other cuts, \num{15} rows containing too many noisy or dead channels, mostly due to faulty connections of the readout cables, are excluded from the analysis, reducing the number of usable rows from \num{84} to \num{69}.
Then, hits containing one or more dead channels are rejected.
Additionally, hits at the module edges, i.e.\ hits containing a pulse on a pad at either end of a row and all hits on the first and last row on each module, are ignored, as are hits on the two rows adjacent to the central bar of the ceramic frame.
If a hit is tagged as a multi\-/hit candidate it is not taken into account as well.
The pulse finder identifies these multi\-/hit candidates based on the presence of a double\-/pulse structure in time, which is equivalent to the drift direction.
Also the hit finding algorithm checks for double hits, both in the drift direction and the transverse plane.
In the drift direction, hits are tagged if any pad contains more than one pulse within the allowed window of arrival time.
In the transverse plane, double\-/hit candidates are identified by looking for a local minimum in the charge distribution between adjacent pads.
After all cuts, an average of \num{56.5} hits per track is used for the \dEdx calculation.
The standard deviation of this value is \num{2.3} hits.

For each of these hits, the charge $Q_\mathrm{hit}$ is normalized to the path length $\Delta s$ of the track segment over the corresponding pad row as an estimate of the energy loss:
\begin{equation}
  \frac{\Delta E}{\Delta x} \propto \frac{Q_\mathrm{hit}}{\Delta s} \enskip.
  \label{eq:dedx_formula}
\end{equation}
To gain a stable estimator for the mean energy loss of a track, several methods have been studied, including truncated and trimmed averages, all yielding very similar results.
In the end, the best resolution was achieved with the truncated mean of \SI{75}{\percent} of the smallest samples, cutting off the tail towards high charge values.

\begin{figure}
  \begin{subfigure}[b]{0.5\textwidth}
    \includegraphics[width=\textwidth]{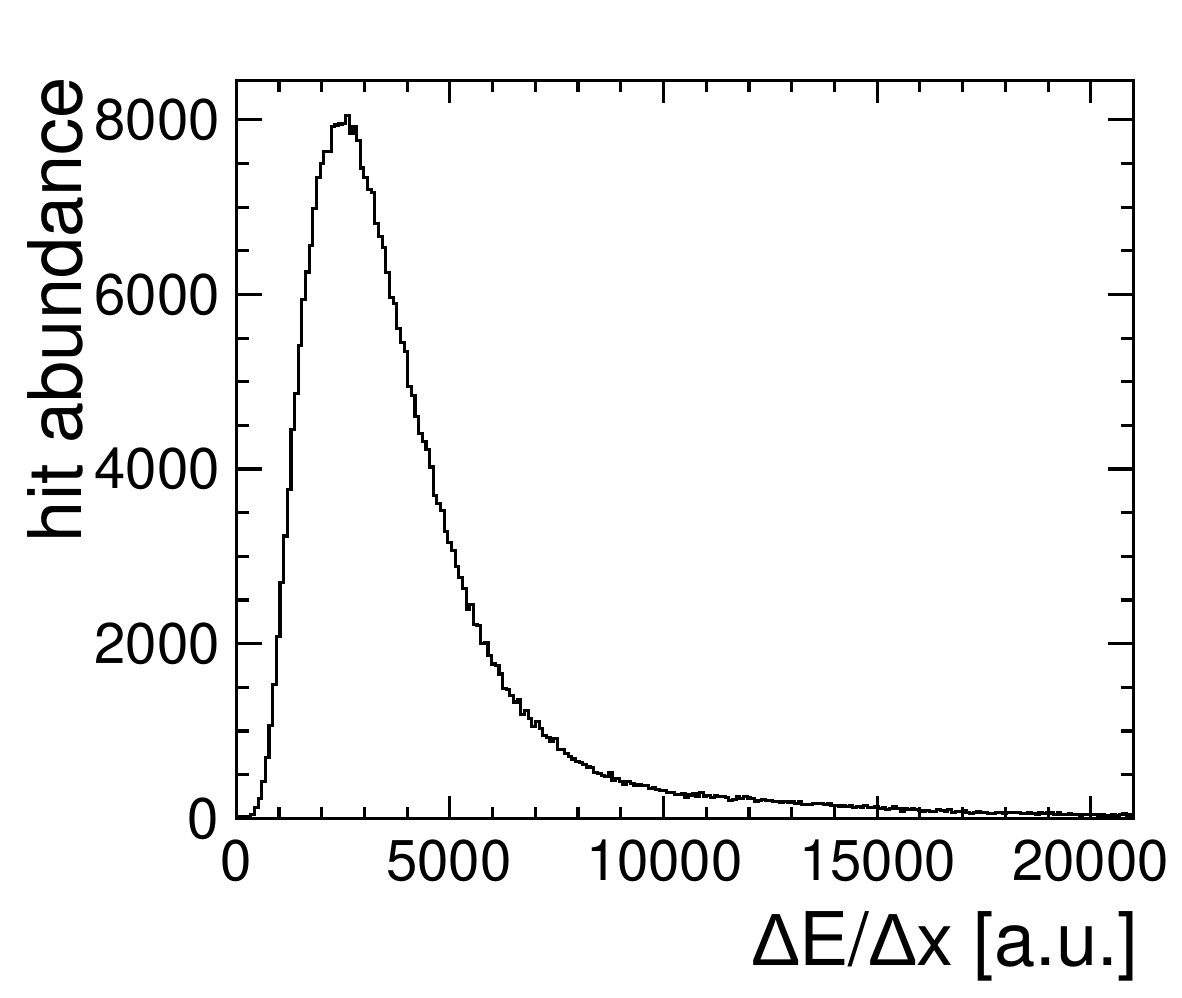}
    \caption{Distribution of $\Delta E / \Delta x$.}
    \label{fig:hit_dedx}
  \end{subfigure}%
  \begin{subfigure}[b]{0.5\textwidth}
    \includegraphics[width=\textwidth]{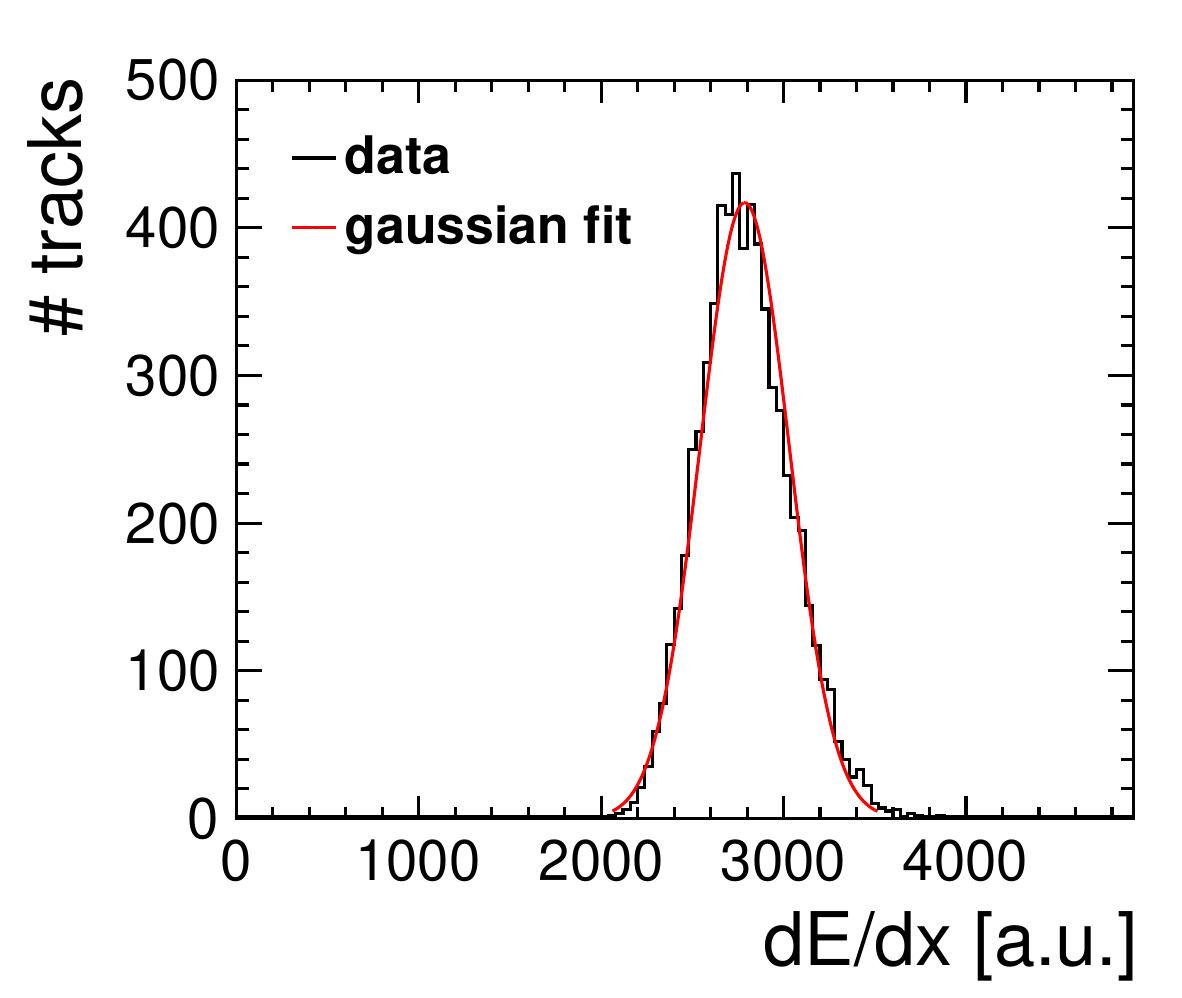}
    \caption{Distribution of \dEdx.}
    \label{fig:track_dedx}
  \end{subfigure}
  \caption{Distributions of the energy loss for one measurement run for
    \subref*{fig:hit_dedx}~individual samples and
    \subref*{fig:track_dedx}~the average of each track.
  }
  \label{fig:dedx_distributions}
\end{figure}

\Cref{fig:hit_dedx} shows the distribution of the energy loss per sample $\Delta E / \Delta x$ for all valid hits of one measurement run.
For each track, the average energy loss is calculated as the truncated mean of all the hits belonging to this track, i.e.\
$\dEdx = {\langle \Delta E / \Delta x \rangle}_{\num{75}}$.
The resulting distribution of \dEdx for all tracks of one run is plotted in \cref{fig:track_dedx}.
As a result of the Landau tail, a slight enhancement of the falling edge of the distribution compared to a Gaussian distribution can be observed.
For this reason, the relative \dEdx resolution is calculated from the RMS and mean of the distribution.
Since the relative resolution is found to be independent of the drift distance, the weighted average of \num{15} consecutive measurement runs at different drift distances is used to obtain a better estimate.
This results in a \dEdx resolution of $\sigma_{\dEdx} = \SI[parse-numbers=false]{(8.95\pm0.02\pm0.14)}{\percent}$\label{page:dEdxRes}, where the first uncertainty is the standard error of the weighted mean of the individual measurements.
The second error is the standard deviation of these measurements, which can be interpreted as a systematic uncertainty.


To extrapolate to the \dEdx resolution achievable at the final ILD TPC, two methods relying exclusively on the test\-/beam data have been applied.
For the first method the measured tracks were artificially shortened by randomly discarding a number of hits until the desired track length was reached.
This method can of course only produce tracks with fewer hits than the average of \num{56.5} hits per track available from the test\-/beam data.
With this method \num{14} sets of shortened tracks with lengths from \num{4} to \num{56} hits were created.
For the ILD TPC with an outer radius of \SI{1770}{\mm}, the foreseen number of pad rows is \num{220}.
Therefore, the second method is designed to acquire pseudo\-/tracks longer than available in the test\-/beam data.
For technical reasons, events can only be accessed one at a time in the analysis.
Thus, if the pseudo\-/track is supposed to contain more hits than available from the track in the current event, all remaining hits are added to the pseudo track.
Then the process is continued with hits from the track in the next event.
This is repeated until the remaining number of hits required for the pseudo track is smaller than the number of hits available from the real track in an event.
Then the hits are selected randomly from the real track until the desired track length is reached.
The latter is also done for a new pseudo track that is shorter than the real track in an event.
This method allows to emulate tracks with an arbitrary number of hits.
Each hit is used only once during this process to avoid introducing correlations.
Twenty\-/five ensembles of pseudo tracks with lengths from \num{4} up to \num{220} hits are created in this way.
For both methods the \dEdx resolution for each track length is extracted in the way described above, including the combination of consecutive measurement runs.

\begin{figure}
  \begin{subfigure}{0.5\textwidth}
    \centering
    \includegraphics[
      width=\textwidth,
      height=0.33\textheight,
      keepaspectratio,
    ]{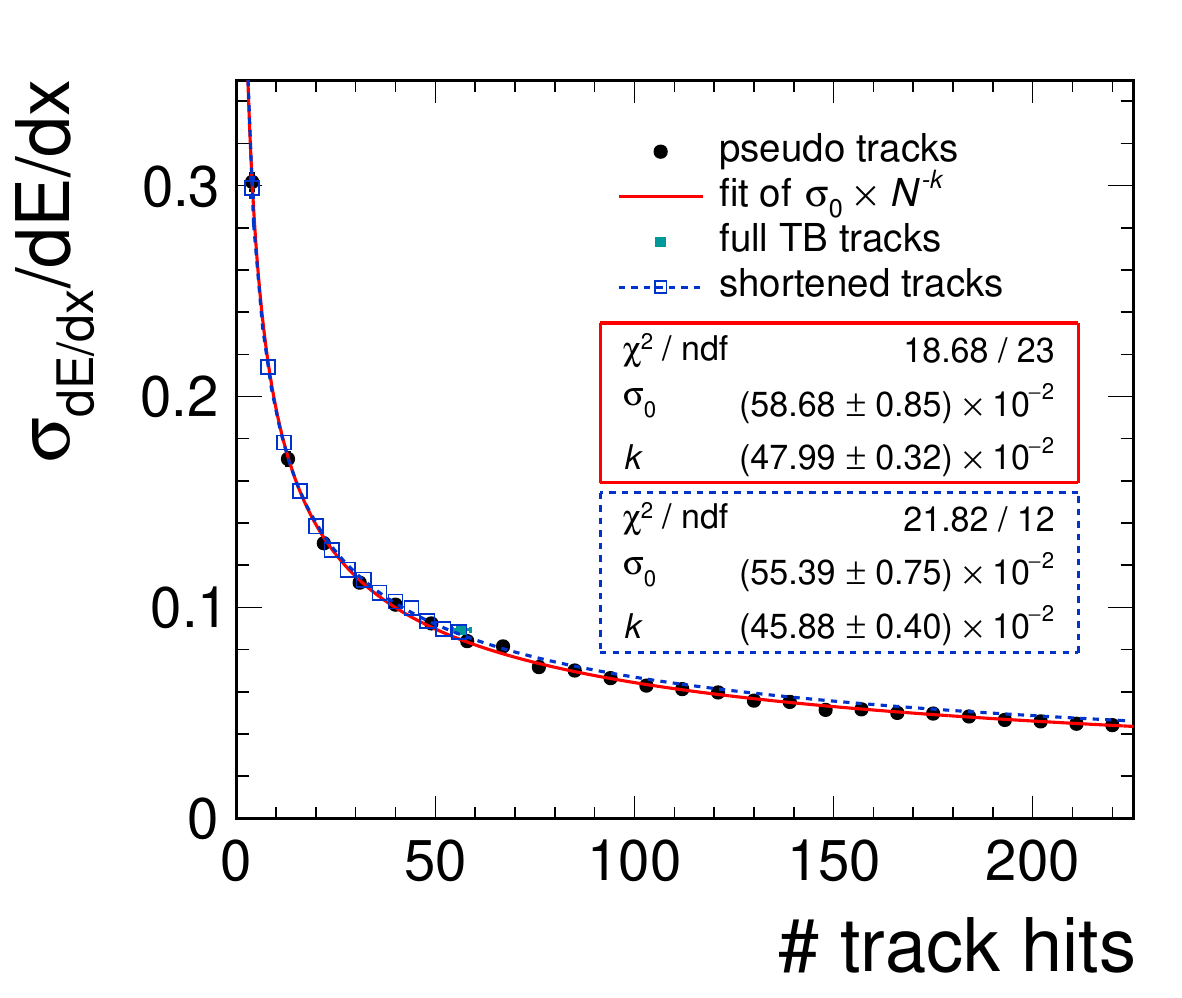}
  \end{subfigure}%
  \begin{subfigure}{0.5\textwidth}
    \centering
    \includegraphics[
      width=\textwidth,
      height=0.33\textheight,
      keepaspectratio,
    ]{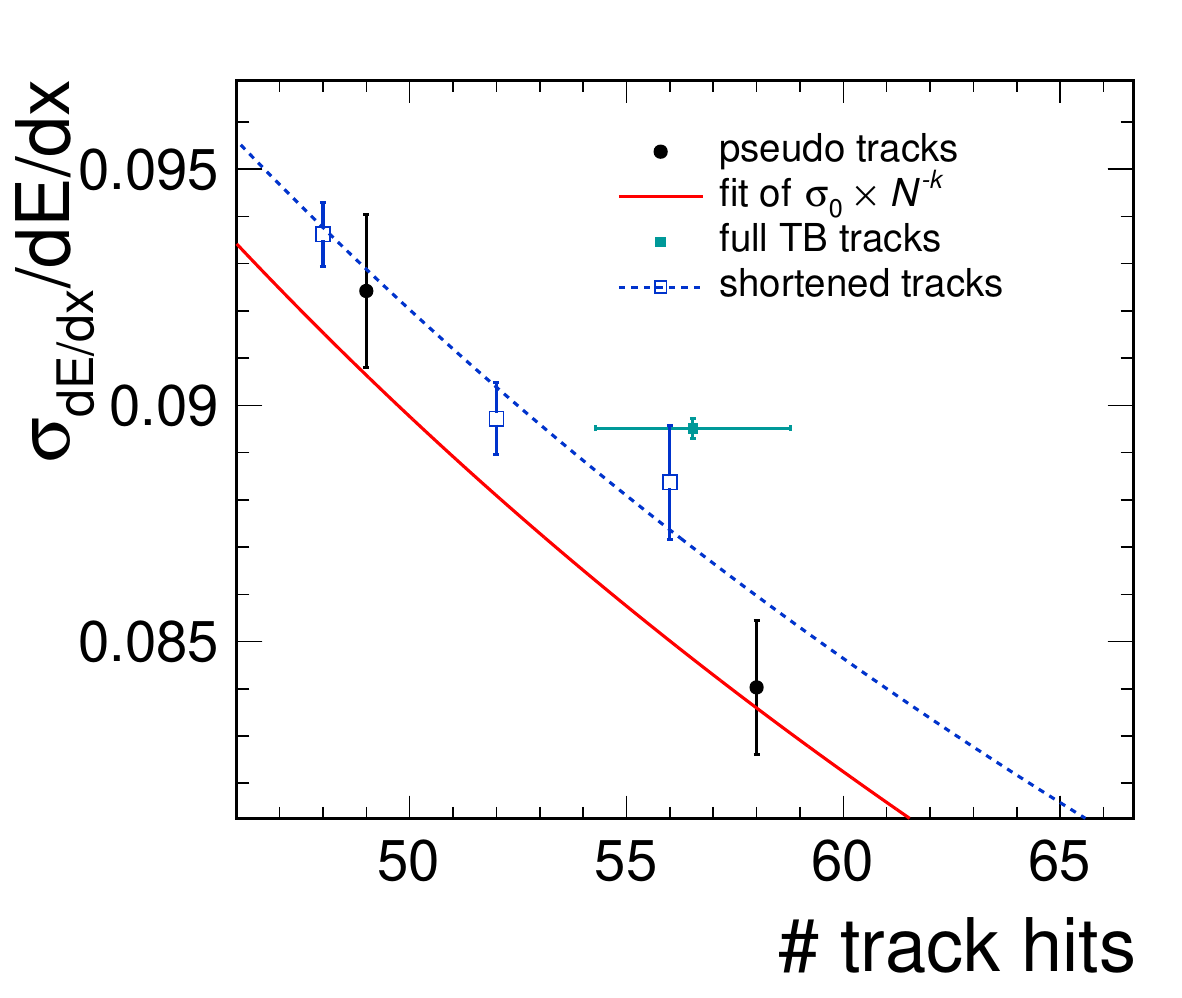}
  \end{subfigure}
  \caption{Extrapolation of the \dEdx resolution to the ILD TPC dimensions, based on test\-/beam data.
    The two extrapolation methods are described in the text.
    The single point labelled \emph{full TB tracks} represents the measurement of real test\-/beam tracks with \num{56.5+-2.3} hits.
    The right figure shows a zoom into the area around this measurement point.
  }
  \label{fig:dedx_reso_ILD}
\end{figure}

The results are shown in \cref{fig:dedx_reso_ILD}.
Naively, one could expect a square\-/root dependence of the resolution on the number of hits.
Instead, former experiments have found that power laws $\sigma_{\dEdx} = \sigma_0 \cdot N^{-k}$ with exponents $k<\num{0.5}$ describe their data best~\cite{allison1980,walenta1979_ieee,walenta1979_nim,opal1987_dedx}.
Therefore, a fit of a power law is performed on the extrapolated resolution, including only the statistical errors, with the exponent $k$ and the constant $\sigma_0$ as free parameters.
For the shortened tracks this results in an exponent of $k = \num{0.459(4)}$.
From the fit, the \dEdx resolution in the ILD TPC with up to \num{220} hits is estimated to be \SI[parse-numbers=false]{(4.66\pm0.04\pm0.12)}{\percent}.
The first error is the uncertainty calculated from the fit results.
As above, the second error is the standard deviation of the extrapolation results gained from the individual measurements, which is interpreted as a systematic uncertainty.
The corresponding values for the pseudo\-/track method are $k = \num{0.480(3)}$
and a resolution with the ILD TPC of \SI[parse-numbers=false]{(4.41\pm0.02\pm0.10)}{\percent}.
The resolution results are only valid for tracks with a polar angle of $\theta=\ang{90}$ with respect to the $z$\=/axis.
For tracks with $\theta\neq\ang{90}$, the effective track length in the TPC increases as $L \propto 1/\sin\theta$ while the number of samples stays fixed.
According to~\cite{allison1980}, this is expected to result in an improved resolution $\sigma_{\dEdx} \propto L^{-k'}$ with $k'=\num{0.32}$.

To asess the validity of the two extrapolation methods, a comparison of the fitted functions in \cref{fig:dedx_reso_ILD} with the measurement of real tracks in the prototype, given on \cpageref{page:dEdxRes}, is performed.
While the resolution at \num{56.5} hits predicted by the fit to the shortened tracks
is in reasonable agreement with the full test\-/beam measurement,
the corresponding resolution using the pseudo track method shows
a discrepancy of about $\SI{3}{\sigma}$ to the full test\-/beam measurement.
This may indicate a bias in this method of extrapolation.
A potential source of this bias may arise from an observed variation of the average measured charge between the individual rows with a standard deviation of \SI{\sim3}{\%}.
Since longer pseudo\-/tracks contain several hits from the each pad row in multiple events, the charge correlation between these hits is larger than for hits on several different rows.
This may reduce the overall variance of the hit charge for these long pseudo tracks, resulting in a better resolution.
Therefore, the slightly worse result from the shortened track method is trusted for now, which still allows to fulfill the goal of a \dEdx resolution of better than \SI{5}{\percent} defined in~\cite{ilc_tdr_detectors}.

\subsection[Optimization of the readout granularity for \dEdx]{Optimization of the readout granularity for \boldmath\dEdx}
\label{sec:dedex_gran}

To explore possible ways to improve the particle identification via the specific energy loss, a simulation study was carried out to investigate the performance of two methods to measure \dEdx, in dependence of the granularity of the readout.
The usual approach to measure \dEdx by determining the number of generated electrons, i.e.\ the total charge, is compared to an alternative approach of counting the number of primary ionizing interactions instead.
In the first approach ---~here called \emph{charge summation}~--- the number of electrons per path length follows a Landau\-/like distribution.
The long tail of this distribution worsens the correlation of the measured average energy loss and the particle species.
The second approach ---~called \emph{cluster counting}~--- relies on the number of ionizing interactions, i.e.\ ionization clusters, of the incident particle, which follows a Poisson distribution with a significantly smaller width.
This results in a better correlation and particle identification power, as shown in a previous simulation study presented in~\cite{DEDXCLUSTER}.

The number of ionizing interactions of the incident particle can be determined by reconstructing and counting charge clusters on a highly granular readout.
If the pad or pixel size and the diffusion are smaller than the typical distance between charge clusters, charge deposits on the readout plane are correlated and can be combined to reconstructed clusters.
Unfortunately, it is not possible to meet these conditions fully, in particular a low enough diffusion, to reach a counting efficiency close to \SI{100}{\percent}.
However, the study in~\cite{DEDXCLUSTER} has shown, that even with a cluster counting efficiency of only \SI{20}{\percent} the resulting separation power is better than by measuring the energy loss by charge summation.
This novel approach has not yet been applied in an experiment, but developments in low\-/noise ASICs and compact packaging may facilitate this in a future detector.

In the present study the performance of the two approaches is compared using the particle separation power.
The separation power $S_{AB}$ for two particle types $A$ and $B$ is defined as
\begin{equation}
    S_{AB} = \frac{\left|\mu_{A} - \mu_{B}\right|}{\sqrt{\frac{1}{2}\left(\sigma^{2}_{A} + \sigma^{2}_{B}\right)}} \enskip.
  \label{eq:sp}
\end{equation}
Here, $\mu_{A/B}$ are the estimators for the energy loss ---~usually the mean \dEdx~--- and $\sigma_{A/B}$ are the corresponding measurement accuracies for particle types $A$ and $B$, respectively.
In the denominator the combined width from $\sigma_A$ and $\sigma_B$ is used since this study is performed without considering any specific experiment background.
Therefore, there is no signal or background particle type and both are compared using an equal number of particles of type $A$ and $B$.

\begin{figure}
  \centering
  \includegraphics[
    width=\textwidth,
    height=0.33\textheight,
    keepaspectratio,
  ]{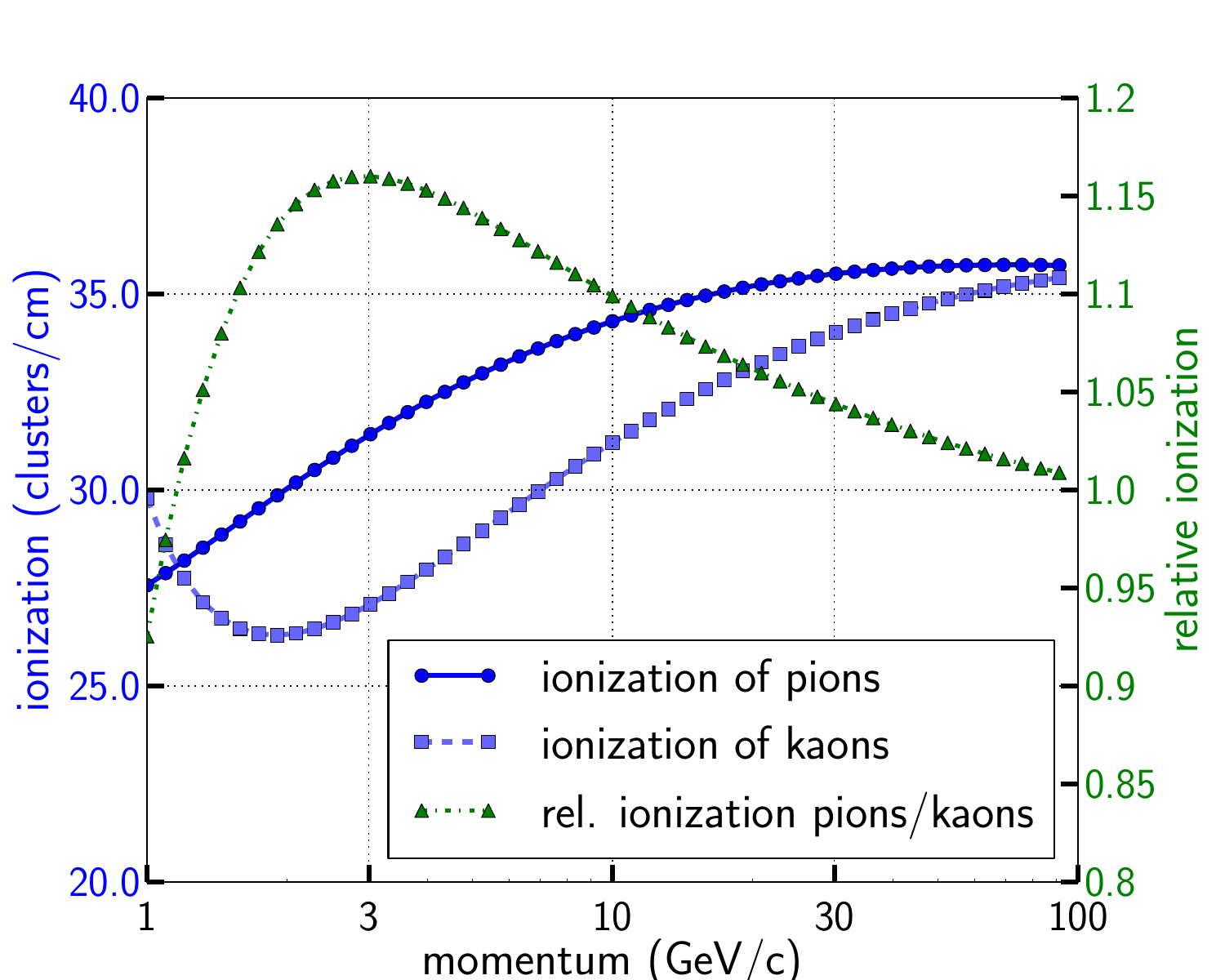}
  \caption{The ionization density of pions and kaons in the simulated argon\-/based gas mixture.
      The third graph shows the relative ionization density of the two particle species on the axis to the right.
  }
  \label{fig:ionization}
\end{figure}

Following \cref{eq:sp}, the performance of particle separation is proportional to the difference of the average ionization.
As shown in \cref{fig:ionization}, the relative ionization of different particle species depends on the momentum.
Throughout the following comparisons, the separation power is evaluated at the maximum of the ionization difference between pions and kaons of about \SI{16}{\percent} at a momentum of \SI{3}{\GeVperc}.
To compare the simulation with the results from test\-/beam measurements with relativistic electrons, a conversion between the measured \dEdx resolution and the pion\-/kaon separation power was performed.
This conversion takes into account the dependence on the number of samples to adjust for the difference in track length, as well as the differences between the ionization densities of electrons, pions and kaons at the respective momenta.
For the test\-/beam measurement presented in this work, this results in a separation power of $S_{\pi K} \approx \num{1.5}$ for the simulated track length of \SI{300}{\mm}.
Equivalent conversions are performed for results with other pad\-/based readout systems~\cite{AsianGEM_dEdx,DEDXMM} as well as a pixel\-/based readout system with a pitch of \SI{55}{\um}~\cite{LIGTENBERG201818}.
The latter is considered twice, once with an anode coverage of \SI{95}{\percent}, comparable to the pad\-/based systems, and once with about \SI{60}{\percent} coverage, comparable to the current pixel\-/based prototypes.


\begin{figure}
  \centering
  \includegraphics[
    width=\textwidth,
    height=0.33\textheight,
    keepaspectratio,
  ]{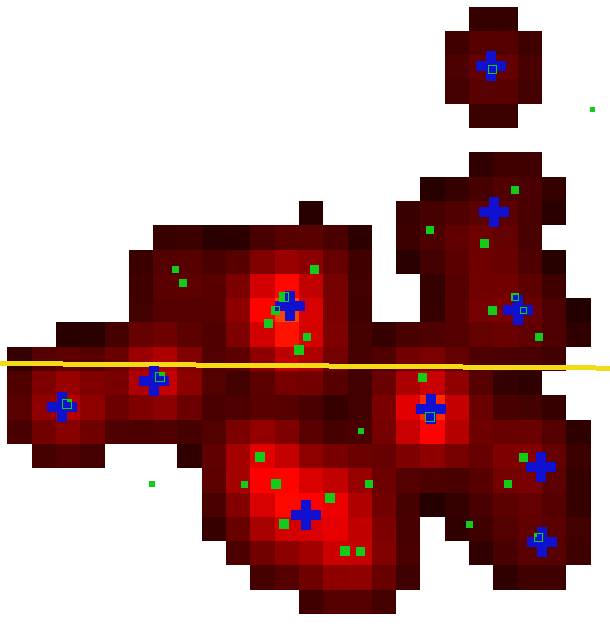}
  \caption{Simulation of the charge deposition in T2K gas at \SI{1}{\tesla} magnetic field on a readout with \SI{220}{\um} wide pads.
    Measured charge values are shown from low to high in dark to bright red.
    The yellow line shows the path of the simulated incident minimum ionizing particle at \SI{1}{\m} drift distance.
    The green squares depict the position of the drifted electrons arriving at the GEM stack, with their respective size being proportional to their randomized amplification.
    The blue crosses represent the cluster centers reconstructed by the source\-/extractor software.
  }
  \label{fig:clusterfind}
\end{figure}

The MarlinTPC framework, introduced in \cref{sec:reco}, has been used for the simulation and the reconstruction.
The detailed simulation operates at the level of individual electrons from primary ionization and entails all steps of the electron propagation in the detecor, i.e.\ ionization, drift, amplification in a triple GEM stack, projection onto the readout plane and finally the digitization in the readout electronics.
For an accurate ionization simulation including so-called delta electrons, an ionization table generated by HEED~\cite{HEED} was used as input.
For the reconstruction of the ionization clusters, the source\-/extractor software package~\cite{SourceExtractor} has been integrated into the MarlinTPC reconstruction.
The simulation is performed for square pads, varying the pad pitch between \SI{55}{\um} and \SI{6}{\mm}.
\Cref{fig:clusterfind} shows an example of the resulting charge deposition on pads with a pitch of \SI{220}{\um}.
Also shown are the incident particle track, the corresponding signal electrons and the reconstructed cluster centers.
In the simulation the gain of the GEM stack was adjusted by varying the GEM voltages between \SI{230}{\V} and \SI{280}{\V} per GEM, resulting in overall gains between \num{\sim500} and \num{62000}.
GEM voltages above \SI{280}{\V} were not used since this is considered the limit of stable operation without frequent discharges.
For each pad size and method, the gain resulting in the best separation power was used.

\begin{figure}
  \centering
  \begin{subfigure}{\textwidth}
    \centering
    \includegraphics[width=\textwidth]{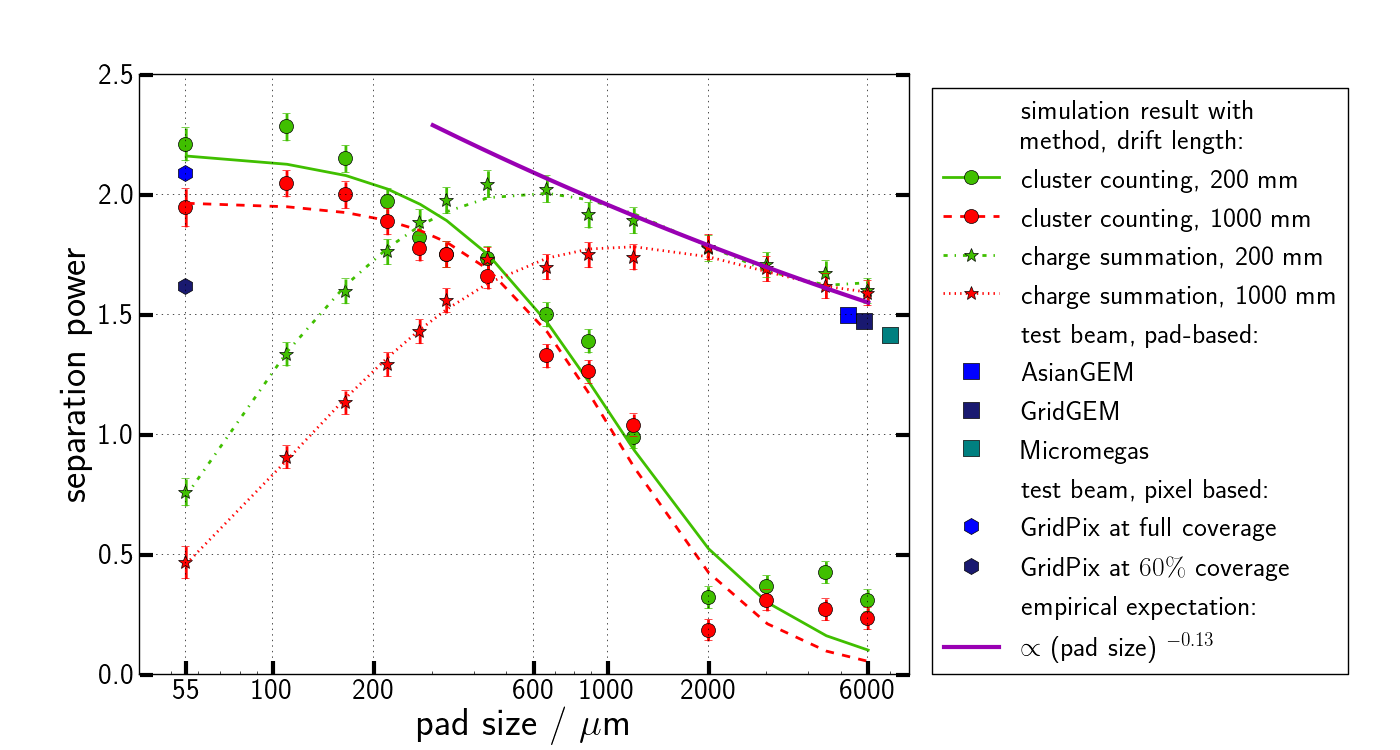}
    \caption{Pion-kaon separation power.}
    \label{fig:dedxcomppadsize_sp}
  \end{subfigure}\\
  \begin{subfigure}{\textwidth}
    \centering
    \includegraphics[width=\textwidth]{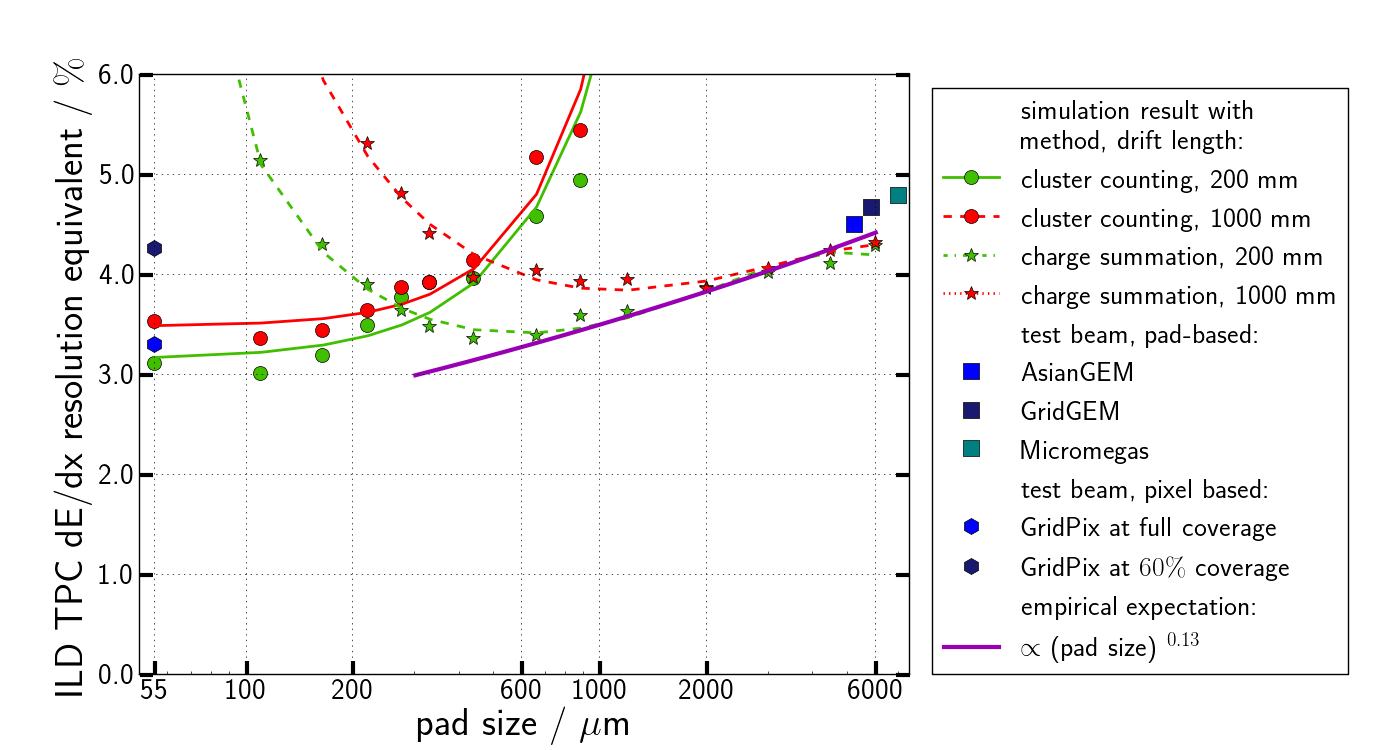}
    \caption{\dEdx resolution.}
    \label{fig:dedxcomppadsize_res}
  \end{subfigure}
  \caption{Results of the simulation study showing the performance of the charge measurement and the cluster counting methods at two drift lengths as a function of the size of the square pads.
    \protect\subref*{fig:dedxcomppadsize_sp}~The separation power between pion and kaon tracks of \SI{300}{\mm} length.
    \protect\subref*{fig:dedxcomppadsize_res}~The \dEdx resolution for relativistic electrons with a track length of \SI{1300}{\mm}.
    Also included are several values from measurements with different systems:
    AsianGEM~\protect\cite{AsianGEM_dEdx}, GridGEM (this work), Micromegas~\protect\cite{DEDXMM}, GridPix~\protect\cite{LIGTENBERG201818}.
  }
  \label{fig:dedxcomppadsize}
\end{figure}

\Cref{fig:dedxcomppadsize} shows the results of the simulations for both \dEdx measurement methods in terms of
\subref{fig:dedxcomppadsize_sp}~the achievable separation power between pion and kaon tracks of \SI{300}{\mm} length and
\subref{fig:dedxcomppadsize_res}~the resolution for electrons with a momentum of \SI{5}{\GeVperc} and a track length of \SI{1300}{\mm}, equivalent to the ILD TPC\@.
Both are shown as a function of the size of the readout pads, once for a drift distance of \SI{200}{\mm} and once for a longer drift distance of \SI{1000}{\mm}, always at a magnetic field of \SI{1}{\tesla}.
These parameters are comparable to the situation at the prototype TPC with a maximum drift length of \SI{570}{\mm} and a magnetic field of \SI{1}{\tesla}.
The overall diffusion is also similar to the conditions at a potential ILD TPC, which would have a larger drift length of over \SI{2000}{\mm} but operates at a higher magnetic field of \SI{3.5}{\tesla}.

For very small readout pads the cluster counting method yields a very good separation power, reaching up to $S=\num{2.3}$ for a drift distance of \SI{200}{\mm} and a pad size of \SI{110}{\um}.
Above \SI{300}{\um} the separation power quickly drops with increasing pad size and becomes nearly independent of the drift distance.

For the charge summation method the separation power rises with decreasing pad size, from about $S=\num{1.5}$ with \SI{6}{\mm} pads to a value of \num{1.75} or \num{2.1}, depending on the drift distance.
At very small pad sizes the separation power decreases again as the charge is distributed over many pads and falls below the noise threshold.
This is compensated to some degree by a small drift distance and thus lower diffusion, resulting in a separation power of \num{2.1} at an optimum pad size of \SI{440}{\um} for a drift distance of \SI{200}{\mm}.
As is presented in~\cite{Blum:2008}, former experiments suggest the separation power measured by charge summation to behave as $S \propto (\text{pad\-/size})^{\alpha}$, with $\alpha$ between \num{-0.12} and \num{-0.15}.
This dependence is included as the purple line in \cref{fig:dedxcomppadsize}, and is compatible with the simulation for large pads, until the aforementioned threshold effect starts to play a role.

The simulation agrees very well with the performance of the test\-/beam measurements with pad\-/based systems with pad sizes around \SI{6}{\mm}.
It describes the increase in separation power with decreasing pad size, including the transition region going from macroscopic pads and charge summation to very small pixels and cluster counting.
At a granularity of \SI{55}{\um} the simulation with cluster counting reaches a separation power similar to the result from the current pixel\-/based readout systems.

This study demonstrates that to perform particle identification using \dEdx, a pad size of about a factor of \num{10} smaller than currently used could improve the separation power by about \SIrange{15}{30}{\percent}, depending on the drift distance.
To do particle identification by cluster counting, the granularity has to be increased by another factor of two to improve the particle separation power, with the benefit of being more independent of the drift distance and the corresponding diffusion.
These results show the direction a further study could take to optimize the particle identification.
In the final pad\-/size optimization, additional factors such as the pattern\-/recognition performance, the spatial resolution and the overall number of channels will have to be taken into account.

\section{Conclusion}
\label{sec:conclusion}

A new generation of readout modules based on a triple GEM stack were operated in a large prototype TPC at the DESY test beam.
The construction procedures for this generation of modules have been improved resulting in a more controlled and reproducible production, while at the same time ensuring a higher quality.
The performance of the new readout modules was studied.
The comparison to measurements with the previous readout generation show a good reproducibility of the performance.
The extrapolation of the spatial single-point resolution confirmed that the requirements set for the ILD TPC can be fulfilled, if the gas contamination is minimized.

Furthermore, two new approaches have been implemented to identify and split hits of close\-/by tracks.
This includes a pad pulse road search as combined hit and track finder as well as a hit splitting algorithm based on fitting pad response functions to the double\-/hit candidates.
The methods were studied on simulated and measured data and compared to the existing method of hit reconstruction.
The combination of both new methods leads to an improvement of the double\-/hit separation power by more than a factor of two, down to a track distance well below \num{2} times the pitch of the readout pads.
For the pad pitch of \SI{1.26}{\mm} in the $r\varphi$ direction, this fulfills the requirement of \SI{2}{\mm} for the ILD TPC\@.

Finally, the resolution of the specific energy loss measurement was determined for the test\-/beam data, resulting in a resolution of \SI[parse-numbers=false]{(8.95\pm0.02\pm0.14)}{\percent}.
The extrapolation of these results to the final ILD TPC yields a resolution of \SI[parse-numbers=false]{(4.66\pm0.04\pm0.12)}{\percent}, which meets the goal of \SI{5}{\percent} for the ILD TPC\@.
A simulation study examining ways to improve the particle identification by specific energy loss was performed.
It indicates that with smaller pads of about \SI{500}{\um} pitch a separation power up to \SI{30}{\percent} better could be achieved.
With even smaller pads below \SI{300}{\um} pitch, a cluster counting method could be employed, further improving the separation power.


\section*{Acknowledgments}

This material is based upon work supported by the National Science Foundation of the U.S.A. under Grant No.~0935316, was supported by the Japan Society for the Promotion of Science (JSPS) KAKENHI Grant No.~23000002 and was funded by the Deutsche Forschungsgemeinschaft (DFG, German Research Foundation) -- 491245950. 
The research leading to these results has received funding from the European Commission under the 6th Framework Programme ``Structuring the European Research Area'', contract no.~RII3-026126, and under the FP7 Research Infrastructures project AIDA, grant agreement no.~262025. 

Special thanks go to Y.~Makida, M.~Kawai, K.~Kasami and O.~Araoka of the KEK IPNS cryogenic group, and A.~Yamamoto of the KEK cryogenic center for their support in the configuration and installation of the superconducting PCMAG solenoid.

The measurements leading to these results have been performed at the Test Beam Facility at DESY Hamburg (Germany), a member of the Helmholtz Association. The authors would like to thank the technical team at the DESY~II accelerator and test beam facility for the smooth operation of the test beam and the support during the test\-/beam campaign.

The contributions to the experiment by the University of Lund, KEK, Nikhef and CEA are gratefully acknowledged.

\section*{Disclaimer}

Any opinions, findings, and conclusions or recommendations expressed in this material are those of the authors and do not necessarily reflect the views of any of the funding agencies involved.
No warranty expressed or implied is made with regard to any information or its use in this paper.

\iftoggle{biblatex}{
  \printbibliography[
  ]
}{
  \bibliographystyle{JHEP}
  \bibliography{references}

@article{GEM,
  author = "Sauli, Fabio",
  title = "GEM: A new concept for electron amplification in gas detectors",
  journaltitle = "Nuclear Instruments and Methods in Physics Research",
  journalsubtitle = "Section A: Accelerators, Spectrometers, Detectors and Associated Equipment",
  shortjournal = "Nucl. Instrum. Methods Phys. Res.",
  series = "A",
  volume = "386",
  number = "2",
  pages = "531-534",
  year = "1997",
  issn = "0168-9002",
  doi = "10.1016/S0168-9002(96)01172-2",
  url = "https://www.sciencedirect.com/science/article/pii/S0168900296011722",
}

@report{MarlinTPC,
  author = "Abernathy, Jason and Dehmelt, Klaus and Diener, Ralf and Hunt, Jim and Janssen, Matthias Enno and Killenberg, Martin and Krautscheid, Thorsten and Münnich, Astrid and Ummenhofer, Martin and Vogel, Adrian and Wienemann, Peter",
  title = "MarlinTPC: A Marlin based common TPC software framework for the LC-TPC Collaboration.",
  institution = "Deutsches Elektronen-Synchrotron DESY",
  type = "LC Note",
  number = "LC-TOOL-2007-001",
  year = "2007",
  eprint = "0709.0790",
  eprinttype = "arXiv",
  eprintclass = "physics.ins-det",
  url = "https://bib-pubdb1.desy.de/record/83142",
}

@mvbook{ilc_tdr,
  publisher = "The Linear Collider Collaboration",
  collaboration = "The Linear Collider Collaboration",
  title = "The International Linear Collider",
  titleaddon = "Technical Design Report",
  volumes = "4",
  year = "2013",
  url = "https://www.linearcollider.org/",
  related = "ilc_tdr_summary,ilc_tdr_physics,ilc_tdr_accelerator1,ilc_tdr_accelerator2,ilc_tdr_detectors",
  relatedtype = "multivolume",
}

@book{ilc_tdr_detectors,
  crossref = "ilc_tdr",
  editor = "Behnke, Ties and Brau, James E. and Burrows, Philip N. and Fuster, Juan and Peskin, Michael and Stanitzki, Marcel and Sugimoto, Yasuhiro and Yamada, Sakue and Yamamoto, Hitoshi",
  title = "Detectors",
  volume = "4",
  isbn = "978-3-935702-78-2",
  eprint = "1306.6329",
  eprinttype = "arXiv",
  eprintclass = "physics.ins-det",
}

@article{Schade,
  author = "Behnke, T. and Dehmelt, K. and Diener, R. and Steder, L. and Matsuda, T. and Prahl, V. and Schade, P.",
  title = "A lightweight field cage for a large TPC prototype for the ILC",
  journaltitle = "Journal of Instrumentation",
  shortjournal = "J. Instrum.",
  volume = "5",
  number = "10",
  pages = "10011",
  year = "2010",
  doi = "10.1088/1748-0221/5/10/P10011",
  eprint = "1006.3220",
  eprinttype = "arXiv",
  eprintclass = "physics.ins-det",
}

@PhdThesis{HallermannPhD,
  author = "Hallermann, Lea",
  title = "Analysis of GEM Properties and Development of a GEM Support Structure for the ILD Time Projection Chamber",
  editor = "Heuer, Rolf-Dieter and Bechtle, Philip",
  editortype = "advisor",
  year = "2010",
  school = "Universität Hamburg",
  doi = "10.3204/DESY-THESIS-2010-015",
  url = "https://bib-pubdb1.desy.de/record/94178",
}

@phdthesis{ZenkerPhD,
  author = "Zenker, Klaus",
  title = "Studies of field distortions in a Time Projection Chamber for the International Linear Collider",
  year = "2014",
  school = "Universität Hamburg",
  url = "https://bib-pubdb1.desy.de/record/207651",
}

@article{ALTROchip,
  author = "Esteve Bosch, R. and Jimenez de Parga, A. and Mota, B. and Musa, L.",
  title = "The ALTRO chip: A 16-channel A/D converter and digital processor for gas detectors",
  journaltitle = "IEEE Transactions on Nuclear Science",
  shortjournal = "IEEE Trans. Nucl. Sci.",
  volume = "50",
  number = "6",
  pages = "2460-2469",
  year = "2003",
  ISSN = "1558-1578",
  doi = "10.1109/TNS.2003.820629",
}

@report{LC-DET-2012-080,
  author = "Hedberg, V. and Jönsson, L. and Lundberg, B. and Mjörnmark, U. and Oskarsson, A. and Österman, L.",
  title = "Development of the Readout System for a TPC at the Future Linear Collider",
  institution = {Lund University},
  type = "LC Note",
  number = "LC-DET-2012-080",
  year = "2012",
  url = "https://flc.desy.de/lcnotes",
}

@article{pcmag:magnet,
  author = "A. Yamamoto and K. Anraku and R. Golden and T. Haga and Y. Higashi and M. Imori and S. Inaba and B. Kimbell and N. Kimura and Y. Makida and H. Matsumoto and H. Matsunaga and M. Motoki and J. Nishimura and M. Nozaki and S. Orito and T. Saeki and J. Suzuki and N. Takimi and K. Tanaka and I. Ueda and N. Yajima and T. Yamagami and A. Yamamoto and T. Yoshida and K. Yoshimura",
  title = "Balloon-borne experiment with a superconducting solenoidal magnet spectrometer",
  journaltitle = "Advances in Space Research",
  shortjournal = "Adv. Space Res.",
  volume = "14",
  number = "2",
  pages = "75-87",
  year = "1994",
  issn = "0273-1177",
  doi = "10.1016/0273-1177(94)90071-X",
  url = "https://www.sciencedirect.com/science/article/pii/027311779490071X",
}

@article{lcsoft,
  author = "Behnke, Ties and Gaede, Frank",
  title = "Software for the international linear collider: Simulation and reconstruction frameworks",
  journaltitle = "Pramana",
  journalsubtitle = "Journal of Physics",
  shortjournal = "Pramana",
  issuetitle = "Proceedings of the Linear Collider Workshop (LCWS06) - Part II",
  volume = "69",
  number = "6",
  pages = "1089-1092",
  date = "2007-12-01",
  issn = "0973-7111",
  doi = "10.1007/s12043-007-0233-z",
}

@article{Marlin,
  author = "Gaede, F.",
  title = "Marlin and LCCD: Software tools for the ILC",
  journaltitle = "Nuclear Instruments and Methods in Physics Research",
  journalsubtitle = "Section A: Accelerators, Spectrometers, Detectors and Associated Equipment",
  shortjournal = "Nucl. Instrum. Methods Phys. Res.",
  series = "A",
  volume = "559",
  pages = "177-180",
  year = "2006",
  doi = "10.1016/j.nima.2005.11.138",
}

@inproceedings{lcio,
  author = "Gaede, Frank and Behnke, Ties and Graf, Norman and Johnson, Tony",
  title = "LCIO: A Persistency framework for linear collider simulation studies",
  booktitle = "Conference Proceedings of the 13th International Conference on Computing in High-Energy and Nuclear Physics",
  series = "eConf",
  number = "C0303241",
  year = "2003",
  //eventtitle = "The 13th International Conference on Computing in High-Energy and Nuclear Physics",
  eventtitleaddon = "CHEP03",
  eventdate = "2003-03-24/2003-03-28",
  venue = "La Jolla, California",
  eprint = "physics/0306114",
  eprinttype = "arXiv",
  eprintclass = "physics",
}

@report{KleinwortHough,
  author = "Kleinwort, Claus",
  title = "A Track Finding Method for a TPC Based on fast Hough Transformation",
  institution = "Deutsches Elektronen-Synchrotron DESY",
  type = "LC Note",
  number = "LC-TOOL-2014-006",
  year = "2014",
  url = "https://bib-pubdb1.desy.de/record/193046",
}

@report{kleinwort_triplet,
  author = "Kleinwort, Claus",
  title = "A Track Finding Method for a TPC Based on Triplet Chains",
  institution = "Deutsches Elektronen-Synchrotron DESY",
  type = "LC Note",
  number = "LC-TOOL-2014-004",
  year = "2014",
  url = "https://bib-pubdb1.desy.de/record/193045",
}

@article{KleinwortGBL,
  author = "Kleinwort, Claus",
  title = "General Broken Lines as advanced track fitting method",
  journaltitle = "Nuclear Instruments and Methods in Physics Research",
  journalsubtitle = "Section A: Accelerators, Spectrometers, Detectors and Associated Equipment",
  shortjournal = "Nucl. Instrum. Methods Phys. Res.",
  series = "A",
  volume = "673",
  pages = "107-110",
  year = "2012",
  doi = "10.1016/j.nima.2012.01.024",
  eprint = "1201.4320",
  eprinttype = "arXiv",
  eprintclass = "physics.ins-det",
}

@article{Magboltz,
  author = "Biagi, S. F.",
  title = "Monte Carlo simulation of electron drift and diffusion in counting gases under the influence of electric and magnetic fields",
  journaltitle = "Nuclear Instruments and Methods in Physics Research",
  journalsubtitle = "Section A: Accelerators, Spectrometers, Detectors and Associated Equipment",
  shortjournal = "Nucl. Instrum. Methods Phys. Res.",
  series = "A",
  volume = "421",
  number = "1",
  pages = "234-240",
  year = "1999",
  publisher = "Elsevier",
  doi = "10.1016/S0168-9002(98)01233-9",
}

@article{YonamineJinst2014,
  author = "R. Yonamine and K. Fujii and K. Ikematsu and A. Ishikawa and T. Fusayasu and P. Gros and Y. Kato and S. Kawada and M. Kobayashi and T. Matsuda and O. Nitoh and R. D. Settles and A. Sugiyama and T. Takahashi and J. Tian and T. Watanabe",
  title = "Spatial resolutions of GEM TPC. A novel theoretical formula and its comparison to latest beam test data",
  journaltitle = "Journal of Instrumentation",
  shortjournal = "J. Instrum.",
  volume = "9",
  number = "03",
  pages = "C03002",
  year = "2014",
  publisher = "IOP Publishing",
  doi = "10.1088/1748-0221/9/03/c03002",
}

@report{AliceUpgradeTDR,
  author = "{The ALICE Collaboration}",
  collaboration = "ALICE",
  title = "Upgrade of the ALICE Time Projection Chamber",
  type = "Technical Design Report",
  institution = "CERN",
  number = "CERN-LHCC-2013-020. ALICE-TDR-016",
  year = "2013",
  url = "https://cds.cern.ch/record/1622286",
}

@article{Heed,
  author = "Smirnov, Igor B.",
  title = "Modeling of ionization produced by fast charged particles in gases",
  journaltitle = "Nuclear Instruments and Methods in Physics Research",
  journalsubtitle = "Section A: Accelerators, Spectrometers, Detectors and Associated Equipment",
  shortjournal = "Nucl. Instrum. Methods Phys. Res.",
  series = "A",
  volume = "554",
  number = "1",
  pages = "474-493",
  year = "2005",
  publisher = "Elsevier",
  doi = "10.1016/j.nima.2005.08.064",
}

@book{Blum:2008,
  author = "Blum, W. and Riegler, W. and Rolandi, L.",
  title = "Particle Detection with Drift Chambers",
  edition = "2",
  year = "2008",
  publisher = "Springer",
  series = "Particle Acceleration and Detection",
  isbn = "978-3-540-76684-1",
  issn = "1611-1052",
  doi = "10.1007/978-3-540-76684-1",
  lccn = "2007940836",
}

@article{Karlen2005,
  author = "D. Karlen and P. Poffenberger and G. Rosenbaum",
  title = "TPC performance in magnetic fields with GEM and pad readout",
  journaltitle = "Nuclear Instruments and Methods in Physics Research",
  journalsubtitle = "Section A: Accelerators, Spectrometers, Detectors and Associated Equipment",
  shortjournal = "Nucl. Instrum. Methods Phys. Res.",
  series = "A",
  volume = "555",
  number = "1",
  pages = "80-92",
  year = "2005",
  issn = "0168-9002",
  doi = "10.1016/j.nima.2005.09.008",
  url = "https://www.sciencedirect.com/science/article/pii/S0168900205018140",
}

@article{Abgrall201125,
  author = "{The T2K ND280 TPC Collaboration}",
  //author = "Abgrall, N. and Andrieu, B. and Baron, P. and Bene, P. and Berardi, V. and Beucher, J. and Birney, P. and Blaszczyk, F. and Blondel, A. and Bojechko, C. and Boyer, M. and Cadoux, F. and Calvet, D. and Catanesi, M. G. and Cervera, A. and Colas, P. and Broise, X. De La and Delagnes, E. and Delbart, A. and Marco, M. Di and Druillole, F. and Dumarchez, J. and Emery, S. and Escudero, L. and Faszer, W. and Ferrere, D. and Ferrero, A. and Fransham, K. and Gaudin, A. and Giganti, C. and Giomataris, I. and Giraud, J. and Goyette, M. and Hamano, K. and Hearty, C. and Henderson, R. and Herlant, S. and Ieva, M. and Jamieson, B. and Jover-Mañas, G. and Karlen, D. and Kato, I. and Konaka, A. and Laihem, K. and Langstaff, R. and Laveder, M. and Coguie, A. Le and Dortz, O. Le and Ross, M. Le and Lenckowski, M. and Lux, T. and Macaire, M. and Mahn, K. and Masciocchi, F. and Mazzucato, E. and Mezzetto, M. and Miller, A. and Mols, J.-Ph. and Monfregola, L. and Monmarthe, E. and Myslik, J. and Nizery, F. and Openshaw, R. and Perrin, E. and Pierre, F. and Pierrepont, D. and Poffenberger, P. and Popov, B. and Radicioni, E. and Ravonel, M. and Reymond, J.-M. and Ritou, J.-L. and Roney, M. and Roth, S. and S´anchez, F. and Sarrat, A. and Schroeter, R. and Stahl, A. and Stamoulis, P. and Steinmann, J. and Terhorst, D. and Terront, D. and Tvaskis, V. and Usseglio, M. and Vallereau, A. and Vasseur, G. and J.Wendland and G.Wikström and Zito, M.",
  title = "Time Projection Chambers for the T2K Near Detectors",
  collaboration = "T2K ND280 TPC",
  journaltitle = "Nuclear Instruments and Methods in Physics Research",
  journalsubtitle = "Section A: Accelerators, Spectrometers, Detectors and Associated Equipment",
  shortjournal = "Nucl. Instrum. Methods Phys. Res.",
  series = "A",
  volume = "637",
  number = "1",
  pages = "25-46",
  year = "2011",
  issn = "0168-9002",
  doi = "10.1016/j.nima.2011.02.036",
  url = "https://www.sciencedirect.com/science/article/pii/S0168900211003421",
  eprint = "1012.0865",
  eprinttype = "arXiv",
  eprintclass = "physics.ins-det",
}

@article{FMueller2017,
  author = "{The LCTPC Collaboration}",
  //author = "Attié, David and Behnke, Ties and Bellerive, Alain and Bezshyyko, Oleg and Bhattacharya, Deb Sankar and Bhattacharya, Purba and Bhattacharya, Sudeb and Caiazza, Stefano and Colas, Paul and Lentdecker, Gilles De and Dehmelt, Klaus and Desch, Klaus and Diener, Ralf and Dixit, Madhu and Fleck, Ivor and Fujii, Keisuke and Fusayasu, Takahiro and Ganjour, Serguei and Gao, Yuanning and Gros, Philippe and Hayman, Peter and Hedberg, Vincent and Ikematsu, Katsumasa and Jönsson, Leif and Kaminski, Jochen and Kato, Yukihiro and Kawada, Shin-ichi and Killenberg, Martin and Kleinwort, Claus and Kobayashi, Makoto and Krylov, Vladyslav and Li, Bo and Li, Yulan and Lundberg, Björn and Lupberger, Michael and Majumdar, Nayana and Matsuda, Takeshi and Mehdiyev, Rashid and Mjörnmark, Ulf and Müller, Felix and Münnich, Astrid and Mukhopadhyay, Supratik and Ogawa, Tomohisa and Oskarsson, Anders and Österman, Lennart and Peterson, Daniel and Riallot, Marc and Rosemann, Christoph and Roth, Stefan and Schade, Peter and Schäfer, Oliver and Settles, Ronald Dean and Shirazi, Amir Noori and Smirnova, Oxana and Sugiyama, Akira and Takahashi, Tohru and Tian, Junping and Timmermans, Jan and Titov, Maksym and Tsionou, Dimitra and Vauth, Annika and Wang, Wenxin and Watanabe, Takashi and Werthenbach, Ulrich and Yang, Yifan and Yang, Zhenwei and Yonamine, Ryo and Zenker, Klaus and Zhang, Fan",
  title = "A Time projection chamber with GEM-Based readout",
  collaboration = "LCTPC",
  journaltitle = "Nuclear Instruments and Methods in Physics Research",
  journalsubtitle = "Section A: Accelerators, Spectrometers, Detectors and Associated Equipment",
  shortjournal = "Nucl. Instrum. Methods Phys. Res.",
  series = "A",
  volume = "856",
  pages = "109-118",
  year = "2017",
  issn = "0168-9002",
  doi = "10.1016/j.nima.2016.11.002",
  url = "https://www.sciencedirect.com/science/article/pii/S0168900216311226",
  eprint = "1604.00935",
  eprinttype = "arXiv",
  eprintclass = "physics.ins-det",
}

@article{KOBAYASHI201941,
  author = "M. Kobayashi and T. Ogawa and A. Shoji and Y. Aoki and K. Ikematsu and P. Gros and T. Kawaguchi and D. Arai and M. Iwamura and K. Katsuki and A. Koto and M. Yoshikai and K. Fujii and T. Fusayasu and Y. Kato and S. Kawada and T. Matsuda and S. Narita and K. Negishi and H. Qi and R.D. Settles and A. Sugiyama and T. Takahashi and J. Tian and T. Watanabe and R. Yonamine",
  title = "Measurement of the electron transmission rate of the gating foil for the TPC of the ILC experiment",
  journaltitle = "Nuclear Instruments and Methods in Physics Research",
  journalsubtitle = "Section A: Accelerators, Spectrometers, Detectors and Associated Equipment",
  shortjournal = "Nucl. Instrum. Methods Phys. Res.",
  series = "A",
  volume = "918",
  pages = "41-53",
  year = "2019",
  issn = "0168-9002",
  doi = "10.1016/j.nima.2018.11.060",
}

@phdthesis{Mueller:301339,
  author = "Müller, Felix Johannes",
  title = "Development of a Triple GEM Readout Module for a Time Projection Chamber $\&$ Measurement Accuracies of Hadronic Higgs Branching Fractions in $\nu\nu$H at a 350 GeV ILC",
  year = "2016",
  editor = "Behnke, Ties and Garutti, Erika",
  editortype = "advisor",
  publisher = "Verlag Deutsches Elektronen-Synchrotron",
  school = "Universität Hamburg",
  issn = "1435-8085",
  doi = "10.3204/PUBDB-2016-02659",
  url = "https://bib-pubdb1.desy.de/record/301339",
}

@article{Allison:2006ve,
  author = "Allison, J. and Amako, K. and Apostolakis, J. and Araujo, H.
  and Dubois, P. Arce and Asai, M. and Barrand, G. and Capra, R. and Chauvie, S.
  and Chytracek, R. and Cirrone, G. A. P. and Cooperman, G. and Cosmo, G.
  and Cuttone, G. and Daquino, G. G. and Donszelmann, M. and Dressel, M.
  and Folger, G. and Foppiano, F. and Generowicz, J. and Grichine, V.
  and Guatelli, S. and Gumplinger, P. and Heikkinen, A. and Hrivnacova, I.
  and Howard, A. and Incerti, S. and Ivanchenko, V. and Johnson, T. and Jones, F.
  and Koi, T. and Kokoulin, R. and Kossov, M. and Kurashige, H. and Lara, V.
  and Larsson, S. and Lei, F. and Link, O. and Longo, F. and Maire, M.
  and Mantero, A. and Mascialino, B. and McLaren, I. and Lorenzo, P. Mendez
  and Minamimoto, K. and Murakami, K. and Nieminen, P. and Pandola, L.
  and Parlati, S. and Peralta, L. and Perl, J. and Pfeiffer, A. and Pia, M. G.
  and Ribon, A. and Rodrigues, P. and Russo, G. and Sadilov, S. and Santin, G.
  and Sasaki, T. and Smith, D. and Starkov, N. and Tanaka, S. and Tcherniaev, E.
  and Tome, B. and Trindade, A. and Truscott, P. and Urban, L. and Verderi, M.
  and Walkden, A. and Wellisch, J. P. and Williams, D. C. and Wright, D.
  and Yoshida, H.",
  title = "Geant4 developments and applications",
  journaltitle = "IEEE Transactions on Nuclear Science",
  shortjournal = "IEEE Trans. Nucl. Sci.",
  volume = "53",
  number = "1",
  pages = "270-278",
  year = "2006",
  issn = "0018-9499",
  doi = "10.1109/TNS.2006.869826",
}

@article{Allison:2016lfl,
  author = "Allison, J. and Amako, K. and Apostolakis, J. and Arce, P. and Asai, M.
  and Aso, T. and Bagli, E. and Bagulya, A. and Banerjee, S. and Barrand, G.
  and Beck, B. R. and Bogdanov, A. G. and Brandt, D. and Brown, J. M. C.
  and Burkhardt, H. and Canal, Ph. and Cano-Ott, D. and Chauvie, S. and Cho, K.
  and Cirrone, G. A. P. and Cooperman, G. and Cortés-Giraldo, M. A. and Cosmo, G.
  and Cuttone, G. and Depaola, G. and Desorgher, L. and Dong, X. and Dotti, A.
  and Elvira, V. D. and Folger, G. and Francis, Z. and Galoyan, A. and Garnier, L.
  and Gayer, M. and Genser, K. L. and Grichine, V. M. and Guatelli, S.
  and Guèye, P. and Gumplinger, P. and Howard, A. S. and H\vrivná\vcová, I.
  and Hwang, S. and Incerti, S. and Ivanchenko, A. and Ivanchenko, V. N.
  and Jones, F. W. and Jun, S. Y. and Kaitaniemi, P. and Karakatsanis, N.
  and Karamitros, M. and Kelsey, M. and Kimura, A. and Koi, T. and Kurashige, H.
  and Lechner, A. and Lee, S. B. and Longo, F. and Maire, M. and Mancusi, D.
  and Mantero, A. and Mendoza, E. and Morgan, B. and Murakami, K. and Nikitina, T.
  and Pandola, L. and Paprocki, P. and Perl, J. and Petrović, I. and Pia, M. G.
  and Pokorski, W. and Quesada, J. M. and Raine, M. and Reis, M. A. and Ribon, A.
  and Fira, A. Ristić and Romano, F. and Russo, G. and Santin, G. and Sasaki, T.
  and Sawkey, D. and Shin, J. I. and Strakovsky, I. I. and Taborda, A.
  and Tanaka, S. and Tomé, B. and Toshito, T. and Tran, H. N. and Truscott, P. R.
  and Urban, L. and Uzhinsky, V. and Verbeke, J. M. and Verderi, M.
  and Wendt, B. L. and Wenzel, H. and Wright, D. H. and Wright, D. M.
  and Yamashita, T. and Yarba, J. and Yoshida, H.",
  title = "Recent developments in \textsc{Geant4}",
  journaltitle = "Nuclear Instruments and Methods in Physics Research",
  journalsubtitle = "Section A: Accelerators, Spectrometers, Detectors and Associated Equipment",
  shortjournal = "Nucl. Instrum. Methods Phys. Res.",
  series = "A",
  volume = "835",
  pages = "186-225",
  year = "2016",
  issn = "0168-9002",
  doi = "10.1016/j.nima.2016.06.125",
  url = "https://www.sciencedirect.com/science/article/pii/S0168900216306957",
}

@article{Agostinelli:2002hh,
  author = "Agostinelli, S. and Allison, J. and Amako, K. and Apostolakis, J.
  and Araujo, H. and Arce, P. and Asai, M. and Axen, D. and Banerjee, S.
  and Barrand, G. and Behner, F. and Bellagamba, L. and Boudreau, J.
  and Broglia, L. and Brunengo, A. and Burkhardt, H. and Chauvie, S. and Chuma, J.
  and Chytracek, R. and Cooperman, G. and Cosmo, G. and Degtyarenko, P.
  and Dell'Acqua, A. and Depaola, G. and Dietrich, D. and Enami, R.
  and Feliciello, A. and Ferguson, C. and Fesefeldt, H. and Folger, G.
  and Foppiano, F. and Forti, A. and Garelli, S. and Giani, S.
  and Giannitrapani, R. and Gibin, D. and Cadenas, J. J. Gómez and González, I.
  and Abril, G. Gracia and Greeniaus, G. and Greiner, W. and Grichine, V.
  and Grossheim, A. and Guatelli, S. and Gumplinger, P. and Hamatsu, R.
  and Hashimoto, K. and Hasui, H. and Heikkinen, A. and Howard, A.
  and Ivanchenko, V. and Johnson, A. and Jones, F. W. and Kallenbach, J.
  and Kanaya, N. and Kawabata, M. and Kawabata, Y. and Kawaguti, M. and Kelner, S.
  and Kent, P. and Kimura, A. and Kodama, T. and Kokoulin, R. and Kossov, M.
  and Kurashige, H. and Lamanna, E. and Lampén, T. and Lara, V. and Lefebure, V.
  and Lei, F. and Liendl, M. and Lockman, W. and Longo, F. and Magni, S.
  and Maire, M. and Medernach, E. and Minamimoto, K. and de Freitas, P. Mora
  and Morita, Y. and Murakami, K. and Nagamatu, M. and Nartallo, R.
  and Nieminen, P. and Nishimura, T. and Ohtsubo, K. and Okamura, M.
  and O'Neale, S. and Oohata, Y. and Paech, K. and Perl, J. and Pfeiffer, A.
  and Pia, M. G. and Ranjard, F. and Rybin, A. and Sadilov, S. and Salvo, E. Di
  and Santin, G. and Sasaki, T. and Savvas, N. and Sawada, Y. and Scherer, S.
  and Sei, S. and Sirotenko, V. and Smith, D. and Starkov, N. and Stoecker, H.
  and Sulkimo, J. and Takahata, M. and Tanaka, S. and Tcherniaev, E.
  and Tehrani, E. Safai and Tropeano, M. and Truscott, P. and Uno, H.
  and Urban, L. and Urban, P. and Verderi, M. and Walkden, A. and Wander, W.
  and Weber, H. and Wellisch, J. P. and Wenaus, T. and Williams, D. C.
  and Wright, D. and Yamada, T. and Yoshida, H. and Zschiesche, D.",
  title = "\textsc{Geant4}---a simulation toolkit",
  journaltitle = "Nuclear Instruments and Methods in Physics Research",
  journalsubtitle = "Section A: Accelerators, Spectrometers, Detectors and Associated Equipment",
  shortjournal = "Nucl. Instrum. Methods Phys. Res.",
  series = "A",
  volume = "506",
  number = "3",
  pages = "250-303",
  year = "2003",
  issn = "0168-9002",
  doi = "10.1016/S0168-9002(03)01368-8",
  url = "https://www.sciencedirect.com/science/article/pii/S0168900203013688",
}

@report{Kleinwort:395416,
  author = "Kleinwort, Claus",
  title = "A combined track and hit finding method for a TPC based on local road search with pad pulses",
  type = "LC Note",
  number = "LC-TOOL-2017-001",
  year = "2017",
  institution = "Deutsches Elektronen-Synchrotron DESY",
  doi = "10.3204/PUBDB-2017-12090",
  url = "https://bib-pubdb1.desy.de/record/395416",
}

@article{DIENER2019265,
  author = "Diener, R. and Dreyling-Eschweiler, J. and Ehrlichmann, H. and Gregor, I.-M. and Kötz, U. and Krämer, U. and Meyners, N. and Potylitsina-Kube, N. and Schütz, A. and Schütze, P. and Stanitzki, M.",
  title = "The DESY II Test Beam Facility",
  journaltitle = "Nuclear Instruments and Methods in Physics Research",
  journalsubtitle = "Section A: Accelerators, Spectrometers, Detectors and Associated Equipment",
  shortjournal = "Nucl. Instrum. Methods Phys. Res.",
  series = "A",
  volume = "922",
  pages = "265-286",
  year = "2019",
  issn = "0168-9002",
  doi = "10.1016/j.nima.2018.11.133",
  eprint = "1807.09328",
  eprinttype = "arXiv",
  eprintclass = "physics.ins-det",
  url = "https://www.sciencedirect.com/science/article/pii/S0168900218317868",
}

@inproceedings{DEDXCLUSTER,
  author = "Hauschild, Michael",
  title = "dE/dx and Particle ID Performance with Cluster Counting",
  eventtitle = "International Linear Collider Workshop",
  eventtitleaddon = "ILC-ECFA and GDE Joint Meeting",
  eventdate = "2006-11-06/2006-11-10",
  venue = "Valencia, Spain",
  //url = "https://ific.uv.es/~ilc/ECFA-GDE2006/",
  url = "https://agenda.linearcollider.org/event/1049/contributions/506/",
}

@article{SourceExtractor,
  author = "Bertin, E. and Arnouts, S.",
  title = "SExtractor: Software for source extraction",
  journaltitle = "Astronomy and Astrophysics",
  journalsubtitle = "Supplemental Series",
  shortjournal = "Astron. Astrophys. Suppl. Ser.",
  //series = "Supplemental Series",
  //series = "Suppl. Ser.",
  volume = "117",
  number = "2",
  pages = "393-404",
  year = "1996",
  doi = "10.1051/aas:1996164",
  url = "https://www.astromatic.net/software/sextractor",
}

@inproceedings{AsianGEM_dEdx,
  author = "Shoji, Aiko",
  title = "Measurement of dE/dx resolution of TPC prototype with gating GEM exposed to an electron beam",
  booktitle = "Proceedings of the International Workshop on Future Linear Colliders",
  series = "eConf",
  number = "C17-10-23.2",
  year = "2018",
  //eventtitle = "International Workshop on Future Linear Colliders",
  eventtitleaddon = "LCWS2017",
  eventdate = "2017-10-23/2017-10-27",
  venue = "Strasbourg, France",
  eprint = "1801.04499",
  eprinttype = "arXiv",
  eprintclass = "physics.ins-det",
}

@article{LIGTENBERG201818,
  author = "Ligtenberg, C. and Heijhoff, K. and Bilevych, Y. and Desch, K. and van der Graaf, H. and Gruber, M. and Hartjes, F. and Kaminski, J. and van der Kolk, N. and Kluit, P.M. and Raven, G. and Scharenberg, L. and Schiffer, T. and Schmidt, S. and Timmermans, J.",
  title = "Performance of a GridPix detector based on the Timepix3 chip",
  journaltitle = "Nuclear Instruments and Methods in Physics Research",
  journalsubtitle = "Section A: Accelerators, Spectrometers, Detectors and Associated Equipment",
  shortjournal = "Nucl. Instrum. Methods Phys. Res.",
  series = "A",
  volume = "908",
  pages = "18-23",
  year = "2018",
  issn = "0168-9002",
  doi = "10.1016/j.nima.2018.08.012",
  url = "https://www.sciencedirect.com/science/article/pii/S0168900218309549",
  eprint = "1808.04565",
  eprinttype = "arXiv",
  eprintclass = "physics.ins-det",
}

@article{Yonamine_2014,
  author = "Yonamine, R. and Fujii, K. and Ikematsu, K. and Ishikawa, A. and Fusayasu, T. and Gros, P. and Kato, Y. and Kawada, S. and Kobayashi, M. and Matsuda, T. and Nitoh, O. and Settles, R. D. and Sugiyama, A. and Takahashi, T. and Tian, J. and Watanabe, T.",
  title = "Spatial resolutions of GEM TPC. A novel theoretical formula and its comparison to latest beam test data",
  journaltitle = "Journal of Instrumentation",
  shortjournal = "J. Instrum.",
  issuetitle = "Proceedings of the 3rd International Conference on Micro Pattern Gaseous Detectors (MPGD2013)",
  volume = "9",
  number = "03",
  pages = "C03002",
  year = "2014",
  month = "3",
  publisher = "IOP Publishing",
  doi = "10.1088/1748-0221/9/03/c03002",
}

@inproceedings{DEDXMM,
  author = "Colas, P.",
  title = "Measurement of dE/dx resolution with resistive Micromegas",
  year = "2018",
  eventtitle = "Asian Linear Collider Workshop",
  eventtitleaddon = "ALCW2018",
  eventdate = "2018-05-28/2018-06-01",
  venue = "Fukuoka, Japan",
  url = "https://agenda.linearcollider.org/event/7826/contributions/41602/",
}

@article{allison1980,
  author = "Allison, W. W. M. and Cobb, J. H.",
  title = "Relativistic Charged Particle Identification by Energy Loss",
  journaltitle = "Annual Review of Nuclear and Particle Science",
  shortjournal = "Annu. Rev. Nucl. Part. Sci.",
  volume = "30",
  number = "1",
  pages = "253-298",
  year = "1980",
  doi = "10.1146/annurev.ns.30.120180.001345",
}

@article{walenta1979_ieee,
  author = "Walenta, A. H.",
  title = "The Time Expansion Chamber and Single Ionization Cluster Measurement",
  journaltitle = "IEEE Transactions on Nuclear Science",
  shortjournal = "IEEE Trans. Nucl. Sci.",
  publisher = "IEEE",
  volume = "26",
  number = "1",
  pages = "73-80",
  year = "1979",
  doi = "10.1109/TNS.1979.4329616",
  ISSN = "1558-1578",
}

@article{walenta1979_nim,
  author = "Walenta, A. H. and Fischer, J. and Okuno, H. and Wang, C. L.",
  title = "Measurement of the ionization loss in the region of relativistic rise for noble and molecular gases",
  journaltitle = "Nuclear Instruments and Methods",
  shortjournal = "Nucl. Instrum. Methods",
  volume = "161",
  number = "1",
  pages = "45-58",
  year = "1979",
  issn = "0029-554X",
  doi = "10.1016/0029-554X(79)90360-4",
}

@article{opal1987_dedx,
  author = "Breuker, H. and Fischer, M. and Hauschild, M. and Hartmann, H. and B. Wünsch and Boerner, H. and Burckhart, J. and Dittmar, M. and R. Hammarström and Heuer, D. and Michelini, A. and Plane, E. and Ö. Runolfsson and Schaile, D. and Weisz, S. and Zankel, K. and Ludwig, J. and Mohr, W. and F. Röhner and Runge, K. and Schaile, O. and Schwarz, J. and Stier, E. and Weltin, A. and Bock, P. and Heintze, J. and Igo-Kemenes, P. and Lennert, P. and Wagner, A.",
  title = "Particle identification with the OPAL jet chamber in the region of the relativistic rise",
  journaltitle = "Nuclear Instruments and Methods in Physics Research",
  journalsubtitle = "Section A: Accelerators, Spectrometers, Detectors and Associated Equipment",
  shortjournal = "Nucl. Instrum. Methods Phys. Res.",
  series = "A",
  volume = "260",
  number = "2",
  pages = "329-342",
  year = "1987",
  issn = "0168-9002",
  doi = "10.1016/0168-9002(87)90097-0",
}

@article{PhysRevD.98.030001,
  //author = "Tanabashi, M. and Hagiwara, K. and Hikasa, K. and Nakamura, K.
  and Sumino, Y. and Takahashi, F. and Tanaka, J. and Agashe, K. and Aielli, G.
  and Amsler, C. and Antonelli, M. and Asner, D. M. and Baer, H. and Banerjee, Sw.
  and Barnett, R. M. and Basaglia, T. and Bauer, C. W. and Beatty, J. J.
  and Belousov, V. I. and Beringer, J. and Bethke, S. and Bettini, A.
  and Bichsel, H. and Biebel, O. and Black, K. M. and Blucher, E.
  and Buchmuller, O. and Burkert, V. and Bychkov, M. A. and Cahn, R. N.
  and Carena, M. and Ceccucci, A. and Cerri, A. and Chakraborty, D.
  and Chen, M.-C. and Chivukula, R. S. and Cowan, G. and Dahl, O.
  and D'Ambrosio, G. and Damour, T. and de Florian, D. and de Gouvêa, A.
  and DeGrand, T. and de Jong, P. and Dissertori, G. and Dobrescu, B. A.
  and D'Onofrio, M. and Doser, M. and Drees, M. and Dreiner, H. K.
  and Dwyer, D. A. and Eerola, P. and Eidelman, S. and Ellis, J. and Erler, J.
  and Ezhela, V. V. and Fetscher, W. and Fields, B. D. and Firestone, R.
  and Foster, B. and Freitas, A. and Gallagher, H. and Garren, L.
  and Gerber, H.-J. and Gerbier, G. and Gershon, T. and Gershtein, Y.
  and Gherghetta, T. and Godizov, A. A. and Goodman, M. and Grab, C.
  and Gritsan, A. V. and Grojean, C. and Groom, D. E. and Grünewald, M.
  and Gurtu, A. and Gutsche, T. and Haber, H. E. and Hanhart, C. and Hashimoto, S.
  and Hayato, Y. and Hayes, K. G. and Hebecker, A. and Heinemeyer, S.
  and Heltsley, B. and Hernández-Rey, J. J. and Hisano, J. and Höcker, A.
  and Holder, J. and Holtkamp, A. and Hyodo, T. and Irwin, K. D.
  and Johnson, K. F. and Kado, M. and Karliner, M. and Katz, U. F.
  and Klein, S. R. and Klempt, E. and Kowalewski, R. V. and Krauss, F.
  and Kreps, M. and Krusche, B. and Kuyanov, Yu. V. and Kwon, Y. and Lahav, O.
  and Laiho, J. and Lesgourgues, J. and Liddle, A. and Ligeti, Z. and Lin, C.-J.
  and Lippmann, C. and Liss, T. M. and Littenberg, L. and Lugovsky, K. S.
  and Lugovsky, S. B. and Lusiani, A. and Makida, Y. and Maltoni, F.
  and Mannel, T. and Manohar, A. V. and Marciano, W. J. and Martin, A. D.
  and Masoni, A. and Matthews, J. and Meißner, U.-G. and Milstead, D.
  and Mitchell, R. E. and Mönig, K. and Molaro, P. and Moortgat, F.
  and Moskovic, M. and Murayama, H. and Narain, M. and Nason, P. and Navas, S.
  and Neubert, M. and Nevski, P. and Nir, Y. and Olive, K. A. and {Pagan Griso}, S.
  and Parsons, J. and Patrignani, C. and Peacock, J. A. and Pennington, M.
  and Petcov, S. T. and Petrov, V. A. and Pianori, E. and Piepke, A.
  and Pomarol, A. and Quadt, A. and Rademacker, J. and Raffelt, G.
  and Ratcliff, B. N. and Richardson, P. and Ringwald, A. and Roesler, S.
  and Rolli, S. and Romaniouk, A. and Rosenberg, L. J. and Rosner, J. L.
  and Rybka, G. and Ryutin, R. A. and Sachrajda, C. T. and Sakai, Y.
  and Salam, G. P. and Sarkar, S. and Sauli, F. and Schneider, O.
  and Scholberg, K. and Schwartz, A. J. and Scott, D. and Sharma, V.
  and Sharpe, S. R. and Shutt, T. and Silari, M. and Sjöstrand, T. and Skands, P.
  and Skwarnicki, T. and Smith, J. G. and Smoot, G. F. and Spanier, S.
  and Spieler, H. and Spiering, C. and Stahl, A. and Stone, S. L.
  and Sumiyoshi, T. and Syphers, M. J. and Terashi, K. and Terning, J.
  and Thoma, U. and Thorne, R. S. and Tiator, L. and Titov, M.
  and Tkachenko, N. P. and Törnqvist, N. A. and Tovey, D. R. and Valencia, G.
  and {Van de Water}, R. and Varelas, N. and Venanzoni, G. and Verde, L.
  and Vincter, M. G. and Vogel, P. and Vogt, A. and Wakely, S. P.
  and Walkowiak, W. and Walter, C. W. and Wands, D. and Ward, D. R.
  and Wascko, M. O. and Weiglein, G. and Weinberg, D. H. and Weinberg, E. J.
  and White, M. and Wiencke, L. R. and Willocq, S. and Wohl, C. G.
  and Womersley, J. and Woody, C. L. and Workman, R. L. and Yao, W.-M.
  and Zeller, G. P. and Zenin, O. V. and Zhu, R.-Y. and Zhu, S.-L.
  and Zimmermann, F. and Zyla, P. A. and Anderson, J. and Fuller, L.
  and Lugovsky, V. S. and Schaffner, P.",
  author = "{The Particle Data Group}",
  collaboration = "Particle Data Group",
  title = "Review of Particle Physics",
  journaltitle = "Physical Review",
  shortjournal = "Phys. Rev.",
  series = "D",
  volume = "98",
  number = "3",
  pages = "030001",
  pagetotal = "1898",
  year = "2018",
  publisher = "American Physical Society",
  doi = "10.1103/PhysRevD.98.030001",
  url = "https://link.aps.org/doi/10.1103/PhysRevD.98.030001",
}

@phdthesis{FedorchPhD,
  author = "Fedorchuk, Oleksiy",
  editor = "Behnke, Ties and Garutti, Erika",
  editortype = "advisor",
  title = "Investigations of the long-term stability of a Gas Electron Multipliers and double hit resolution for the highly granular Time-Projection Chamber",
  school = "Universität Hamburg",
  publisher = "Staats- und Universitätsbibliothek Hamburg ``Carl von Ossietzky''",
  year = "2019",
  url = "https://ediss.sub.uni-hamburg.de/handle/ediss/8760",
}
}

\end{document}